\documentstyle[12pt,aaspp4]{article}
\newcommand{\pld}{$\phi_{\rm LD}$}
\newcommand{\pud}{$\phi_{\rm UD}$}
\newcommand{\teff}{T$_{\rm eff}$}

\slugcomment{to appear in the March 2000 issue of The Astronomical Journal}
\begin{document}

\title{Improved Color--Temperature Relations and Bolometric Corrections for Cool Stars}
\author{M. L. Houdashelt,\altaffilmark{1} R. A. Bell}
\affil{Department of Astronomy, University of Maryland, College Park, MD 20742-2421}
\authoremail{mlh@pha.jhu.edu; roger@astro.umd.edu}

\and

\author{A. V. Sweigart}
\affil{Code 681, NASA/Goddard Space Flight Center, Greenbelt, MD 20771}
\authoremail{sweigart@gsfc.nasa.gov}

\altaffiltext{1}{Current address: Department of Physics \& Astronomy, Johns Hopkins University, 3400 North Charles Street, Baltimore, MD 21218}

\begin{abstract}

We present new grids of colors and bolometric corrections for F--K stars having
4000~K~$\leq$~\teff~$\leq$~6500~K, 0.0~$\leq$~log~g~$\leq$~4.5 and
--3.0~$\leq$~[Fe/H]~$\leq$~0.0.  A companion paper extends these calculations
into the M~giant regime (3000~K~$\leq$~\teff~$\leq$~4000~K).  Colors are
tabulated for Johnson U--V and B--V; Cousins V--R and V--I; Johnson-Glass V--K,
J--K and H--K; and CIT/CTIO V--K, J--K, H--K and CO.
We have developed these color-temperature relations by convolving synthetic
spectra with the best-determined, photometric filter-transmission-profiles.  The
synthetic spectra have been computed with the SSG spectral synthesis code using
MARCS stellar atmosphere models as input.  Both of these codes have been
improved substantially, especially at low temperatures, through the
incorporation of new opacity data.  The resulting synthetic colors have been
put onto the observational systems by applying color calibrations derived from
models and photometry of field stars which have effective
temperatures determined by the infrared-flux method.
These color calibrations have zero points which change most of the original
synthetic colors by less than 0.02 mag, and
the corresponding slopes generally alter the colors by less than 5\%.
The adopted temperature scale (Bell \& Gustafsson 1989) is
confirmed by the extraordinary agreement between the predicted and observed
angular diameters of these field stars, indicating that the differences between
the synthetic colors and the photometry of the field stars are not due to errors
in the effective temperatures adopted for these stars.
Thus, we have derived empirical color-temperature relations from the field-star
photometry, which we use as one test of our calibrated, theoretical,
solar-metallicity, color-temperature relations.  Except for the coolest dwarfs
(\teff~$<$~5000~K), our calibrated model colors
are found to match these relations, as well as the empirical relations of
others, quite well,
and our calibrated, 4~Gyr, solar-metallicity isochrone also provides a good
match to color-magnitude diagrams of M67.  We regard this 
as evidence that our calibrated colors can be applied to many astrophysical
problems, including modelling the integrated light of galaxies.
Because there are indications that the dwarfs cooler than 5000~K may require
different optical color calibrations than the other stars, we present
additional colors for our coolest dwarf models which account for this
possibility.

\end{abstract}

\keywords{stars: fundamental parameters --- stars: late-type --- stars: atmospheres --- stars: evolution --- infrared: stars}

\section{Introduction}

Color-temperature (CT) relations and bolometric corrections (BCs) are often
used to infer the physical characteristics of stars from their photometric
properties and, even more commonly, to translate
isochrones from the theoretical (effective temperature, luminosity) plane
into the observational (color, magnitude) plane.  The latter
allows the isochrones to be compared to observational data to estimate
the ages, reddenings and chemical compositions of star clusters and to test
the theoretical treatment of such stellar evolutionary phenomena as convection
and overshooting.  Isochrones are also used in evolutionary synthesis to model
the integrated light of simple stellar populations, coeval groups of stars 
having the same (initial) chemical composition, and accurate CT relations and
BCs are needed to reliably predict the observable properties of these stellar
systems.

Theoretical color-temperature relations are generally produced by convolving
models of photometric filter-transmission-profiles with synthetic spectra of
stars having a range of effective temperatures, surface gravities, and/or
chemical compositions.  Stellar atmosphere models are an integral part of this
process because the synthetic spectra are either produced as part of the model
atmosphere calculations themselves or come from later computations in which
the model atmospheres are used.  Indeed, recent tabulations of theoretical
color-temperature relations, such as those of \cite{bk92} and Bessell
et al.~(1998), differ mainly in the details of the model
atmosphere calculations (input physics, opacities, equation of state, etc.),
although there are differences in the adopted filter profiles and other
aspects of the synthetic color measurements as well.

VandenBerg \& Bell (1985; hereafter \cite{vb85}) and Bell \& Gustafsson
(1989; hereafter \cite{bg89}) have published
color-temperature relations for cool dwarfs and cool giants, respectively,
which were derived from a combination of MARCS model atmospheres (\cite{gben},
\cite{begn}) and synthetic spectra computed with the SSG spectral synthesis
code (Bell \& Gustafsson 1978; hereafter \cite{bg78};
\cite{gb79}; \cite{bg89}).  However,
since the work of \cite{vb85} and \cite{bg89}, substantial improvements have
been made to the MARCS and SSG computer codes, the opacity data employed by
each (especially at low temperatures) and the spectral line lists
used in SSG.  In addition, some of the photometric filter-transmission-profiles
used in the synthetic color measurements have been replaced by more recent
determinations, and the calibration of the synthetic colors has been improved.
Thus, as part of our evolutionary synthesis program, which will be fully
described in a forthcoming paper (Houdashelt et al.~2001; hereafter
\cite{houdy3}), we have calculated an improved, comprehensive set of MARCS/SSG
color-temperature relations and bolometric corrections for stars cooler than
6500~K.

We present here the colors and BCs for a grid of F, G~and K~stars having
4000~K~$\leq$~\teff$~\leq$~6500~K, 0.0~$\leq$~log~g~$\leq$~4.5 and 
--3.0~$\leq$~[Fe/H]~$\leq$~0.0.  We discuss the improvements made to
the stellar modelling and compare our synthetic spectra of the Sun and Arcturus
to spectral atlases of these stars at selected infrared wavelengths.  We
also describe the measurement and calibration of the synthetic
colors and show the good agreement between the resulting CT relations and their
empirical counterparts derived from observations of field stars.
A cooler grid, representing the M~giants, is presented in a companion paper
(Houdashelt et al.~2000; hereafter \cite{houdy2}).

The content of this paper is structured as follows.  In Section 2, the
computation of the stellar atmosphere models and synthetic spectra is described,
emphasizing the updated opacity data and other recent improvements in these
calculations.
Section 3 discusses the calibration of the synthetic colors.  We first reaffirm
the effective temperature scale adopted by \cite{bg89} and then use it to derive
the color calibrations required to put the synthetic colors onto the
observational systems. The significance of these calibrations is illustrated
through
comparisons of a 4 Gyr, solar-metallicity isochrone and photometry of M67.
In Section 4, we present the improved grid of color-temperature relations and
compare them to observed trends and to previous MARCS/SSG results.
A summary is given in Section 5.

\section{The Models}

To model a star of a given effective temperature, surface
gravity and chemical composition, a MARCS stellar atmosphere is calculated and
is then used in SSG to produce a synthetic spectrum.
In this section, we describe these calculations more fully and how they
improve upon previous MARCS/SSG work.  Because we display and discuss some
newly-calculated isochrones later in this
paper, we start by briefly describing the derivation of these isochrones and
the stellar evolutionary tracks used in their construction.

\subsection{The Stellar Evolutionary Tracks and Isochrones}

The stellar evolutionary tracks were calculated with the computer code and
input physics described by \cite{sweigart} and references therein.
The tracks were calibrated by matching a 4.6~Gyr, 1~M$_\odot$ model to the known
properties of the Sun.  To simultaneously reproduce
the solar luminosity, radius and Z/X ratio of \cite{grevesse} required 
Z$_\odot$~=~0.01716, Y$_\odot$~=~0.2798 and $\alpha$~=~1.8452, where X, Y and
Z are the mass fractions of hydrogen, helium and metals, respectively, and
$\alpha$ is the convective mixing-length-to-pressure-scale-height ratio.
The surface pressure boundary condition was specified by assuming a scaled,
solar T($\tau$) relation.
Diffusion and convective overshooting were not included in these calculations.

The solar-metallicity isochrones shown later in this paper are taken from the
set of isochrones we calculated for use in our evolutionary synthesis program.
They were produced by interpolating among stellar
evolutionary tracks with masses ranging from 0.2~M$_\odot$ through
1.5~M$_\odot$ using the method of equivalent-evolutionary points (see e.g.,
\cite{bvb92}).  Further details of the derivation of these isochrones and
evolutionary tracks will be given in a future paper presenting our
evolutionary synthesis models (\cite{houdy3}).

\subsection{The MARCS Stellar Atmosphere Models}

{\sc MARCS} computes a flux-constant, homogeneous, plane-parallel atmosphere
assuming hydrostatic equilibrium and LTE.
The continuous opacity sources used in the Maryland version of this program
include H$^{-}$, \ion{H}{1}, H$_2^{-}$, H$_2^+$,
He$^{-}$, \ion{C}{1}, \ion{Mg}{1}, \ion{Al}{1}, \ion{Si}{1}, \ion{Fe}{1},
electron scattering, and Rayleigh scattering by \ion{H}{1} and H$_2$.
In addition, for $\lambda <$ 7200 \AA, the opacity from atomic lines, as well
as that due to molecular lines of MgH, CH, OH, NH, and the violet system of CN,
is included in the form of an opacity distribution function (ODF).
At longer wavelengths, an ODF representing the molecular lines of CO and the red
system of CN supplements the continuous opacities.  The main improvements made
to the MARCS opacity data since the work of \cite{bg89} are the use
of the H$^-$ free-free opacity of \cite{bb87}, replacing that of \cite{df66};
the addition of continuous opacities from the Opacity Project for
\ion{Mg}{1}, \ion{Al}{1} and \ion{Si}{1} and from \cite{dragon} for \ion{Fe}{1};
and the use of detailed ODFs over the entire wavelength range from
900--7200~\AA\ (earlier MARCS models used only schematic ODFs between 900
and~3000~\AA).

For all of the MARCS models, a value of 1.6 was used for the
mixing-length-to-pressure-scale-height ratio, and the {\it y} parameter, which
describes the transparency of convective bubbles (\cite{henyey}), was taken to
be 0.076.  In general, we have calculated an ODF of the appropriate metallicity
to use in computing the model atmospheres, and we have adopted solar abundance
ratios (but see Sections~3.2.2 and~2.4 for exceptions to these two guidelines,
respectively).

\subsection{The SSG Synthetic Spectra}

Unless otherwise specified, the synthetic spectra discussed in this
paper have been calculated at 0.1~\AA\ resolution
and in two pieces, optical and infrared (IR).  The optical portion of the
spectrum covers wavelengths from 3000--12000~\AA, and the IR section
extends from 1.0--5.1~$\mu$m (the overlap is required for calculating J-band
magnitudes).  In addition, the microturbulent velocity, $\xi$, used to 
calculate each synthetic spectrum has been
derived from the star's surface gravity using the field-star relation,
$\xi$~=~2.22~--~0.322~log~g (Gratton et al.~1996), and the chemical composition
used in the spectral synthesis was always the same as that adopted for the
corresponding MARCS model atmosphere.

The spectral computations used a version of SSG
which has been continually updated since the work
of \cite{bg89}.  Here, the \cite{bb87} H$^-$ free-free opacity data replaced
that of \cite{bkm75}, although the differences are small.  The continuous
opacities of \ion{Mg}{1}, \ion{Al}{1}, \ion{Si}{1} and \ion{Fe}{1} described
for the MARCS models have also been incorporated into SSG, as have 
continuous opacities for OH and CH (\cite{kur87}).

We used an updated version of the atomic and molecular spectral line list
denoted as the Bell ``N'' list by Bell et al.~(1994; hereafter \cite{bpt94}).
This line list has been improved through further detailed comparisons of
synthetic and empirical, high-resolution spectra.
For $\lambda <$~1.0~$\mu$m, it is supplemented by
spectral lines of Ca, Sc, Ti, V, Cr, Mn, Fe, Co and Ni which have been culled
from the compilation of \cite{k91} in the manner described in \cite{bpt94}.
Molecular data for the vibration-rotation bands of CO were taken from
\cite{goor}.  We omit H$_2$O lines from our calculations, but molecular lines
from the $\alpha$, $\beta$, $\gamma$, $\gamma^{\prime}$, $\delta$, $\epsilon$
and $\phi$ bands of TiO have been included in all of the synthetic spectra;
the latter have been given special consideration so that the observed
relationship between TiO band depth and spectral type in M~giants is reproduced
in the synthetic spectra.
A complete explanation of the sources of the TiO line data and the treatment
of the TiO bands in general is given in our companion paper presenting
synthetic spectra of M~giants (\cite{houdy2}).

Spectral line list improvements have also been determined by comparing
a synthetic spectrum of Arcturus ($\alpha$ Boo) to the Arcturus atlas
(\cite{arcatlas}) and by comparing
a synthetic solar spectrum to the solar atlases of
\cite{del73}, its digital successors from the National Solar Observatory,
and the solar atlas obtained by the ATMOS experiment aboard the space
shuttle (\cite{fn89}).
Identification of the unblended solar spectral lines, especially in the J and H
bandpasses, was aided by the compilations of \cite{sol90} and \cite{rams95};
the relevant atomic data for these lines was taken from \cite{bie85a}
\cite{bie85b} \cite{bie86}.  \cite{g92} has also identified many of
the lines in the ATMOS
spectrum, but few laboratory oscillator strengths are available for these
lines.  In addition, \cite{jl90} (1990)
reported about 360 new \ion{Fe}{1} lines in the infrared, identifying them as
transitions between the
3d$^6$4s($^6$D)4d and 3d$^6$4s($^6$D)4f states.  More than 200 of these lines coincide in
wavelength with lines in the solar spectrum, but only 16 of them are included
in the laboratory gf measurements of \ion{Fe}{1} lines by \cite{ob91}.
While some of the line identifications may be in error, \cite{jl90} have
checked their
identifications by comparing the line intensities from the laboratory source
with those in the solar spectrum.  They found that only four of their lines
were stronger in the solar spectrum than inferred from the laboratory data,
indicating that coincidence in wavelength implies a high probability of correct
identification.

Additional sources of atomic data were \cite{nave94} for \ion{Fe}{1}, \cite{litz93}
for \ion{Ni}{1}, \cite{dav78} for \ion{V}{1}, \cite{tak90} for \ion{Mn}{1},
\cite{fors91} for \ion{Ti}{1}, and \cite{k91}.  Opacity Project gf-value
calculations were used for lines of \ion{Na}{1}, \ion{Mg}{1}, \ion{Al}{1},
\ion{Si}{1}, \ion{S}{1} and \ion{Ca}{1}.
However, in view of the overall dearth of atomic data, ``astrophysical'' gf
values have been
found for many lines by fitting the synthetic and observed solar spectra.

Probably the greatest uncertainty remaining in the synthetic spectra is the
``missing ultraviolet (UV) opacity problem,''  which has been known to exist
for some time
(see e.g., \cite{gb79}).  \cite{holweger} speculated that this missing opacity
could be caused by a forest of weak Fe lines in the UV, and in fact, \cite{bk92}
have claimed to have ``solved''
the problem through the use of a new, larger spectral line list.  However,
this claim rests solely on the fact that
the models calculated using this new list produce
better agreement between the synthetic and empirical color-color relations of
field stars.  While this improvement is evident and
the UV opacity has definitely been enhanced by the new line list, it is also
clear that the
missing UV opacity has not been ``found.''  \cite{bpt94} have shown that
many of the spectral lines which appear in the new list are either undetected or
are much weaker in the observed solar spectrum than they are in the solar
spectrum synthesized from this line list.

\cite{bb98} have recently examined the solar abundance of Be, which has been
claimed to be depleted with respect to the meteoritic abundance.  They argue
that the OH lines of the A--X system, which appear in the same spectral region
as the \ion{Be}{2} lines near 3130 \AA, should yield oxygen abundances for the
Sun which match those derived from the vibration-rotation lines of OH in the
near-infrared.  To produce such agreement requires an increase in the
continuous opacity in the UV corresponding, for example, to a thirty-fold 
increase in
the bound-free opacity of \ion{Fe}{1}.  Such an opacity enhancement not only
brings the solar Be abundance in line with the meteoritic value, but it also
improves the agreement of the model fluxes with the limb-darkening behavior
of the \ion{Be}{2} lines and the solar fluxes measured by the Solstice
experiment (\cite{woods}).  However, this is such a large opacity discrepancy
that we have not included it in the models presented here.  Thus, we expect our
synthetic U--V (and possibly B--V) colors to show the effects of insufficient
UV opacity. For this reason, we recommend that the U--V colors presented in
this paper be used with caution.

After this paper was written, it was found that recent \ion{Fe}{1}
photoionization cross-sections calculated by \cite{bautista} are much larger
than those of Dragon \& Mutschlecner (1980), which we have used in our models.
In the region of the spectrum encompassing the aforementioned \ion{Be}{2}
lines ($\sim$3130 \AA), these new cross-sections cause a reduction of 15\% in
the solar continuous flux; this represents about half of the missing UV
opacity at these wavelengths.  Further work
is being carried out using Bautista's opacity data, with the expectation of
improving both our model atmospheres and synthetic spectra.

\subsection{Mixing}

The observed abundance patterns in low-mass red giants having approximately
solar metallicity indicate that these stars can mix CNO-processed material
from the deep interior outward into the stellar atmosphere.  This mixing is
in addition to that which accompanies the first dredge-up at the beginning of
the red-giant-branch (RGB) phase of evolution.
We have included the effects of such mixing in both the stellar atmosphere
models and the synthetic spectra of the brighter red giants
by assuming [C/Fe]~=~--0.2, [N/Fe]~=~+0.4 and
$^{12}$C/$^{13}$C~=~14 (the ``unmixed'' value is 89) for these stars.  These
quantities are deduced from abundance determinations in G8--K3~field giants
(\cite{kgwh}) and in field M~giants (\cite{sl90}), as well as from carbon
isotopic
abundance ratios in open cluster stars (e.g., \cite{gilroy} for M67 members).

In our evolutionary synthesis program, we incorporate mixing for all stars
more evolved than the ``bump'' in the RGB luminosity function which occurs
when the hydrogen-burning shell, moving outward in mass, encounters the chemical
composition discontinuity produced by the deep inward penetration of the
convective envelope during the preceding first-dredge-up.  \cite{sweigart79}
have argued that, prior to this point, mixing would be inhibited by the mean
molecular weight barrier caused by this composition discontinuity.  The M67
observations mentioned above and the work of \cite{c94} \cite{c95} and
\cite{c98} also indicate that this extra mixing first appears near the RGB bump.
For simplicity, we assume that the change in composition due to mixing occurs
instantaneously after a star has evolved to this point. 

When modelling a group of stars of known age and metallicity, it is
straightforward to determine where the RGB bump occurs along the
appropriate isochrone and then to incorporate mixing in the models of the stars
more evolved than the bump.  However, for a field star of unknown age, this
distinction is not nearly as clear because the position of the RGB bump
in the HR diagram is a function of both age and metallicity.  To determine
which of our field-star and grid models should include mixing effects, we
have used our solar-metallicity isochrones as a guide.

The RGB bump occurs near log~g~=~2.38 on our 4~Gyr,
solar-metallicity isochrone and at about log~g~=~2.46 on the corresponding
16~Gyr isochrone.  Based upon these gravities, we only include mixing effects
in our field-star and grid models having log~g~$\leq$~2.4.  Using this gravity
threshold should be reasonable when modelling the field stars used to
calibrate the synthetic colors (see Section 3.2) because the majority of these
stars have approximately solar metallicities.  However, this limit may not be
appropriate for all of our color-temperature grid models (see Section 4).
Our isochrones show that the surface gravity at which the RGB bump occurs
decreases with decreasing metallicity at a given age.  Linearly extrapolating
from these isochrones, which to-date only encompass
metallicities greater than about --0.5 dex in [Fe/H], it appears that
perhaps the models having log~g~=~2.0 and [Fe/H]~$\leq$~--2.0 should have
included the effects of mixing as well.

Overall, including mixing in our models only marginally affects the
resulting broad-band photometry (see Section 3.3) but has a more noticeable
influence on the synthetic spectra and some narrow-band colors.

\subsection{Spectra of the Sun and Arcturus}

\begin{figure}[p]
\epsfxsize=6.0in
\vspace*{-0.5in}
\hspace*{0.25in}
\epsfbox{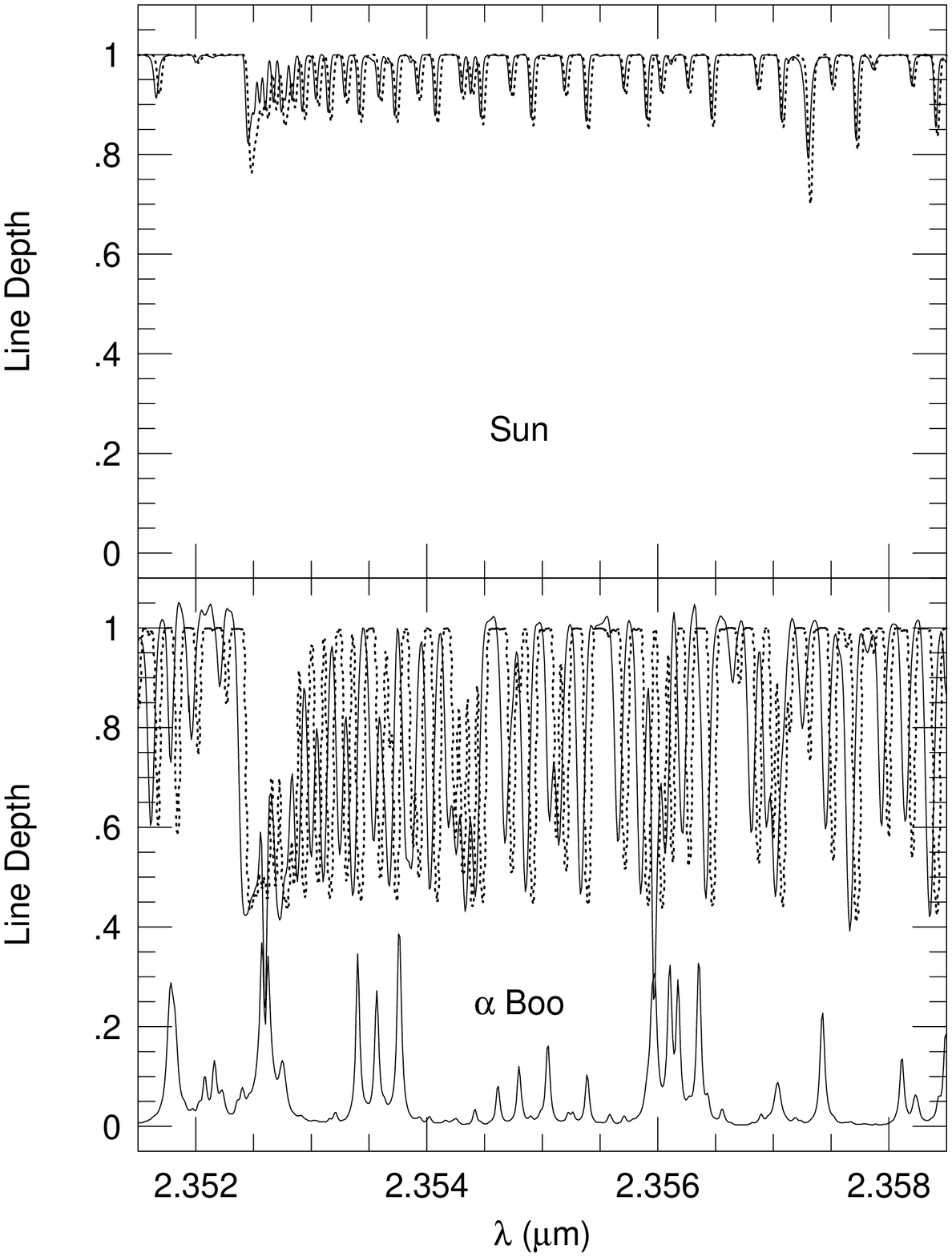}
\vspace*{-0.5in}
\caption{Comparison of our synthetic spectra of the Sun
and Arcturus ($\alpha$~Boo) and the observed spectra of these stars near the
bandhead of the $^{12}$CO(4,2) band.  Our synthetic spectra are shown as dotted
lines in each panel.  The observational data is shown as solid lines and has
been taken from Farmer \& Norton (1989) for the Sun (upper panel) and from
Hinkle et al. (1995) for Arcturus (lower panel).  The latter is ground-based,
and the solid line along the
lower part of the lower panel is the (substantial) amount of telluric
absorption in this part of the spectrum (Hinkle et al. 1995); it is drawn in
emission and to half-scale.}
\label{solararccomp1}
\end{figure}

\begin{figure}[p]
\epsfxsize=6.0in
\vspace*{-0.5in}
\hspace*{0.25in}
\epsfbox{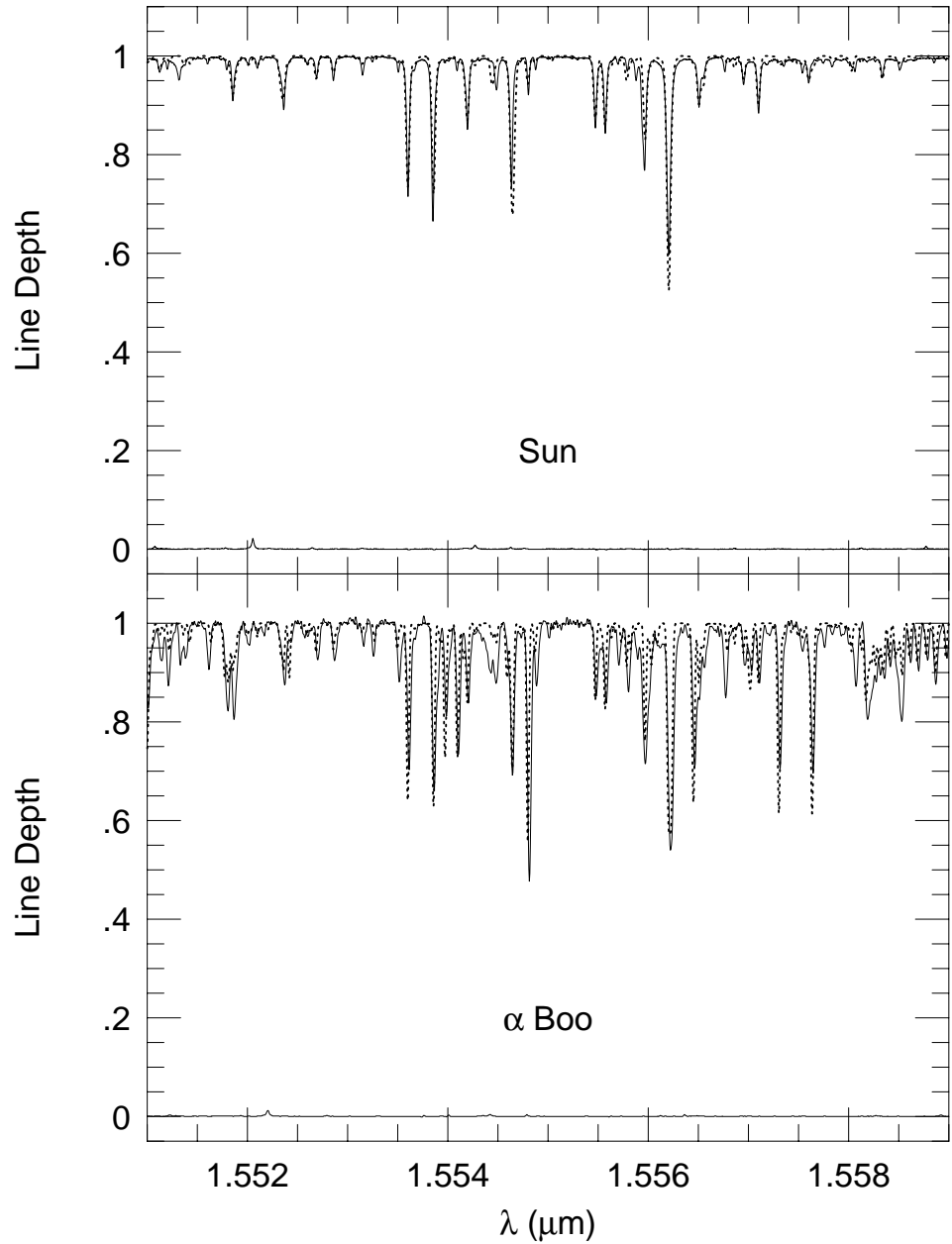}
\vspace*{-0.5in}
\caption{Comparison of our synthetic spectra of the Sun
and Arcturus ($\alpha$~Boo) and the observed spectra of these stars near the
center of the H band.  Our synthetic spectra are shown as dotted
lines in each panel.  The observational data is shown as solid lines and has
been taken from Livingston \& Wallace (1991) for the Sun (upper panel) and from
Hinkle et al. (1995) for Arcturus (lower panel).  The solid lines along the
lower part of each panel are the telluric absorption in this part of the
spectrum; it is taken from the same sources as the empirical data and is
drawn in emission and to half-scale.}
\label{solararccomp2}
\end{figure}

To judge the quality of the synthetic spectra on which 
our colors are based, we present some comparisons of
observed and synthetic spectra of the Sun and Arcturus in the near-infrared.
At optical wavelengths, the agreement between our synthetic solar
spectrum and the observed spectrum of the Sun is similar to that presented by 
\cite{bpt94}, \cite{briley} and \cite{bt95}.
Refinements of the near-infrared line lists are much more recent, and some
examples of near-IR fits can be found in \cite{bell97} and in
Figures~\ref{solararccomp1} and~\ref{solararccomp2}, which compare our
synthetic spectra of the Sun (5780~K, log~g~=~4.44) and Arcturus (4350~K,
log~g~=~1.50, [Fe/H]~=~--0.51, [C/H]~=~--0.67, [N/H]~=~--0.44, [O/H]~=~--0.25,
$^{12}$C/$^{13}$C~=~7)
to data taken from spectral atlases of these stars.  In each figure, the
synthetic spectra are shown as dotted lines, and the observational data are
represented by solid lines; the former have been calculated at 0.01~\AA\ 
resolution and convolved to the resolution of the empirical spectra.

The upper panel of Figure~\ref{solararccomp1} shows our synthetic solar
spectrum and the ATMOS spectrum of the Sun (\cite{fn89}) near the bandhead of
the $^{12}$CO(4,2) band.  The lower panel of this figure is a similar plot for
Arcturus, but the observed spectrum is taken from NOAO (\cite{arcatlas}).
In Figure~\ref{solararccomp2}, NOAO spectra of the Sun (\cite{lw91}) and
Arcturus (\cite{arcatlas}) are compared to the corresponding synthetic spectra
in a region of the H~band.  The NOAO data shown in these two figures
are ground-based and have been corrected
for absorption by the Earth's atmosphere.  To allow the reader to judge its
importance, the telluric absorption at these wavelengths is also displayed,
in emission and to half-scale, across the bottoms of the appropriate panels;
these telluric spectra have been taken from the same sources as the
observational data.
The slight wavelength difference between the empirical and synthetic spectra
in the lower panel of Figure~\ref{solararccomp1} occurs because the observed
spectrum is calibrated using vacuum wavelengths, while the spectral
line lists used to calculate the synthetic spectra assume wavelengths in air;
this offset has been removed in the lower panel of Figure~\ref{solararccomp2}.

Note that the oxygen abundance derived for the Sun from the near-infrared
vibration-rotation lines of OH is dependent upon the model atmosphere adopted.
The \cite{hm74} (1974) model gives a result 0.16 dex higher than the OSMARCS
model of \cite{edvard}.  We use a logarithmic, solar oxygen abundance of
8.87~dex (on a scale where H~=~12.0 dex), only 0.04 dex smaller than the value
inferred from the \cite{hm74} model.  Presumably, this is
part of the reason that our CO lines are slightly too strong in our synthetic
solar spectrum while being about right in our synthetic spectrum of Arcturus
(see Figure~\ref{solararccomp1}).
Overall, though, the agreement between the empirical and synthetic spectra
illustrated in Figures~\ref{solararccomp1} and~\ref{solararccomp2} is
typical of that achieved throughout the J,~H~and K~bandpasses.

\section{The Calculation and Calibration of the Synthetic Colors}

\begin{figure}[p]
\epsfxsize=6.0in
\vspace*{-0.9in}
\hspace*{0.25in}
\epsfbox{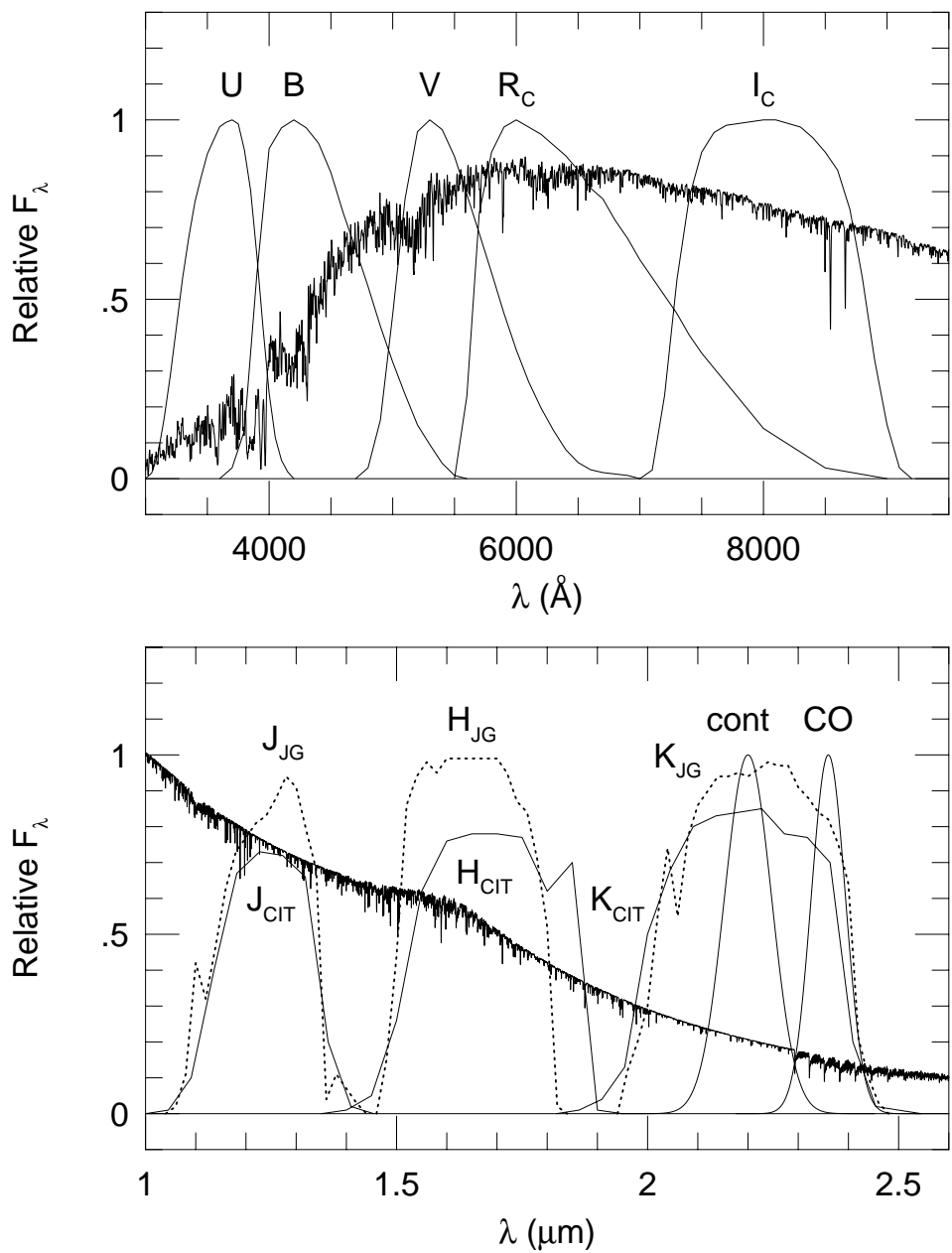}
\vspace*{-0.5in}
\caption{The filter-transmission-profiles used to
measure the synthetic colors are shown atop our synthetic spectrum of Arcturus
(convolved to 2~\AA\ resolution).
The UBVR$_{\rm C}$I$_{\rm C}$ filter profiles are those of Bessell (1990).  We display only
one of Bessell's two B~filter profiles, that which is used in the B--V
calculations; it differs slightly
from his BX~filter used for U--B colors.  The Johnson-Glass (JG) JHK filters are
taken from Bessell \& Brett (1988) and shown as dotted profiles in the lower
panel of the figure.  Persson (1980) and Frogel et al.~(1978) supplied the
JHK and CO filter profiles of the CIT/CTIO system, respectively.  These
are shown as solid lines in the lower panel, and the
continuum filter used for the CO index is labeled ``cont.''}
\label{filterplot}
\end{figure}

Colors are measured from the synthetic spectra by convolving the spectra with
filter-transmission-profiles.  To put the synthetic colors
onto the respective photometric systems, we then apply zero-point corrections
based upon models of the A-type standard star, $\alpha$ Lyrae (Vega), since
the colors of Vega are well-defined in most of these systems.
In this paper, we measure synthetic colors on the following systems using
the filter-transmission-profiles taken from the cited references --
Johnson~UBV and Cousins~VRI: \cite{bsl90};
Johnson-Glass~JHK: \cite{bb88}; CIT/CTIO~JHK: \cite{pers};
and CIT/CTIO CO: \cite{fpam}.
Gaussian profiles were assumed for the CO filters.
In Figure~\ref{filterplot}, these filter-transmission-profiles overlay
our synthetic spectrum of Arcturus.

The synthetic UBVRI colors are initially put onto the observed Johnson-Cousins
system using the \cite{hayes} absolute-flux-calibration data for Vega.  Since
Hayes' observations begin at 3300~\AA, they must be supplemented by the fluxes
of a Vega model (9650~K, log~g~=~3.90, [Fe/H]~=~0.0; \cite{db80}) shortward of
this.  The colors 
calculated from the Hayes data are forced to match the observed colors of Vega
by the choice of appropriate zero points; we assume U--B~=~B--V~=~0.0 for Vega
and take V--R~=~--0.009 and V--I~=~--0.005, as observed by \cite{bess83}.
The resulting zero points are then
applied to all of the synthetic UBVRI colors.  While this results in UBVRI
colors for our Vega model which are slightly different than those observed,
we prefer
to tie the system to the Hayes calibration, which is fixed, rather than to
the Vega model itself because the latter will change as our modelling improves.
The use of our model Vega fluxes shortward of 3300~\AA\ introduces only a small
uncertainty in the U--B zero points.

Since there is no absolute flux calibration of Vega as a function of wavelength 
in the infrared, the VJHK and CO colors are put onto the synthetic
system by applying zero point corrections which force the near-infrared
colors of our
synthetic spectrum of Vega to all become 0.0.  This is consistent with the
way in which the CIT/CTIO and Johnson-Glass systems are defined by \cite{elias}
and \cite{bb88}, respectively.

\subsection{Testing the Synthetic Colors}

As a test of our new isochrones and synthetic color calculations, we decided
to compare our 4~Gyr, solar-metallicity isochrone to color-magnitude diagrams
(CMDs) of the Galactic open cluster M67.  We chose this cluster because it has
been extensively observed and
therefore has a very well-determined age and metallicity.  Recent metallicity
estimates for M67 (\cite{js84}, \cite{nissen}, \cite{hobbs}, \cite{mont},
\cite{fan}) are all solar or slightly subsolar, and the best age estimates
of the cluster each lie in the range of
3--5~Gyr, clustering near 4~Gyr (\cite{nissen}, \cite{hobbs}, \cite{dem92},
\cite{mont}, \cite{meynet}, \cite{din95}, \cite{fan}).
Thus, we expect this particular isochrone to be a good match
to all of the CMDs of M67.

We translated the theoretical isochrone to the
color-magnitude plane by calculating synthetic spectra for effective
temperature/surface gravity combinations lying along it and then
measuring synthetic colors from these spectra.
Absolute V-band and K-band magnitudes were derived
assuming M$_{{\rm V},\odot}$~=~+4.84 and BC$_{{\rm V},\odot}$~=~--0.12.
After doing this, we 
found that our isochrone and the M67 photometry differed
systematically in some colors (see Section 3.2.3),
which led us to more closely examine the
synthetic color calculations.  By modelling field stars with relatively
well-determined physical properties, we learned that, after applying the Vega
zero-point corrections, some of the synthetic
colors were still not on the photometric systems of the observers.
Using these field star models, we have derived the linear color
calibrations which are needed to put the synthetic colors onto the
observational systems.  In the following sections, we describe the
derivation of these color calibrations and show how the application of
these relations removes most of the disagreement between the 4~Gyr
isochrone and the M67 colors, especially near the main-sequence turnoff.

Note, however, that the location of the isochrone in the HR diagram does 
depend upon the details of the stellar interior calculations, such as the
surface boundary conditions and the treatment of convection, particularly at
cooler temperatures.  Near the main-sequence turnoff, its detailed morphology
could also be sensitive to convective overshooting (see \cite{dem92} and
\cite{nord97}), which we have omitted.  The primary purpose in making
comparisons between this isochrone and the M67 photometry
is to test the validity of using this and older isochrones
to model the integrated light of galaxies.

\subsection {The Color Calibrations}

It should not be too surprising, perhaps, that the colors measured
directly from the synthetic spectra are not always on the systems defined
by the observational data.
Bessell et al.~(1998) present an excellent
discussion of this topic in Appendix E of their paper.   They
conclude that, ``As the standard systems have been established from natural
system colors using linear and non-linear corrections of at least a few percent,
we should not be reluctant to consider similar corrections to synthetic
photometry to achieve good agreement with the standard system across the whole
temperature range of the models.''
Thus, we will not explore why the synthetic colors need to be calibrated to put
them onto the observational systems but instead will focus upon the best way
to derive the proper calibrations.

\cite{pb91}, \cite{bpt94} and \cite{tb91} have also looked at the calibration of
synthetic colors.  They
determined the relations necessary to make the colors measured
from the Gunn \& Stryker (1983; hereafter \cite{gs83}) spectral scans match
the observed photometry of the \cite{gs83} stars.
This approach has the advantage that it is model-independent, i.e., it is not
tied in any way to the actual spectral synthesis, and should therefore
primarily be sensitive to
errors in the filter-transmission-profiles.  Of course, it is {\it directly}
dependent upon the accuracy of the flux calibration of the \cite{gs83} scans.

The \cite{gs83} scans extend from 3130--10680~\AA\ and are comprised of
blue and red sections;  these pieces have 20~\AA\ and 40~\AA\ resolution,
respectively, and are joined at about 5750~\AA.
We have made comparisons of the \cite{gs83} scans of stars of similar spectral
type and found the relative flux levels of the blue portions to be very
consistent from star to star; however, there appear to be
systematic differences, often quite large, between the corresponding red
sections.  This ``wiggle'' in the red part of the \cite{gs83} scans usually
originates near 5750 \AA, the point at which the blue and red parts
are merged.  Because of these discrepancies and other suspected problems with
the flux calibration of the \cite{gs83} scans (see e.g., \cite{ruf88},
\cite{tj90}, \cite{worthey}), we have chosen instead to calibrate the synthetic
colors by calculating
synthetic spectra of a group of stars which have well-determined physical
properties and comparing the resulting synthetic colors to photometry of these
stars.  The drawback of this method is that the calibrations are now
model-dependent.  However, since we have decoupled the color calibrations
and the \cite{gs83}
spectra, this allows us to calibrate both the optical and infrared colors
in a consistent manner.  In addition, we use more recent determinations of the
photometric filter-transmission-profiles than the previous calibrations
cited above.

\subsubsection{Effective Temperature Measurements and the Field Star Sample}

To calibrate the synthetic colors, we have chosen a set of field stars which
have well-determined effective temperatures, since \teff\ is the most critical
determinant of stellar color.  Most stars in this set have
surface gravity and metallicity measurements as well, although we have not
always checked that the log~g and [Fe/H] values were found using effective
temperatures similar to those which we adopt.

The most straightforward way to derive the effective temperature of a star is
through measurement of its angular diameter and apparent bolometric flux.
The effective temperature can then be estimated from the angular diameter
through the relation,
\begin{equation}
{\rm T}_{\rm eff} \propto \left( {\rm f}_{\rm bol} \over \phi^2 \right)^{0.25}\ ,
\end{equation}
where $\phi$ is the limb-darkened angular diameter of the star, hereafter
denoted \pld, and f$_{\rm bol}$ is its apparent bolometric flux.
Obviously, this method can only be used for nearby stars.

The infrared flux method
(IRFM) of \cite{bs77} is regarded as one of the more reliable ways to estimate
\teff\ for stars which are not near enough to have their angular
diameters measured.  It relies on the temperature sensitivity of the ratio of
the apparent bolometric flux of a star to the apparent flux in an infrared bandpass
(usually the K band).  Model atmosphere calculations are employed to predict the
behavior of this flux ratio with changing \teff, and the resulting calibration
is then used to infer effective temperatures from observed fluxes.
IRFM temperatures have been published by \cite{black90}, Blackwell \&
Lynas-Gray (1994), Saxner \& Hammarb$\ddot{\rm a}$ck
(1985; hereafter \cite{sh85}), \cite{aam96} (1996), and 
\cite{bg89}, among others, although \cite{bg89} used their color calculations
to adopt final temperatures which are systematically 80~K cooler than their
IRFM estimates.  \cite{bg89} also calculated angular diameters for the stars
in their sample from the apparent bolometric fluxes that they used in the IRFM
and their ``adopted'' effective temperatures.  We can check the accuracy of
these \cite{bg89} \teff\ values by comparing their angular diameter predictions
to recent measurements.

Pauls et al.~(1997; hereafter \cite{pauls}) have used the U.S.~Navy
Prototype Optical Interferometer to make observations in 20 spectral channels
between 5200~\AA\ and 8500~\AA\ and have measured
\pld\ for two of the stars with IRFM temperatures quoted by \cite{bg89}.
In addition, \cite{moz91} and Mozurkewich (1997) have presented uniform-disk
angular diameters (\pud), measured with the Mark~III Interferometer
at 8000~\AA, for 20 of \cite{bg89}'s stars, including the two observed by
\cite{pauls}; these data, which we will hereafter refer to collectively as
\cite{moz97}, must be converted to
limb-darkened diameters before comparing them to the other estimates.

To determine the uniform-disk-to-limb-darkened conversion factor, we performed
a linear, least-squares fit between the ratio \pld/\pud\ and spectral type for
the giant stars listed in Table~3 of \cite{moz91}, using their angular diameters
measured at 800~nm.  In deriving this correction,
$\delta$ And was omitted; \pld\ for $\delta$ And (4.12 mas) does not
follow the trend of the other giants, possibly due to a typographical error
(4.21 mas would fit the trend).
We applied the resulting limb-darkening correction:
\begin{equation}
\phi_{\rm LD}/\phi_{\rm UD}~=~1.078~+~0.002139~\times~\rm{SP}\ ,
\end{equation}
where SP is the spectral type of the star in terms of its M~subclass (i.e.,
SP~=~0 for an M0~star, SP~=~--1 for a K5~star, etc.), to the \cite{moz97}
angular diameters to derive \pld\ for these
stars, keeping in mind that the predicted correction factor may not be
appropriate for all luminosity classes; for example, the above equation gives
\pld/\pud~=~1.033 for $\alpha$~CMi, an F5~IV--V star, while \cite{moz91}
used 1.047.

\begin{figure}[p]
\epsfxsize=5.5in
\vspace*{-0.6in}
\hspace*{0.5in}
\epsfbox{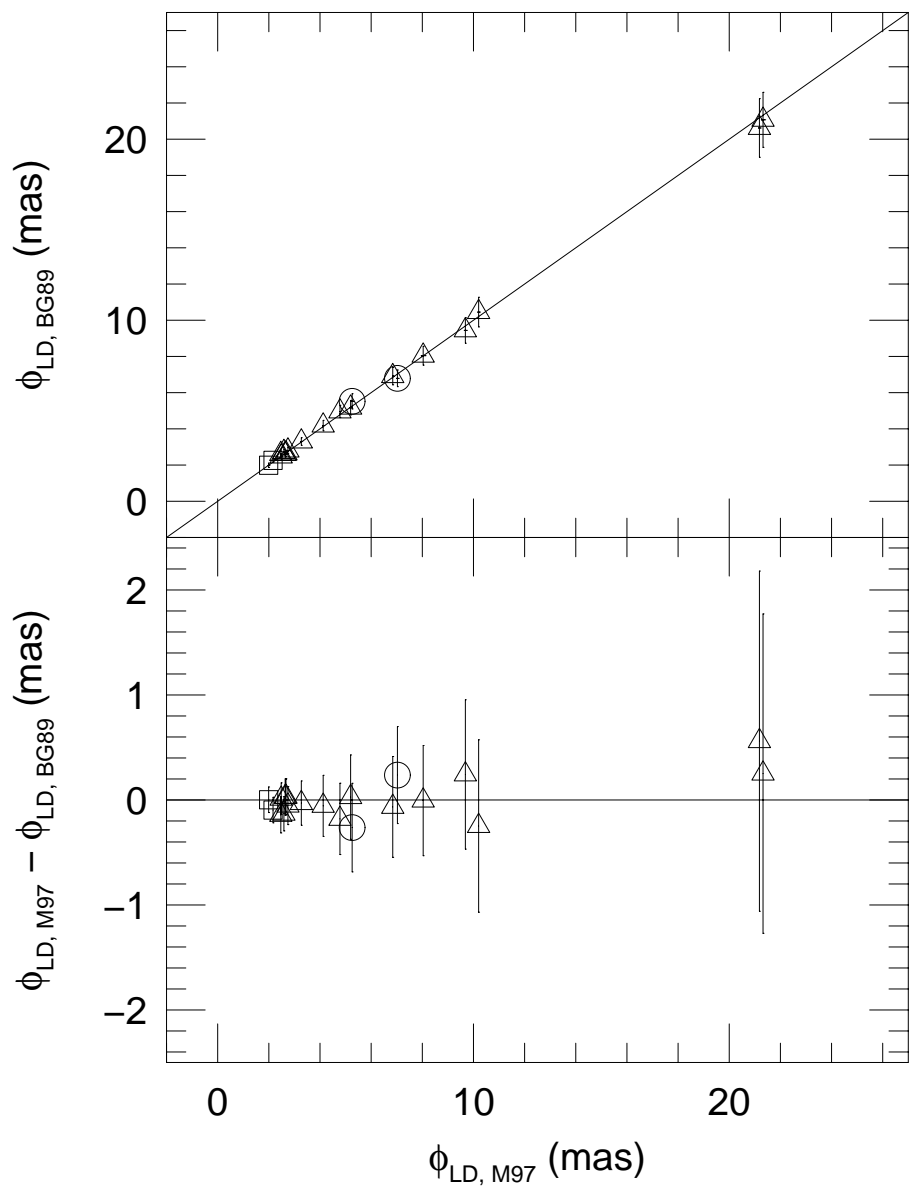}
\vspace*{-0.4in}
\caption{Comparison of the angular diameters measured by
Mozurkewich et al. (1991) and Mozurkewich (1997; M97 collectively) and those
estimated from the ``adopted'' effective temperatures of Bell \&
Gustafsson (1989; BG89).  All angular diameters are limb-darkened values;
the M97 data plotted here were derived from his uniform-disk
diameters as described in the text.  Error bars are those reported by the
original authors; the M97 uncertainties are approximately the size
of the symbols in the upper panel.  Measurements for G--K
supergiants and giants are shown as open circles and open triangles,
respectively; subgiants are shown as open squares.
The upper panel presents the angular diameter comparison, and
the solid line represents equality of the BG89 and M97
diameters.  The lower panel shows the absolute differences in the angular
diameter estimates of the individual stars.}
\label{angdiams}
\end{figure}

The limb-darkened angular diameters of \cite{moz97} and \cite{pauls} are
compared to the \cite{bg89} estimates in Table~\ref{angdiamtable}; the
\cite{moz97} and
\cite{bg89} comparison is also illustrated in Figure~\ref{angdiams}.
Note that the \cite{pauls} value for $\alpha$~UMa is in much better
agreement with the \cite{bg89} estimate than is the \cite{moz97} diameter.
The average absolute and percentage differences between the \cite{bg89}
and \cite{moz97} angular diameters, in the sense \cite{bg89}~--~\cite{moz97},
are only --0.006~mas and 0.87\%, respectively.  Translated into effective
temperatures, these differences become --23~K and 0.44\%.  For the
\cite{pauls} data, the analogous numbers are +0.03~mas (0.51\%) and +13~K
(0.30\%), albeit for only two stars.  This exceptional agreement between
the IRFM-derived and observed angular diameters is a strong confirmation of
the ``adopted'' effective
temperatures of \cite{bg89} and gives us faith that any errors in the synthetic
colors which we measure for the stars taken from this source are not
dominated by uncertainties in their effective temperatures.

Since we are confident that the \cite{bg89} ``adopted'' \teff\ scale for giant
stars is accurate, we
have used a selection of their stars as the basis of the synthetic color
calibrations.  Specifically, we chose all of the stars for which they
derived IRFM temperatures.
However, \cite{bg89} included only G~and K~stars in their
work.  For this reason, we have supplemented the \cite{bg89} sample with
the F~and G~dwarfs studied by \cite{sh85}; before using the latter data, of course,
we must determine whether
the IRFM temperatures of \cite{sh85} are consistent with those of \cite{bg89}.

There is only one star, HR~4785, for which both \cite{bg89} and \cite{sh85}
estimated \teff.  The IRFM temperature of \cite{sh85} is 5842~K,
while \cite{bg89} adopt 5861~K.  In addition, \cite{moz97} has measured an
angular diameter for one of the \cite{sh85} stars, $\alpha$~CMi (HR~2943).
Applying the limb-darkening correction used by
\cite{moz91} for this star and using the bolometric flux estimated by
\cite{sh85} gives \teff~=~6569~K; \cite{sh85} derive 6601~K.  Based upon these
two stars, it appears that the \cite{sh85} effective temperature scale agrees
with the \cite{bg89} scale to within $\sim$20--30~K, 
which is well within the expected uncertainties in the IRFM effective
temperatures.  Thus, we conclude that the two sets of measurements are in
agreement and adopt the IRFM temperatures of \cite{sh85} as given.

Table~\ref{paramtable} lists the stars for which we have calculated synthetic
spectra for use in calibrating the synthetic color measurements and gives the
effective temperatures, surface gravities and metallicities used to model each.
From \cite{bg89}, we used the ``adopted'' temperatures and metallicities
given in their Table~3.  In most cases, we used the surface gravities and
metallicities from this table as well.  However, for seven of the stars, we
took the more recent log~g determinations of \cite{bb93a}, \cite{bb93b}.
For the \cite{sh85} stars, we assumed the average effective temperatures
presented in their Table~7, omitting stars HR~1008 and HR~2085, which may be
peculiar (see \cite{sh85}); the surface gravities and metallicities of these
stars were taken from Table~5 of \cite{sh85}.
We adopted the \cite{bg89} parameters for HR~4785.
For those stars with unknown surface gravities, log~g was estimated from our
new isochrones, taking into account the metallicity and luminosity class of
the star, or from other stars of similar effective temperature
and luminosity class.  Because most of the stars
with well-determined chemical compositions have [Fe/H]~$\sim$~0.0, solar
abundances were assumed for those stars for which [Fe/H] had not been measured.

\begin{figure}[p]
\epsfxsize=6.0in
\vspace*{-0.9in}
\hspace*{0.25in}
\epsfbox{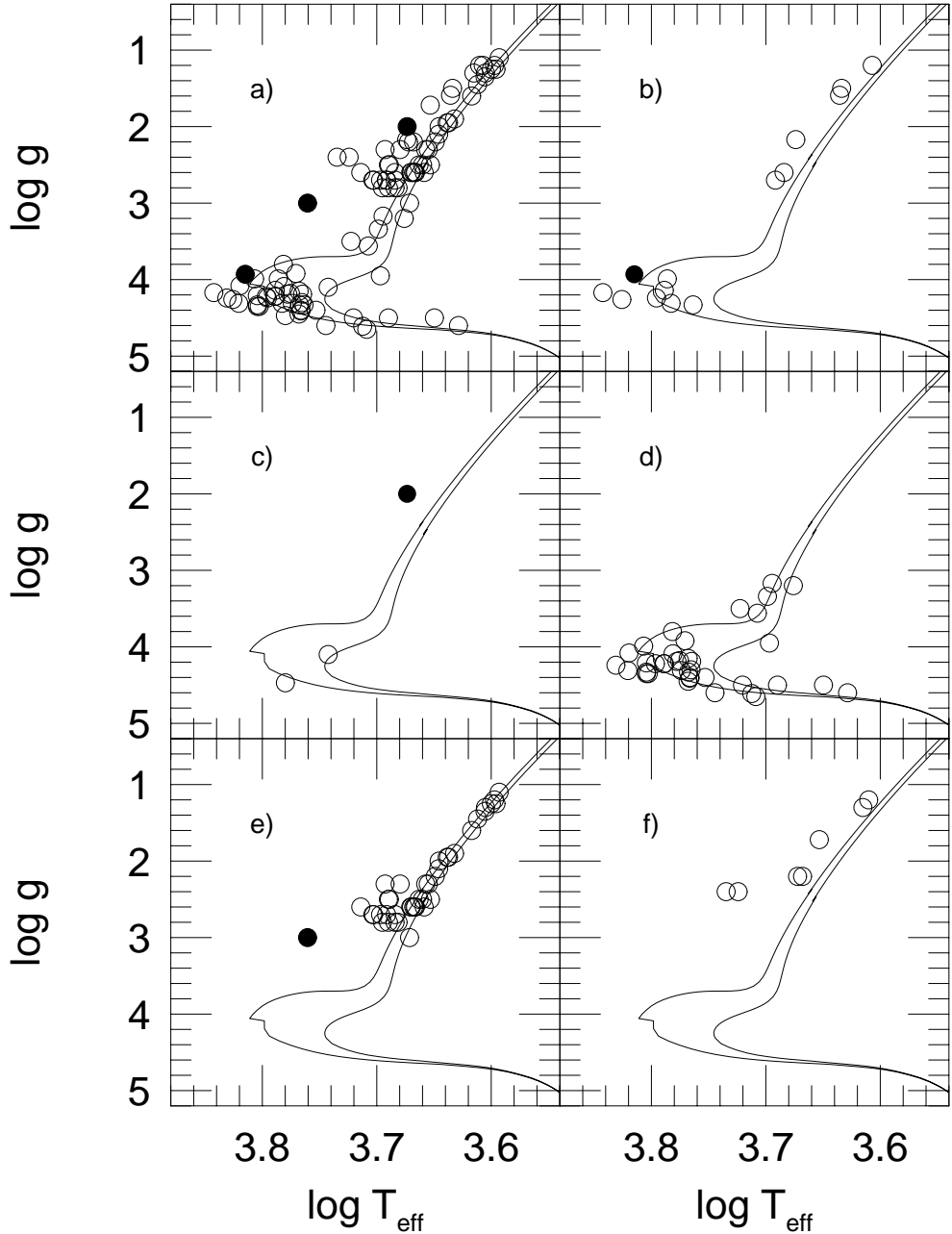}
\vspace*{-0.5in}
\caption{The field stars used to calibrate the synthetic colors
are plotted in the H-R diagram.  The solid lines in each panel are our 3~Gyr
and 16~Gyr, solar-metallicity isochrones.  The six panels show a) the entire
sample of stars, b) those having [Fe/H]~$<$~--0.3, c) the stars with
[Fe/H]~$>$~+0.2, d) the solar-metallicity (--0.3~$\leq$~[Fe/H]~$\leq$~+0.2)
dwarfs and subgiants (classes~IV and~V), e) the solar-metallicity normal giants
(class III), and f) the solar-metallicity bright giants (classes I and II).  The
three stars which lie in unexpected positions in the H-R diagram are shown as
filled points and are discussed in the text -- they are $\gamma$~Tuc, a
metal-poor star with T$_{\rm eff}$~=~6530~K; 72~Cyg, a metal-rich, 4715~K
giant; and 31~Com, a (peculiar) solar-metallicity giant having
T$_{\rm eff}$~=~5761~K.}
\label{hrds}
\end{figure}

In Figure~\ref{hrds}, we show where the stars used to calibrate the synthetic
colors lie in the log~\teff, log~g plane.  Our 3~Gyr and 16~Gyr,
solar-metallicity isochrones are also shown in this figure for reference.
Figure~\ref{hrds}a shows the entire sample of stars from Table~\ref{paramtable};
the other panels present interesting subsets of the larger group.  In panels
b) and c), the metal-poor ([Fe/H]~$<$~--0.3)
and metal-rich ([Fe/H]~$>$~+0.2) stars are shown, respectively.
In panels d), e) and f),
the solar metallicity (--0.3~$\leq$~[Fe/H]~$\leq$~+0.2) stars are broken up
into the dwarfs and subgiants (luminosity classes III--IV, IV, IV--V and V),
the normal giants (classes II--III and III), and the brighter giants (classes
I and II), respectively.

Even though many of the field stars in 
Figure~\ref{hrds}a do not fit the solar-metallicity isochrones as well as one
might expect, especially along the red-giant
branch, examining the subsets of these stars shows that the ``outliers''
are mostly brighter giants and/or metal-poor stars; at a given surface gravity,
these stars generally lie at higher temperatures than the solar-metallicity
isochrones, as would be expected from stellar-evolution theory and
observational data.  The stars located near log~\teff~=~3.7, log~g~=~2.7
in panel e) of Figure~\ref{hrds} (the solar-metallicity giants)
are probably clump stars.
In fact, there are only three stars in Figure~\ref{hrds} which lie very far
from the positions expected from their physical properties or spectral
classification, and even their locations are not unreasonable.  The three stars
in question are represented by filled symbols and are the
metal-poor star, $\gamma$~Tuc; the metal-rich giant, 72~Cyg; and the
solar-metallicity giant, 31~Com.  $\gamma$~Tuc is unusual in that SIMBAD assigns
it the spectral type F1~III, yet it lies very near the turnoff of the 3~Gyr,
Z$_\odot$ isochrone.  The Michigan Spectral Survey (\cite{mss}) classifies
this star as an F3~IV--V star, and \cite{sh85} listed it as an F-type
dwarf, which appears to be appropriate for its temperature and surface gravity,
indicating that the SIMBAD classification is probably incorrect.  72~Cyg is
suspect only because it is hotter than the solar-metallicity isochrones while
having a supersolar metallicity.  Since it could simply be younger (i.e., more
massive) than the other stars in the sample, we do not feel justified in
altering the parameters adopted for this star.  The third star, 31~Com, lies
further from the isochrones than any other star, but it is classified as a
G0~III peculiar star, so perhaps its log~g and \teff\ are not unreasonable
either.

Overall, Figure~\ref{hrds} indicates that the temperatures and
gravities assigned to the stars in Table~\ref{paramtable} are good 
approximations to those one would infer from their spectral types and
metallicities.

\subsubsection {Derivation of the Color Calibrations}

We have calculated a MARCS stellar atmosphere model and an SSG synthetic
spectrum for each of the stars in Table~\ref{paramtable}.   However, rather
than compute a separate ODF for each star, we chose to employ only three
ODFs in these model atmosphere calculations -- the same ODFs used in our
evolutionary synthesis work (\cite{houdy3}).  For stars with [Fe/H]~$<$~--0.3,
a Z~=~0.006 ([Fe/H]~=~--0.46) ODF was used; stars having
--0.3~$\leq$~[Fe/H]~$\leq$~+0.2 were assigned a solar-metallicity (Z~=~0.01716)
ODF; and a Z~=~0.03 ([Fe/H]~=~+0.24) ODF was used for stars with
[Fe/H]~$>$~+0.2.  The synthetic spectra were then calculated using the
metallicities of the corresponding ODFs.

To calibrate the synthetic colors, photometry of the field stars in
Table~\ref{paramtable} has been compiled from the literature.
Since all of these stars are relatively nearby, no
reddening corrections have been applied to this photometry.
The UBV photometry comes from \cite{merm91} and \cite{j66}.  The VRI data is
from \cite{cuz80} and \cite{j66}; the latter were transformed to the Cousins
system using the color transformations of \cite{bess83}.  The VJHK photometry
has been derived from colors observed on the Johnson system by \cite{j66},
\cite{j68}, \cite{lee70} and \cite{eng81} and on the SAAO system by \cite{g74};
the color transformations given by \cite{bb88} were used to calculate the
observed Johnson-Glass and CIT/CTIO colors.  Table~\ref{paramtable} cites the
specific sources of the photometry used for each field star.
Unfortunately, we were unable to locate CO observations for a sufficient
number of these stars to be able to calibrate this index. 

\begin{figure}[p]
\epsfxsize=6.0in
\vspace*{-0.5in}
\hspace*{0.25in}
\epsfbox{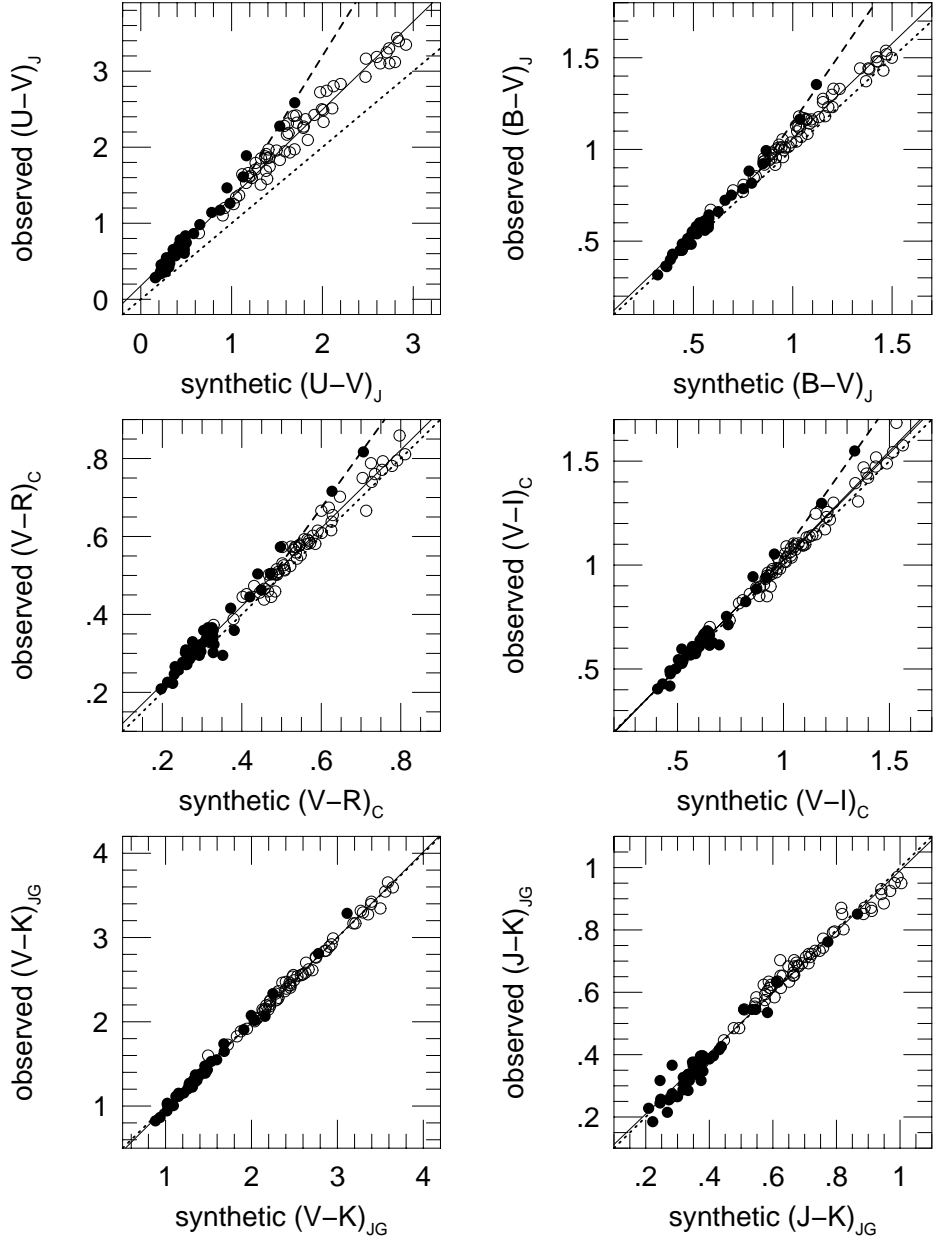}
\vspace*{-0.5in}
\caption{Some of the color calibrations required to put the
synthetic colors onto the observational
systems are illustrated.  Solid lines are the linear, least-squares fits to
the data, excluding the three coolest dwarfs; dashed lines are the cool-dwarf
color calibrations described in the text.  Dotted lines show equality of
the synthetic and photometric colors.
Filled symbols represent dwarfs and subgiants
(luminosity classes IV, V), and open symbols are giants and supergiants
(luminosity classes I, II, III).  Sources of the photometry are described in the
text and listed in Table~2.}
\label{colorfits}
\end{figure}

Although a few colors show a hint that a higher-order polynomial may be
superior, we have fit simple linear, least-squares relations to the field-star
data, using the photometric color as the dependent variable and the synthetic
color as the independent variable in each case and omitting the three coolest
dwarfs.  Figure~\ref{colorfits} shows some of these color-calibration fits
(solid lines), and the zero points and
slopes of all of the resulting relations are given in
Table~\ref{colorcaltable}; the $\sigma$ values listed there are the 1$\sigma$
uncertainties in the coefficients of the least-squares fits.
The number of stars used to derive each calibration equation, n, and the
colors spanned by the photometric data are also given in Table~\ref{colorcaltable}.

For the optical colors, the coolest dwarfs appear to follow a different color
calibration than the other field stars, so we have also calculated a separate
set of color calibrations for cool dwarfs.  This was done by fitting a
linear relation to the colors of the three reddest field dwarfs but forcing
this fit to intersect the ``main'' calibration relation of
Table~\ref{colorcaltable} at the color corresponding to a 5000~K dwarf.  These
cool-dwarf color calibrations are shown as dashed lines in
Figure~\ref{colorfits}.  However,
we wish to emphasize that the cool-dwarf color calibrations are
very uncertain, being derived from
photometry of only three stars (HR~8085, HR~8086 and HR~8832); the choice
of 5000~K as the temperature at which the cool dwarf colors diverge is
also highly subjective.  Thus,
since it is not clear to us whether the use of different
optical color calibrations for cool dwarfs is warranted, we will proceed by
discussing and illustrating the calibrated colors of the cool dwarf models
which result when both the color calibrations of Table~\ref{colorcaltable} and
the separate cool-dwarf calibrations are adopted.
In Section 4.1.1, we further discuss the coolest
dwarfs in our sample and possible explanations for their erroneous colors.

Due to the small color range of the H--K photometry and the
uncertainties in the Johnson/SAAO H--K photometry and color transformations,
the calibrations derived for the H--K colors are
highly uncertain; application of the corresponding
calibrations may not improve the agreement with the observational data
significantly.  It is also apparent that
our U--V calculations are not very good, but we expect them to improve
substantially after the aforementioned \ion{Fe}{1} data of \cite{bautista} have
been incorporated into both MARCS and SSG.  Additionally, we want to emphasize
that the mathematical form of the color calibrations is presented here only to
illustrate their significance.  We caution others against the use of these
specific relations to calibrate synthetic colors in general because the
coefficients are dependent, at least in part, on our models.

The poor agreement between the synthetic and observed CIT/CTIO J--K colors,
contrasted with the relatively good fit for Johnson-Glass J--K, is probably
due to better knowledge of the J-band filter-transmission-profile of the latter
system.  This is suggested by the ratio of the scale factors (slopes) in
Table~\ref{colorcaltable} (JG/CIT = 0.976/0.895 = 1.090), which is very close 
to the slope of the color transformation (1.086) between the two systems
found by \cite{bb88}.
In other words, our synthetic colors for the two systems are
very similar, while the Bessell \& Brett observational transformation
indicates that they should differ.

Even though most of our color calibrations call for only small corrections to
the synthetic colors, we have chosen to apply each of them because every
relation has either a slope or a zero point which is significant at greater
than the 1$\sigma$ level.
We believe that using these calibration equations will help us to reduce
the uncertainties in the integrated galaxy colors predicted by our evolutionary
synthesis models (\cite{houdy3}).  Indeed, the importance of calibrating the
synthetic colors when modelling the integrated light of galaxies and other
stellar aggregates becomes apparent when the uncalibrated and calibrated colors
of our 4~Gyr, solar-metallicity isochrone are compared to color-magnitude
diagrams of M67. 

\subsubsection{The M67 Color-Magnitude Diagrams}

\begin{figure}[p]
\epsfxsize=6.0in
\vspace*{-0.6in}
\hspace*{0.25in}
\epsfbox{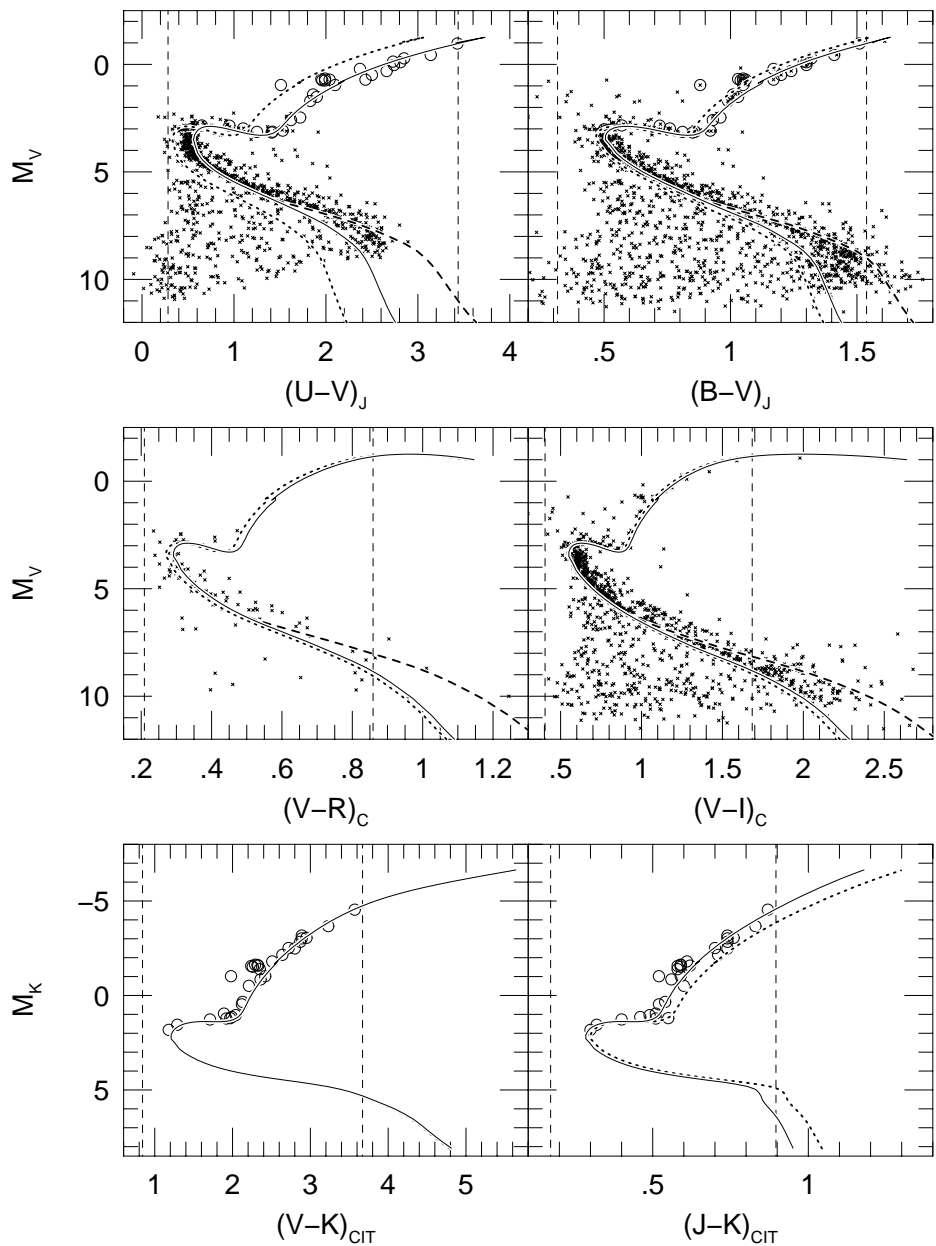}
\vspace*{-0.6in}
\caption{Photometry of stars in the Galactic open
cluster M67 is compared to our 4 Gyr, solar-metallicity isochrone.  Small
crosses and large open circles represent photometry taken from
Montgomery et al.~(1993) and Houdashelt et al.~(1992), respectively.  The
dotted line is the uncalibrated isochrone, while the solid line is the
calibrated isochrone which is produced by applying the
color calibrations given in Table~3 to the uncalibrated isochrone
colors; the dashed-line portion of the isochrone results when the cool-dwarf
color calibrations are used instead for the optical colors of the dwarf models
having T$_{\rm eff}$~$<$~5000~K.  Vertical dashed lines show the color range of
the field stars used to derive the color calibrations.}
\label{m67cmds}
\end{figure}

The (uncalibrated) colors of our 4 Gyr, solar-metallicity isochrone were
determined by calculating synthetic spectra for effective temperature/surface
gravity combinations lying along it, measuring synthetic colors from these
spectra and then applying the Vega-based zero-point corrections to these colors.
In Figure~\ref{m67cmds}, we show the effects of applying the color
calibrations to the synthetic colors of the 4~Gyr, Z$_{\odot}$ isochrone.
In each panel of this figure, the uncalibrated
isochrone is shown as a dotted line, while the calibrated isochrones
are represented by a solid line (Table~\ref{colorcaltable} color calibrations
throughout) and a dashed line (Table~\ref{colorcaltable} relations + cool-dwarf
calibrations for optical colors of main-sequence stars cooler than 5000~K),
respectively.  The M67 UBVRI photometry of Montgomery
et al.~(1993) is shown as small crosses in Figure~\ref{m67cmds}, and the
giant-star data tabulated by \cite{houdy1} are represented by open circles.

We have corrected the M67 photometry
for reddening using E(B--V)~=~0.06 and the reddening ratios of a K~star as
prescribed by \cite{cohen}; we also assumed E(U--V)/E(B--V)~=~1.71.
This reddening value is on the upper end of the range of recent estimates
for M67 (\cite{js84}, \cite{nissen}, \cite{hobbs}, \cite{mont}, \cite{meynet},
\cite{fan}), but we found that a reddening this large produced the best overall
agreement between the 4~Gyr, solar-metallicity isochrone and the M67 photometry
in the region of the main-sequence turnoff for the complete set of CMDs which
we examined.
This reddening leaves the isochrone turnoff a bit bluer than the midpoint of
the turnoff-star color distribution in V--I and a bit redder than the analogous 
M67 photometry in U--V, but it gives a reasonable fit for the other colors.
Because the turnoff stars in the M$_{\rm V}$, U--V CMD suggest a smaller
reddening is more appropriate, we also
examined the possibility that the Montgomery et al.~(1993) V--I colors are
systematically in error; however, they are consistent with the photometry
of M67 reported by \cite{chev} and by \cite{jt90}.  Most importantly, we simply
trust our synthetic V--I colors (and the corresponding reddening ratios) more
than those in U--V, given the aforementioned
uncertainties in the bound-free \ion{Fe}{1} opacity, so we accepted the poorer
turnoff fit in the U--V CMD.
We derived the absolute magnitudes of the M67 stars, after correcting
for extinction, assuming 
(m--M)$_0$~=~9.60 (\cite{nissen}, \cite{mont}, \cite{meynet},
\cite{din95}, \cite{fan}).

The agreement between the isochrone and the M67 observations is improved in
all of the CMDs (except perhaps in the H--K diagram, which is not shown)
after the isochrone colors have been calibrated to the photometric systems
using the color calibrations derived in Section 3.2.2.  In fact, if we
concentrate upon the main-sequence-turnoff region, there is no further
indication of problems with the synthetic colors, with the possible exception
of U--V, which again is sensitive to opacity uncertainties.
The level of agreement between the photometry and the isochrone
along the red-giant branch strengthens this conclusion.  Of course, more
R-band and deeper near-infrared photometry of M67 members would be helpful
in verifying this result for the respective colors.

We also note that the detailed fit of the isochrone to the M67 turnoff could possibly
be improved by including convective overshooting in our stellar interior models.
While there is no general agreement regarding the importance of convective
overshooting in M67 (see \cite{dem92}, \cite{meynet}, and
\cite{din95} for competing views), solar-metallicity turnoff stars
only slightly more massive than those in M67 do show evidence for
significant convective overshooting (\cite{nord97}, \cite{rv98}).  To help us
examine this further, 
\cite{vb99} has provided us with a 4 Gyr, solar-metallicity isochrone which
includes the (small) amount of convective overshooting which he considers to be
appropriate for M67.  This isochrone is in excellent agreement with ours
along the lower main-sequence but contains a ``hook'' feature at the
main-sequence turnoff, becoming cooler than our 4 Gyr, Z$_\odot$ isochrone at
M$_{\rm V}$=3.5 and then hooking back to be hotter than ours at
M$_{\rm V}$=3.0; this ``hook''
does appear to fit the M67 photometry slightly better.  Still, we do not expect
convective overshooting to significantly affect our evolutionary synthesis
models of elliptical galaxies because it
will have an even smaller impact upon the isochrones older than 4 Gyr.

In those colors for which the deepest photometry is available, the calibrated
isochrone is still bluer than the M67 stars on the lower main-sequence when
only the color calibrations of Table~\ref{colorcaltable} are applied.  Using
separate calibrations for the cool dwarfs makes the faint part of the calibrated
isochrone agree much better with the U--V and B--V colors of the faintest M67
stars seen in Figure~\ref{m67cmds} but appears to overcorrect their V--I colors.
This improved fit between the calibrated isochrone and the M67 photometry
supports the use of different color calibrations for the optical colors of
the cool dwarfs, although
there is a hint in Figure~\ref{m67cmds} that perhaps the cool-dwarf
calibrations should apply at an effective temperature slightly hotter
than 5000~K. Nevertheless, we do not find lower
main-sequence discrepancies as worrisome as color differences near the turnoff
would be, since the lowest-mass stellar interior models are sensitive to the
assumed low-temperature opacities, equation of state and surface pressure
boundary treatment.  In addition, these faint, lower-main-sequence
stars make only a small contribution to the integrated light of our
evolutionary synthesis models when a Salpeter IMF is assumed.  Consequently,
small errors in their colors will not generally have a detectable effect
on the integrated light of the stellar population represented by the entire
isochrone.

\subsection{Uncertainties in the Synthetic Colors}

While the color calibrations are technically appropriate only for stars with
near-solar metallicities, there is no evidence from the photometry of the field
star sample that the synthetic color calibrations depend upon
chemical composition.  In fact, it is likely that we are at least partially
accounting for
metallicity effects by applying color corrections as a function of {\it color}
rather than effective temperature.  If the differences between the uncalibrated,
synthetic colors and the photometry are due to errors in the synthetic spectra
caused by missing opacity, for example, then different color corrections
would be
expected for two stars having the same temperature but different metallicities;
the more metal-poor star should require a smaller correction.  This is in
qualitative agreement with the calibrations derived here, since this star
would be bluer than its more metal-rich counterpart.

In this section, we present the sensitivities of the colors to uncertainties
in some of the
model parameters: effective temperature, surface gravity, metallicity,
microturbulent velocity ($\xi$) and mixing.
We estimate the uncertainties in these stellar properties to be
$\pm$80~K in \teff, $\pm$0.3 dex in log~g, $\pm$0.25~dex in [Fe/H], and
$\pm$0.25~km~s$^{-1}$ in $\xi$.
To examine the implications of these uncertainties, we have varied the
model parameters of the coolest dwarf (61 Cyg B), the coolest giant
($\alpha$ Tau) and the hottest solar-metallicity dwarf (HR 4102) listed in
Table~\ref{paramtable} by the estimated uncertainties, constructing new model
atmospheres and synthetic spectra of these stars.
We also constructed an $\alpha$~Tau model which neglected mixing.
In Table~\ref{errortable}, we present the changes in the uncalibrated synthetic colors
which result when each of the model parameters is varied as described above.
The color changes presented in this table are the average changes produced
by parameter variations in the positive and negative directions.
Since the VJHK color changes were almost always identical for colors on the
Johnson-Glass and CIT/CTIO systems, the $\Delta$(V--K), $\Delta$(J--K) and
$\Delta$(H--K) values given in Table~\ref{errortable} are applicable to either
system.

If our uncertainty estimates are realistic, then it is clear from
Table~\ref{errortable} that the variations in most colors are dominated by
the uncertainties in one parameter.
For the V--R, V--I, V--K, J--K and H--K colors, this parameter is \teff, with the
exception being the V--R color of the cool dwarf, which appears to be especially
metallicity-sensitive.  The behavior of the U--V and B--V colors is more
complicated.  Gravity and metallicity uncertainties dominate U--V for the cool
giant and the hot dwarf, but effective temperature uncertainties have the
greatest influence on the U--V color of the cool dwarf.  Uncertainties in
each of the model parameters have a similar significance in determining the
B--V color of the cool giant, while those in \teff\ dominate for the cool
dwarf, and \teff\ and [Fe/H] uncertainties are most important in
the hot dwarf.  Overall, we conclude that estimating the formal uncertainties of
the synthetic colors needs to be done on a star-by-star and color-by-color
basis, which we have not attempted to do.  However, it is encouraging to
see that \teff\ uncertainties dominate so many of the color determinations,
since our effective temperature scale is well-established by the angular
diameter measurements discussed in Section~3.2.1.

\section{The New Color-Temperature Relations and Bolometric Corrections}

From the comparisons of the calibrated, 4~Gyr, solar-metallicity isochrone
and the CMDs of M67, we conclude that the color calibrations that were
derived in Section 3.2.2 and presented in Table~\ref{colorcaltable} generally
put the synthetic colors onto the photometric systems of the observers but
leave some of the optical colors of cool dwarfs too blue.
These calibrations, coupled with the previously
described improvements in the model atmospheres and
synthetic spectra, have encouraged us to calculate a new grid of
color-temperature relations and bolometric corrections.  The bolometric
corrections are calculated after calibrating the model colors, assuming
BC$_{{\rm V},\odot}$~=~--0.12 and M$_{{\rm V},\odot}$~=~+4.84;
when coupled with the calibrated color of our solar model,
(V--K)$_{\rm CIT}$~=~1.530, we derive
M$_{{\rm K},\odot}$~=~+3.31 and BC$_{{\rm K},\odot}$~=~+1.41.  However,
keep in mind that the color calibrations have been derived from Population~I
stars, so the colors of the models having [Fe/H]~$\lesssim$~--0.5 should be
used with some degree of caution (but see Section~3.3).

In Table~\ref{gridtable}, we present a grid of calibrated 
colors and K-band bolometric corrections for stars having
4000~K~$\leq$~\teff~$\leq$~6500~K and 0.0~$\leq$~log~g~$\leq$~4.5 at five
metallicities between [Fe/H]~=~--3.0 and solar metallicity; the optical colors
of the dwarf (log~g~=~4.5) models resulting from the use of the cool-dwarf
color calibrations discussed in Section 3.2.2 are enclosed in parentheses,
allowing the reader to decide which color calibrations to adopt for these
models.  We will now proceed to
compare our new, theoretical color-temperature relations to empirical
relations of field stars and to previous MARCS/SSG results.

\subsection{Comparing Our Results to Other Empirical Color-Temperature Relations}

In the following sections, we use the photometry and effective temperatures
of the stars listed in Table~\ref{paramtable} to derive empirical
color-temperature relations for field giants and field dwarfs.  We then
compare our models and our empirical,
solar-metallicity color-temperature relations to the CT relations of field
stars which have been derived by Blackwell \& Lynas-Gray (1994; hereafter
\cite{blg94}), Gratton et al.~(1996; hereafter \cite{gcc96}), Bessell (1979;
hereafter \cite{bess79}), Bessell (1995; hereafter \cite{bess95}), Bessell et
al.~(1998; hereafter \cite{bcp98}) and Bessell (1998; hereafter \cite{bsl98}).

\subsubsection{The Color-Temperature Relations of Our Field Stars}

\begin{figure}[p]
\epsfxsize=6.5in
\vspace*{-3.0in}
%\hspace*{0.25in}
\epsfbox{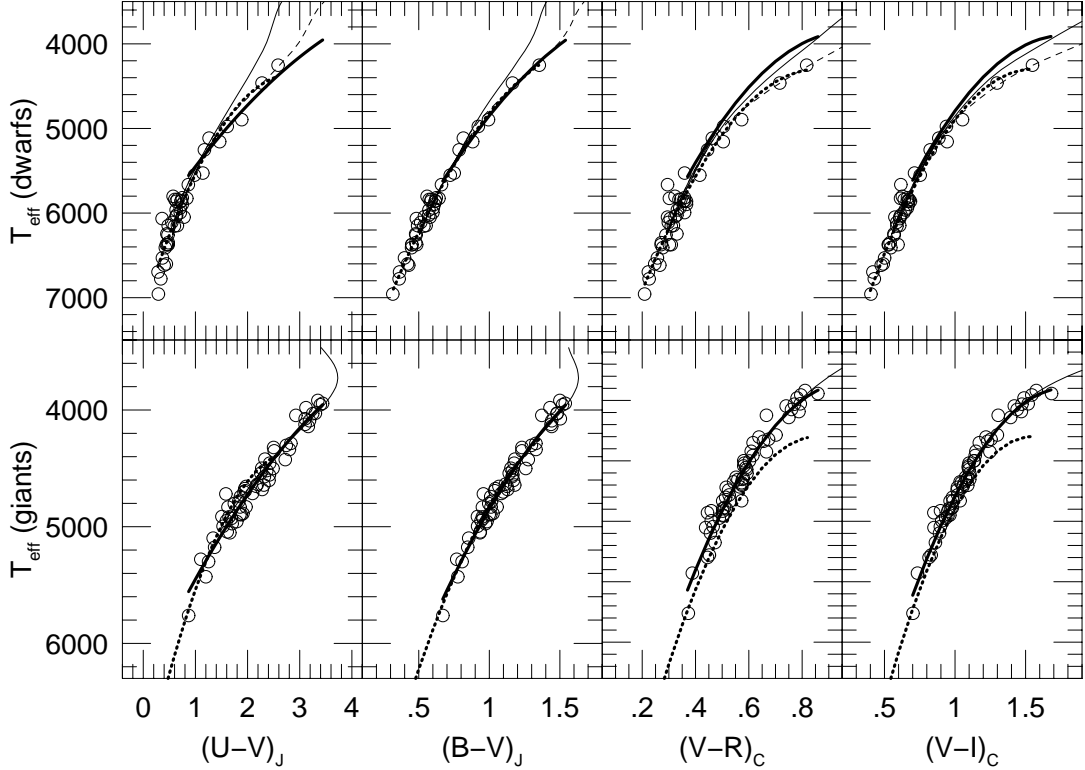}
\vspace*{-0.3in}
\caption{The empirical, optical color-temperature relations of
our sample of field dwarfs (upper panels) and field giants (lower panels) are
shown.  The bold, solid and bold, dotted lines are quadratic fits to the
field-giant and field-dwarf data, respectively; the coefficients of these
relations are given in Table~6.  The solid line is our calibrated,
4~Gyr, Z$_\odot$ isochrone (using the color calibrations of Table~3); the
dashed line results when the cool-dwarf
color calibrations are used instead for the optical colors of the dwarf models
having T$_{\rm eff}$~$<$~5000~K.}
\vspace*{1.0in}
\label{opfield}
\end{figure}

\begin{figure}[p]
\epsfxsize=6.5in
%\vspace*{0.2in}
%\hspace*{0.25in}
\epsfbox{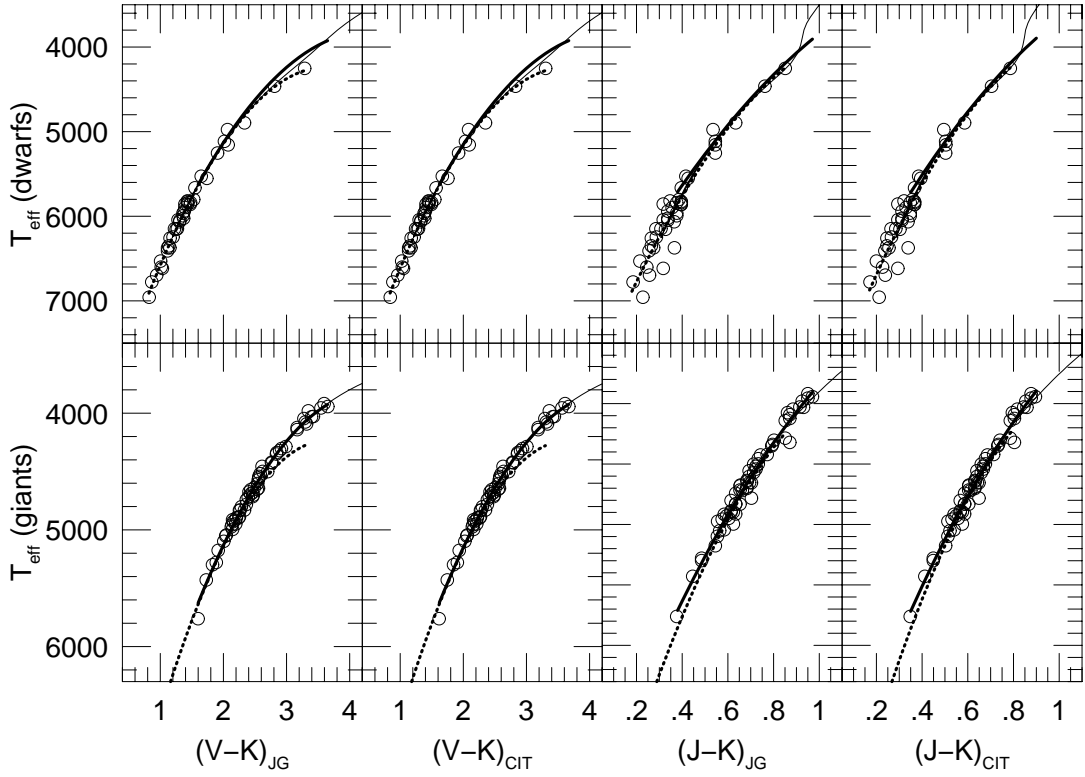}
\vspace*{-0.3in}
\caption{The empirical, near-infrared color-temperature relations of
our sample of field dwarfs (upper panels) and field giants (lower panels) are
shown.  The bold, solid and bold-dotted lines are quadratic fits to the
field-giant and field-dwarf data, respectively; the coefficients of these
relations are given in Table~6.  The solid line is our calibrated,
4~Gyr, Z$_\odot$ isochrone.}
\label{irfield}
\end{figure}

The color and temperature data for our set of color-calibrating field stars
are plotted in Figures~\ref{opfield} and~\ref{irfield}, where we have split
the sample into giants (log~g~$\leq$~3.6) and dwarfs (log~g~$>$~3.6), shown
in the lower and upper panels of the figures, respectively.
We have fit quadratic relations to the effective temperatures of the dwarfs
and giants separately as a function of color to derive empirical,
solar-metallicity color-temperature relations for each; the resulting CT
relations of the giants and dwarfs are shown as bold, solid lines and bold,
dotted lines, respectively, in these figures.  The coefficients of the fits
(and their 1$\sigma$ uncertainties) are given in Table~\ref{fieldrelns}.  The
calibrated colors of our M67
isochrone (4~Gyr, Z$_\odot$) are also plotted in Figures~\ref{opfield}
and~\ref{irfield} as solid lines; the dashed lines in the upper panels of
Figure~\ref{opfield} show the calibrated
isochrone when the cool-dwarf color calibrations are applied.
Comparisons with our other isochrones indicate that the CT relation of the
solar-metallicity isochrones is not sensitive to age.

Although we have divided our field-star sample into giants and dwarfs, the
empirical CT relations of the two groups only appear to differ significantly
in (V--R)$_{\rm C}$ and (V--I)$_{\rm C}$.  In B--V and J--K, the CT relations
of the dwarfs and giants are virtually identical, and it is only the color
of the coolest dwarf that prevents the same from being true in V--K.  The
differences in the U--V color-temperature relations of the dwarfs and giants
can largely be attributed to the manner in which we chose to perform the
least-squares fitting.  Therefore, we want to emphasize that our tabulation of
separate CT relations for dwarfs and giants does not necessarily infer that the
two have significantly different colors at a given \teff.

It is apparent that the calibrated isochrone matches the properties of the
field giants very nicely in Figures~\ref{opfield} and~\ref{irfield} and, in
fact, is usually indistinguishable from the empirical CT relation.
However, using the color calibrations of Table~\ref{colorcaltable}, the dwarf
and giant portions of the isochrone diverge cooler than $\sim$5000~K in
all colors;
this same behavior is not always seen in the stellar data.
This discrepancy at cool temperatures means that one of the following
conditions must apply: 1) the
field star effective temperatures of the cool dwarfs are systematically too
hot; 2) the isochrones predict the wrong \teff/gravity relation for cool
dwarfs; or 3) the cool dwarf temperatures and gravities are correct but the
model atmosphere and/or synthetic spectrum calculations are wrong for these
stars.
Only the latter condition would lead to different
color calibrations being required to put the colors of cool dwarfs and cool
giants onto the photometric systems.

The previously-mentioned comparison between our isochrones and those of 
\cite{vb99} lead us to believe that the lower main-sequence of our 4~Gyr,
solar-metallicity isochrone is essentially correct, so it would appear that
the synthetic colors of the coolest
dwarf models are bluer than the color calibrations of Table~\ref{colorcaltable}
would predict because either we have adopted incorrect effective temperatures
for these stars or there is some error in the model atmosphere or synthetic
spectrum calculations, such as a missing opacity source.

Unpublished work on CO band strengths in cool dwarfs suggests
that part of the problem could lie with the BG89 temperatures derived
for these stars.  In addition, some IRFM work (\cite{aam96} 1996) indicates
that the effective temperatures adopted for our K~dwarfs cooler than 5200~K
may be too hot by $\sim$100~K and infers an even larger discrepancy (up to
400~K) for HR~8086.  A recent angular diameter measurement (\cite{pauls99})
also predicts a hotter \teff\ than we adopted for HR~1084.  If our K-dwarf
effective temperatures are too warm, this would imply that the
cool-dwarf color calibrations should not be used because the errors lie in the
model parameters adopted for the cool dwarfs and not in the calculation of their
synthetic spectra.  However, \teff\ errors of 100~K
are not sufficiently large to make the three cool field dwarfs lie on
the optical color calibrations of the other stars (see Table~\ref{errortable}).
In addition, the similarities of the V--K and J--K colors of our two coolest
dwarfs and those of the calibrated isochrone at the assumed
effective temperatures argue that the \teff\ estimates are essentially correct
but do not preclude temperature changes of the magnitude implied by the IRFM
and angular diameter measurements.  Thus, some combination of \teff\ errors
and model uncertainties may conspire to produce the color effects seen
here, and more angular diameter measurements of K~dwarfs
would be very helpful in sorting out these possibilities.  Nevertheless,
the use of the cool-dwarf color calibrations improves the agreement
between the models and the empirical data so remarkably well in
Figures~\ref{m67cmds} and~\ref{opfield} that a good case can be made for
their adoption.

As discussed in the Section 3.3, the model B--V colors of giants are
noticeably dependent on gravity at low temperatures.
The agreement between the observed B--V colors and the corresponding calibrated,
synthetic colors for both
the sample of field stars and the cool M67 giants suggests that the
gravity-temperature relation of the isochrone truly represents the field stars,
even at cool temperatures.  While this is certainly anticipated, the fact that
it occurs indicates that the surface boundary conditions and mixing-length
ratio used for the interior
models are satisfactory and gives us confidence that these newly-constructed
isochrones are reasonable descriptions of the stellar populations they are
meant to represent in our evolutionary synthesis program.

\subsubsection{Other Field-Star Color-Temperature Relations}

\cite{blg94} estimated effective temperatures for 80 stars using the IRFM
and derived V--K, \teff\ relations for the single stars, known binaries and total
sample which they studied.  Because these three relations are very similar, we
have adopted their single star result for our comparisons, after 
converting their V--K colors from the Johnson to the Johnson-Glass system using the
color transformation given by \cite{bb88}.  We expect the \cite{blg94}
effective temperature scale to be in good agreement with that adopted here
for the field stars
used to derive the color calibrations -- for the 19 stars in common between
\cite{blg94} and \cite{bg89}, \cite{blg94} report that the average difference
in \teff\ is 0.02\%.  For later discussion, we note that the coolest dwarf
included in \cite{blg94}'s sample is BS~1325, a K1 star, with \teff~=~5163~K.

\cite{gcc96} derived their CT relations from
photometry of about 140 of the \cite{bg89} and \cite{blg94} stars.  They
adjusted \cite{bg89}'s IRFM \teff\ estimates to put them onto the
\cite{blg94} scale and then fit polynomials to the effective temperatures
as a function of color to determine CT relations in Johnson's B--V,
V--R\footnote{\cite{gratton} has advised us that the a3 coefficient of the
\teff, V--R relation of the bluer class~III stars should be +85.49; it is
given as --85.49 in Table~1 of \cite{gcc96}.},
R--I, V--K and J--K colors.  In each
color, they derived three relations -- one represents the dwarfs, and the
other two apply to giants bluer and redder than some specific color.
For these relations, we have used \cite{bess79}'s color transformations to
convert the Johnson V--R and V--I colors to the Cousins system; \cite{bb88}
again supplied the V--K and J--K transformations to the Johnson-Glass system.

\cite{bcp98} combined the IRFM effective temperatures of \cite{blg94} and
temperatures estimated from the angular diameter measurements of \cite{dib87},
\cite{dbbr} and \cite{perrin} to derive a polynomial relation between V--K
and \teff\ for giants.
The color-temperature relations provided by \cite{bsl98} are
presumably derived in exactly the same manner from this same group of stars. 
For cool dwarfs, \cite{bcp98} also determined a CT relation, this time in V--I, 
from the IRFM effective temperatures of \cite{blg94} and \cite{aam96} (1996).

\cite{bess79} merged his V--I, V--K color-color relation for field giants and
the V--K, \teff\ relation of \cite{rjww} to come up with a V--I, \teff\ relation.
He then combined this with V--I, color trends to define CT relations for giants
in other colors.  By assuming that the same V--I, \teff\ relation applied to
giants and dwarfs with 4000~K~$\leq$~\teff~$\leq$~6000~K, \cite{bess79} 
derived dwarf-star CT relations in a similar manner.  For the coolest dwarfs,
updated versions of \cite{bess79}'s V--R, \teff\ and V--I, \teff\ relations
were given by \cite{bess95}.

\subsubsection{Comparisons of the Color-Temperature Relations}

\begin{figure}[p]
\epsfxsize=5.5in
\vspace*{-0.9in}
\hspace*{0.5in}
\epsfbox{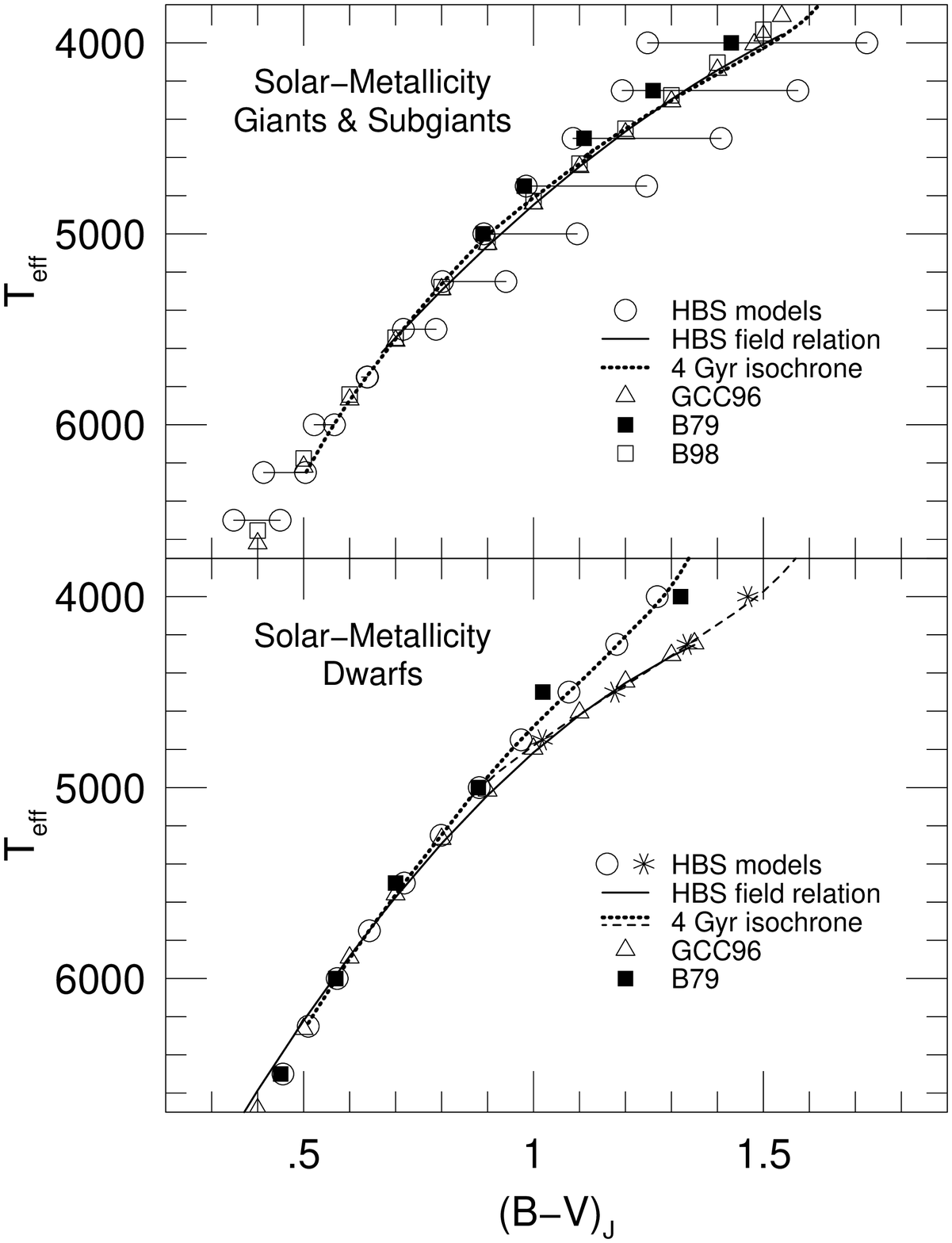}
\vspace*{-0.5in}
\caption{The Johnson B--V colors of the solar-metallicity models
are compared to (B--V) vs.~effective temperature relations of field stars.  The
upper panel presents the comparison for giants; the lower panel shows dwarfs.
In both panels, our model colors (HBS) are shown as open circles
(connected by solid lines in the giant-star panels), and
the dotted lines are the color-temperature relations predicted by our 4~Gyr,
Z$_{\odot}$ isochrone when the color calibrations of Table~3 are used; the
dashed line and asterisks are the lower main-sequence of the calibrated
isochrone and the HBS models, respectively, when the cool-dwarf color
calibrations are used instead for the dwarf models having
T$_{\rm eff}$~$<$~5000~K.
The solid lines are the empirical relations derived here from
the photometry of the field stars used to calibrate the synthetic colors.
The field relations of Gratton et al.~(1996; GCC96), Bessell (1979; B79) and
Bessell (1998; B98) are represented by
open triangles, filled squares and open squares, respectively.}
\label{teffbv}
\end{figure}

\begin{figure}[p]
\epsfxsize=5.5in
\vspace*{-0.9in}
\hspace*{0.5in}
\epsfbox{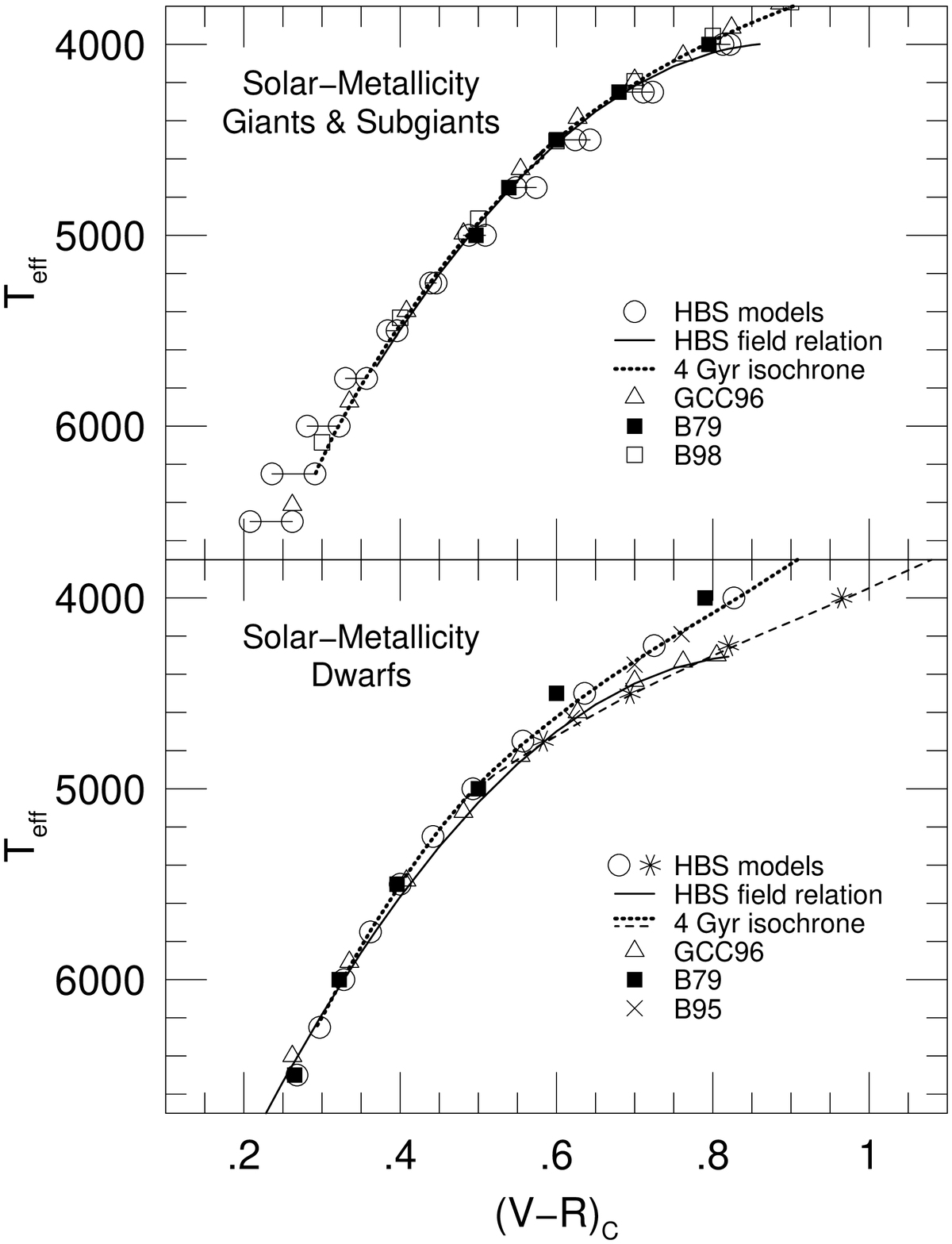}
\vspace*{-0.5in}
\caption{The Cousins V--R colors of the solar-metallicity models
are compared to (V--R) vs.~effective temperature relations of field stars.  The
upper panel presents the comparison for giants; the lower panel shows dwarfs.
In both panels, our model colors (HBS) are shown as open circles
(connected by solid lines in the giant-star panels), and
the dotted lines are the color-temperature relations predicted by our 4~Gyr,
Z$_{\odot}$ isochrone when the color calibrations of Table~3 are used; the
dashed line and asterisks are the lower main-sequence of the calibrated
isochrone and the HBS models, respectively, when the cool-dwarf color
calibrations are used instead for the dwarf models having
T$_{\rm eff}$~$<$~5000~K.
The solid lines are the empirical relations derived here from
the photometry of the field stars used to calibrate the synthetic colors.
The field relations
of Gratton et al.~(1996; GCC96), Bessell (1979; B79), Bessell (1998; B98)
and Bessell (1995; B95) are represented by open triangles, filled squares,
open squares and crosses, respectively.}
\label{teffvr}
\end{figure}

\begin{figure}[p]
\epsfxsize=5.5in
\vspace*{-0.9in}
\hspace*{0.5in}
\epsfbox{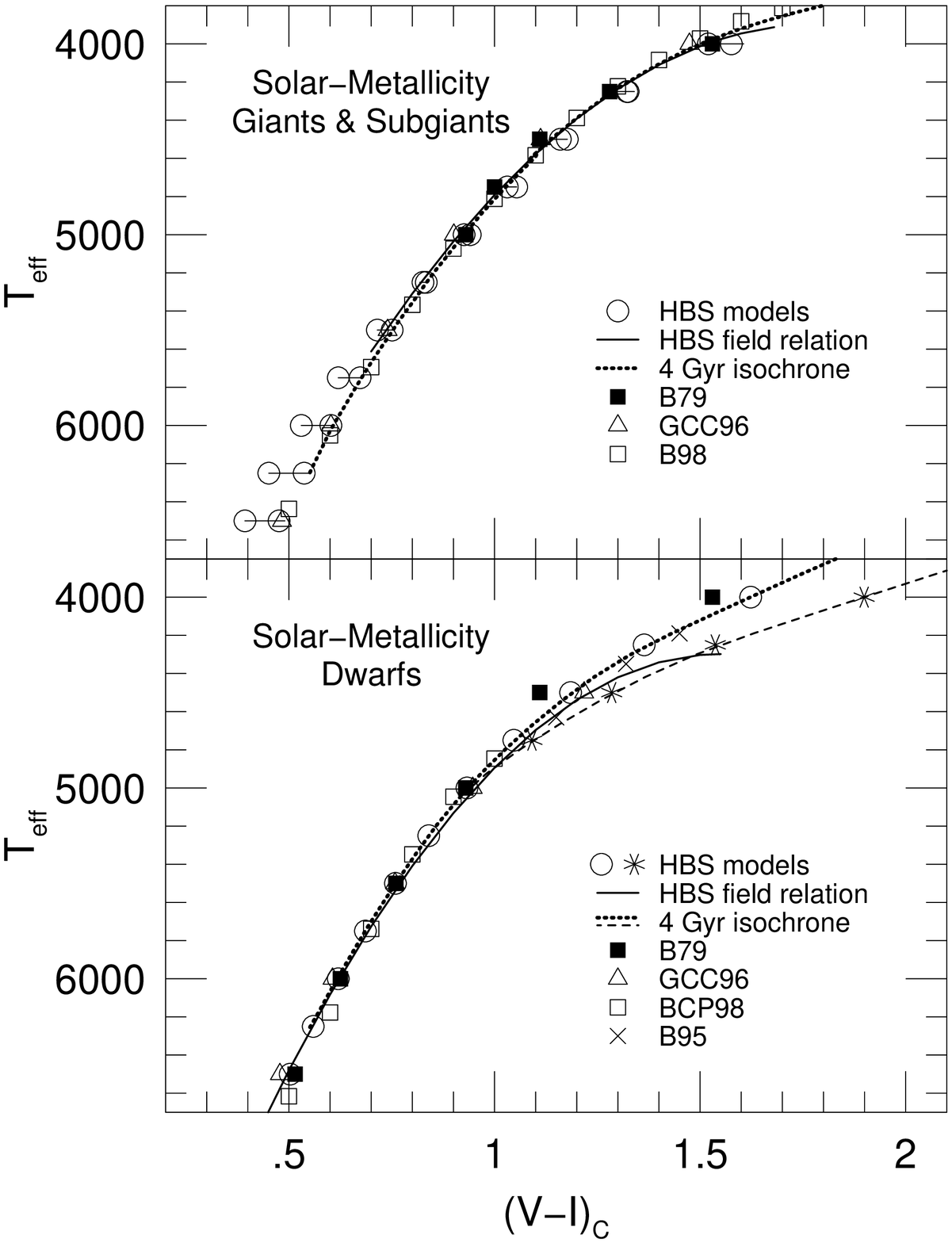}
\vspace*{-0.5in}
\caption{The Cousins V--I colors of the solar-metallicity models
are compared to (V--I) vs.~effective temperature relations of field stars.  The
upper panel presents the comparison for giants; the lower panel shows dwarfs.
In both panels, our model colors (HBS) are shown as open circles
(connected by solid lines in the giant-star panels), and
the dotted lines are the color-temperature relations predicted by our 4~Gyr,
Z$_{\odot}$ isochrone when the color calibrations of Table~3 are used; the
dashed line and asterisks are the lower main-sequence of the calibrated
isochrone and the HBS models, respectively, when the cool-dwarf color
calibrations are used instead for the dwarf models having
T$_{\rm eff}$~$<$~5000~K.
The solid lines are the empirical relations derived here from
the photometry of the field stars used to calibrate the synthetic colors.
The field relations
of Gratton et al.~(1996; GCC96), Bessell (1979; B79), Bessell (1998; B98),
Bessell et al.~(1998; BCP98) and Bessell (1995; B95) are represented by open
triangles, filled squares, open squares (giants), open squares (dwarfs) and
crosses, respectively.}
\label{teffvi}
\end{figure}

\begin{figure}[p]
\epsfxsize=6.0in
\vspace*{-0.9in}
\hspace*{0.25in}
\epsfbox{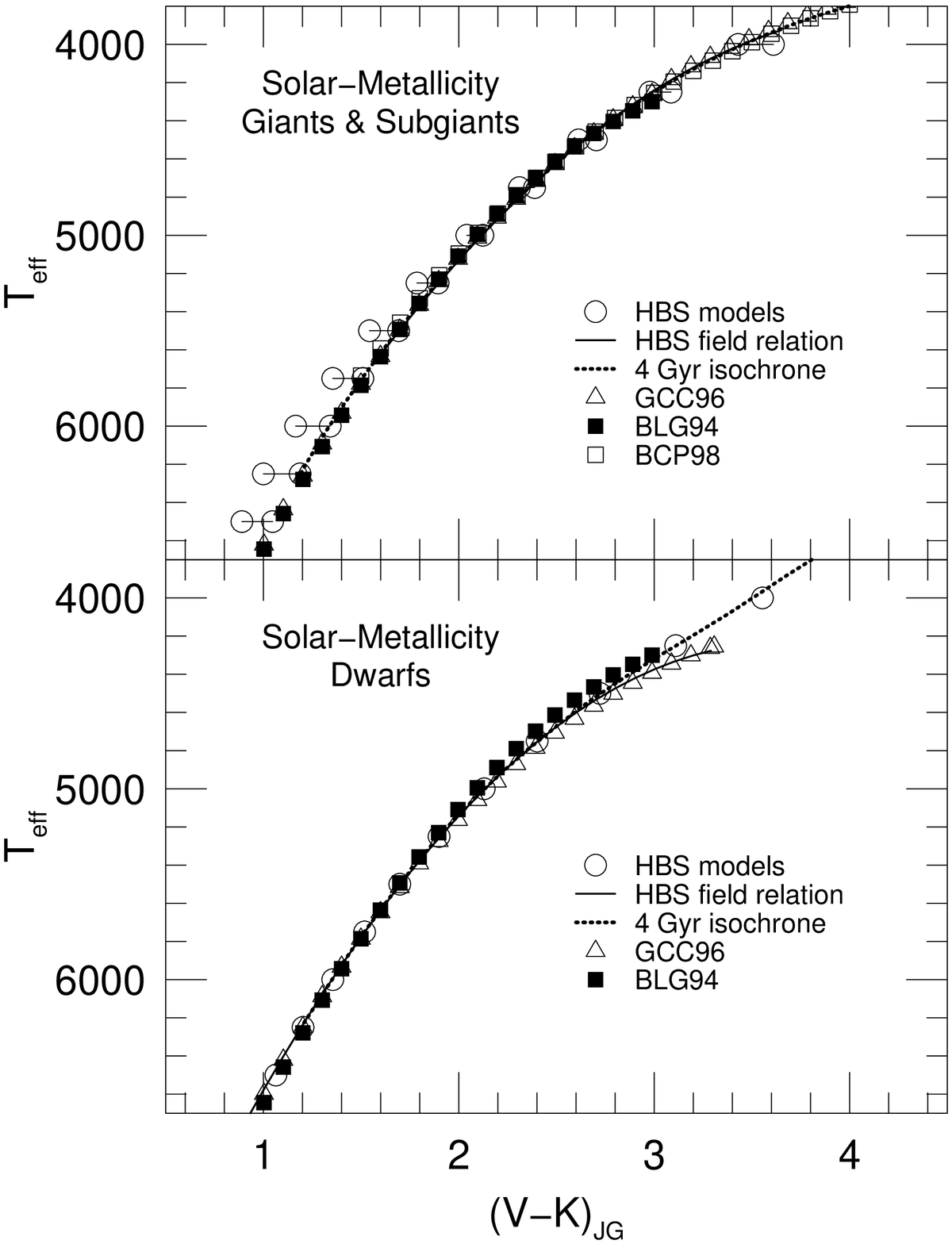}
\vspace*{-0.5in}
\caption{The Johnson-Glass V--K colors of the solar-metallicity
models are compared to (V--K) vs.~effective temperature relations of field
stars.  The upper panel presents the comparison for giants; the lower panel
shows dwarfs.  In both panels, our model colors (HBS) are shown as open circles
(connected by solid lines in the giant-star panels), and
the dotted lines are the color-temperature relations predicted by our 4~Gyr,
Z$_{\odot}$ isochrone.
The solid lines are the empirical relations derived here from
the photometry of the field stars used to calibrate the synthetic colors.
The field relations
of Gratton et al.~(1996; GCC96), Bessell et al.~(1998; BCP98) and Blackwell
\& Lynas-Gray (1994; BLG94) are represented by open triangles, open squares
and filled squares, respectively.}
\label{teffvk}
\end{figure}

\begin{figure}[p]
\epsfxsize=6.0in
\vspace*{-0.7in}
\hspace*{0.25in}
\epsfbox{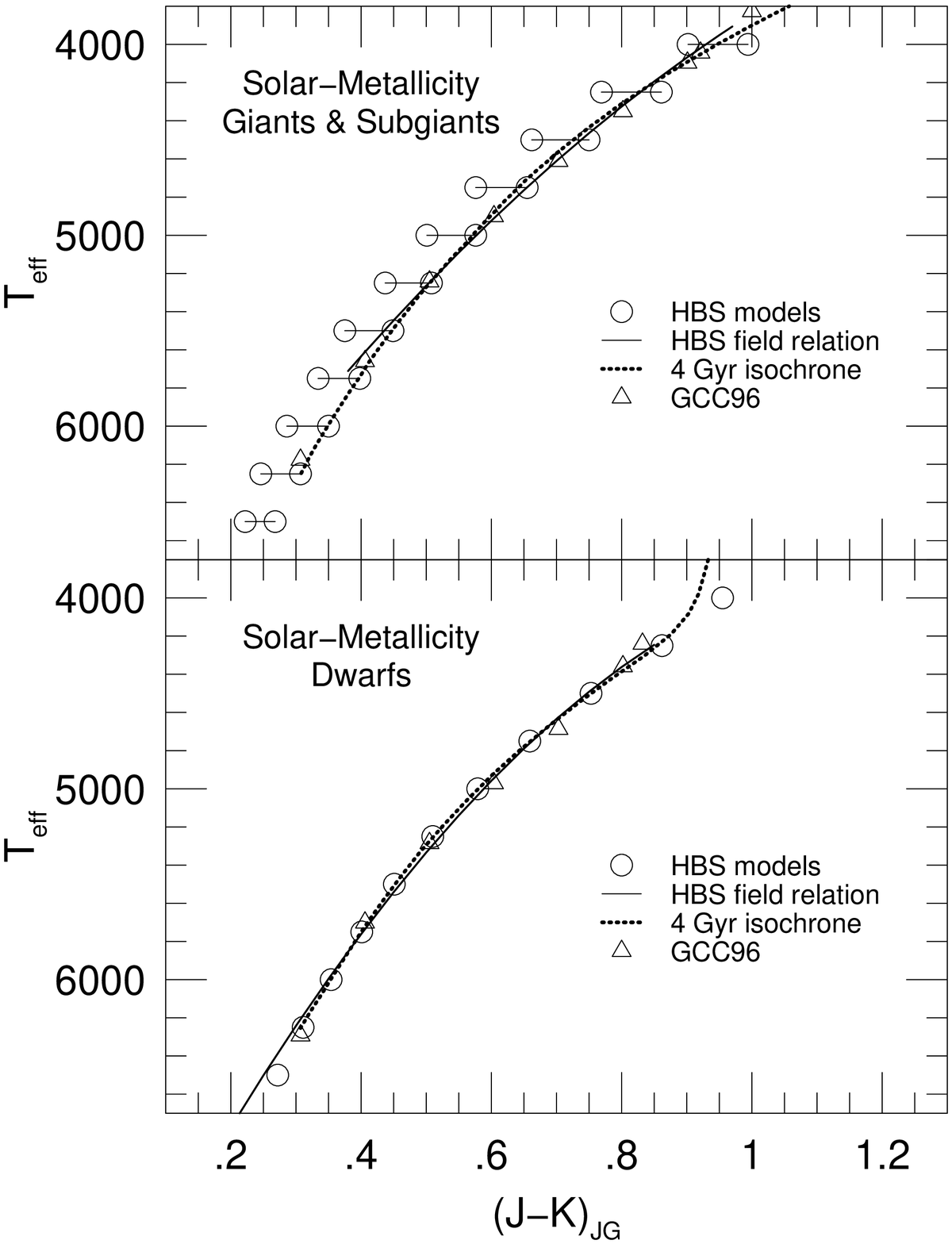}
\vspace*{-0.5in}
\caption{The Johnson-Glass J--K colors of the solar-metallicity
models are compared to (J--K) vs.~effective temperature relations of field
stars.  The upper panel presents the comparison for giants; the lower panel
shows dwarfs.  In both panels, our model colors (HBS) are shown as open circles
(connected by solid lines in the giant-star panels), and
the dotted lines are the color-temperature relations predicted by our 4~Gyr,
Z$_{\odot}$ isochrone.
The solid lines are the empirical relations derived here from
the photometry of the field stars used to calibrate the synthetic colors.
The field relations
of Gratton et al.~(1996; GCC96) are represented by open triangles.}
\label{teffjk}
\end{figure}

We compare our calibrated, solar-metallicity color grid to empirical
determinations of color-temperature relations of field stars in
Figures~\ref{teffbv}--\ref{teffjk}.  For these comparisons, we break our
grid up into giants \& subgiants, which we equate with models having
0.0~$\leq$~log~g~$\leq$~4.0, and dwarfs (log~g~=~4.5).  In all of the
figures in this section, the giant-star CT relations appear in the upper
panels, and the lower panels give the dwarf relations.  The calibrated
colors of our solar-metallicity grid are shown as open circles (or as asterisks
when the cool-dwarf color calibrations have been used), but
to relieve crowding in the giant-star plots, we connect the model colors at a
given \teff\ by a solid line and plot only the highest and lowest gravity
models as open circles.  Our empirical color-temperature relations
(Table~\ref{fieldrelns})
are shown as solid lines in Figures~\ref{teffbv}--\ref{teffjk};
the CT relations taken from the literature
are represented by the symbols indicated in the figures.  In addition, our
calibrated, solar-metallicity, 4~Gyr isochrone is shown as a dotted line; the
dashed line in the dwarf-star panels of Figures~\ref{teffbv}--\ref{teffvi} is
the calibrated isochrone which results when the cool-dwarf color calibrations
have been used.  Comparisons with our other solar-metallicity
isochrones shows that this CT relation is insensitive to age, so it should
agree closely with the field-star relations.  We omit the U--V and H--K
color-temperature relations from these comparisons because we
are not aware of any recent determinations of field-star relations in these
colors, which nevertheless are less well-calibrated than the others.

For the giant stars, our empirical CT relations are in excellent agreement
with the field relations of \cite{blg94}, \cite{gcc96}, \cite{bcp98} and
\cite{bsl98}, with the temperature differences at a given color usually much
less than 100~K.  The 4~Gyr, Z$_\odot$ isochrone also matches the field-giant
relations quite well, as would be expected from Figures~\ref{opfield}
and~\ref{irfield}.  In fact, most of the (small) differences between our
empirical CT relations and the isochrone are probably due to our decision to
use quadratic relations to represent the field stars; a quadratic function may
not be a good representation of the true color-temperature relationships,
and such a fit often introduces a bit of excess curvature at the color extremes
of the observational data.  The \cite{bess79} color-temperature relations for
giants are also seen to be in good agreement with the other empirical and
theoretical, giant-star data in V--R and V--I, but his B--V, \teff\ relation
is bluer than the others; we have not attempted to determine the reason for
this discrepancy.

Overall, since the colors of the field giants lie within the bounds of our
solar-metallicity grid at a given effective temperature and also show good
agreement with the 4~Gyr isochrone, our calibrated models are accurately
reproducing the colors of solar-metallicity, F--K field
giants of a specific temperature and surface gravity.
Unfortunately, the dwarf-star color-temperature relations do not show the same
level of agreement as those of the giants, overlying one another for
\teff~$\gtrsim$~5000~K but diverging at cooler effective temperatures.
Therefore, we will examine each of the CT relations of the dwarfs individually.

In B--V (Figure~\ref{teffbv}), the field stars from
Table~\ref{paramtable} produce a CT relation which is in good agreement with
that of \cite{gcc96}.  When using the color calibrations 
of Table~\ref{colorcaltable}, our isochrone and grid models of cool dwarfs
are bluer than the empirical relations at a given \teff, which is consistent
with the comparison to M67 in Figure~\ref{m67cmds}, indicating that these
calibrated B--V colors are too blue.  On the other hand, the
agreement between these models and the CT relation of \cite{bess79} suggests
that these grid and isochrone colors are essentially correct.
Similar quandaries are posed by the color-temperature relations of the dwarfs
in V--R and V--I (Figures~\ref{teffvr} and~\ref{teffvi}).  Here, the
\cite{bess79} relations predict slightly bluer colors than the analogous
isochrone and grid models at a given temperature, but \cite{bess95}'s updated
data for the coolest dwarfs agrees with our calibrated isochrone and models
quite well.  Our empirical CT relations
for dwarfs and the \cite{gcc96} data, on the other hand, lie to the red
of these models and isochrones, again showing the same kind of discrepancy seen
in the M67 comparisons.  Unfortunately, the \cite{bcp98} V--I, \teff\ 
relation does not extend to cool enough temperatures to help resolve the
situation.  Supplementing the color calibrations of Table~\ref{colorcaltable}
with the (optical) cool-dwarf color calibrations, however, brings the
solar-metallicity isochrone and models into agreement with all of the
field-dwarf CT relations of \cite{gcc96} instead.

In Figure~\ref{teffvk}, we also see that there is one empirical CT
relation (\cite{blg94}) which indicates that our cool-dwarf models are
essentially correct
at a given \teff, while another (\cite{gcc96}) predicts that they are too blue,
although the magnitude of disagreement in this figure is much smaller than that
seen in Figures~\ref{teffbv}--\ref{teffvi}.  Since we are not aware of any
near-infrared photometry for lower-main-sequence stars in M67, we cannot use
Figure~\ref{m67cmds} to bolster any of the V--K CT relations.
The only color in which all of the theoretical
and empirical data are in relative agreement for the dwarfs is J--K; this is
shown in Figure~\ref{teffjk}.

We hesitate to emphasize the similarities between our empirical
color-temperature relations and the others plotted in the dwarf-star panels of
Figures~\ref{teffbv}--\ref{teffjk}.  Our V--K CT relation for field stars only
differs from that of \cite{blg94} because they neglected to treat dwarfs
and giants separately; if we do the same, these differences disappear.  Even
so, the coolest dwarf in the \cite{blg94} study was
BS~1325, a K1~V star with \teff~=~5163~K, which is hotter than the temperature
at which the dwarf and giant CT relations begin to bifurcate, so it is
somewhat misleading to extend \cite{blg94}'s relation to redder colors in the
dwarf-star panel of
Figure~\ref{teffvk}.  Likewise, it is not surprising that our CT relations
and \cite{gcc96}'s relations are so compatible -- \cite{gcc96} used the
\cite{blg94} and \cite{bg89} data to define their relations, so the cool end
of their dwarf relations are defined by the same cool dwarfs included in our
sample.  The poor agreement between the other empirical
CT relations and the dwarf-star relations of \cite{bess79} is probably due to
the manner in
which \cite{bess79} derived his field-star relations.  He assumed a similar
V--I, \teff\ relation for cool dwarfs and cool giants, an assumption which
Figure~\ref{opfield} suggests to be inappropriate.

Overall, then, if we assume that the effective temperatures and photometry
that we have adopted for the field stars of Table~\ref{paramtable} and
the members of M67 are all correct, we can account for the similarities
and the differences between our color-temperature relations and those taken
from the literature.  In this case, it appears that the
CT relations of \cite{gcc96} are the most representative of field dwarfs, while
each of those which we have examined are probably equally reliable for the field
giants (except \cite{bess79}'s B--V, \teff\ relation).  This also produces a consistent picture in which both the field-star
CT relations and the M67 photometry indicate that our model colors for cool
dwarfs are too blue at a given effective temperature if only the color
calibrations of Table~\ref{colorcaltable} are employed; adopting the cool-dwarf
color calibrations for \teff~$<$~5000~K then brings the models and the
observational data into close agreement.

However, we can envision an alternate scenario in which the coolest three (or
more) field dwarfs have been assigned effective temperatures which are too hot.
By lowering \teff\ by $\sim$100~K for HR~8085 and HR~8832 and by $\sim$200~K
for HR~8086, these stars would lie along the the same color calibrations as
the other field stars in all colors except perhaps U--V and B--V.  This would
shift our empirical CT relations and those of \cite{gcc96} to bluer colors for
the cool dwarfs and thus resolve the disagreement with \cite{bess95}'s relations
and the models calibrated using the Table~\ref{colorcaltable} color
calibrations.  It would also leave V--R and V--I as the only two colors in which
the dwarf-star and giant-star CT relations differ significantly, which is
understandable, since the V, R and I bands contain many molecular features
(e.g., TiO, CN) which are gravity-sensitive.  However, the discrepancies between
the calibrated, 4~Gyr, solar-metallicity isochrone (solid line in
Figure~\ref{m67cmds}) and the U--V and B--V photometry of the lower
main-sequence of M67 would likely remain.

Since it is not clear to us which of these situations applies, we have
presented the calibrated model colors for each case here.  The essential pieces
of information needed to disentangle these two possibilities are
angular diameter measurements of K~dwarfs; these would certainly aid us in
testing the \cite{bg89} temperatures of the cool dwarfs in our sample and
confirming their CT relations.  However, for the time-being, the small number of
cool dwarfs with effective temperature estimates and the fact that such stars
make only a small contribution to the integrated light of a galaxy have led us
to adopt the calibration relations given in Table~\ref{colorcaltable} for the
synthetic colors of all of our evolutionary synthesis models.

\subsection{Comparisons to Previous MARCS/SSG Color-Temperature Relations}

In Figures~\ref{opmodels} and~\ref{irmodels}, we compare our new,
solar-metallicity CT relations to the previous MARCS/SSG results published
by \cite{vb85}
and \cite {bg89}; the models of the former and latter are shown as open and
filled circles, respectively.  In each panel of these figures,
we plot log~g vs. color and use dotted lines to represent isotherms resulting
from the uncalibrated colors of our models; the isotherms produced by our
improved, calibrated colors (Table~\ref{colorcaltable} calibrations only)
are shown as solid lines.  Thus, differences
between the older colors and the appropriate dotted line can be identified
with improvements in the models and color measurements, and differences
between the dotted and solid lines of the same temperature can be equated
with the effects of calibrating the synthetic colors.

\subsubsection{The Optical Color-Temperature Relations}

\begin{figure}[p]
\epsfxsize=6.0in
\vspace*{-0.5in}
\hspace*{0.25in}
\epsfbox{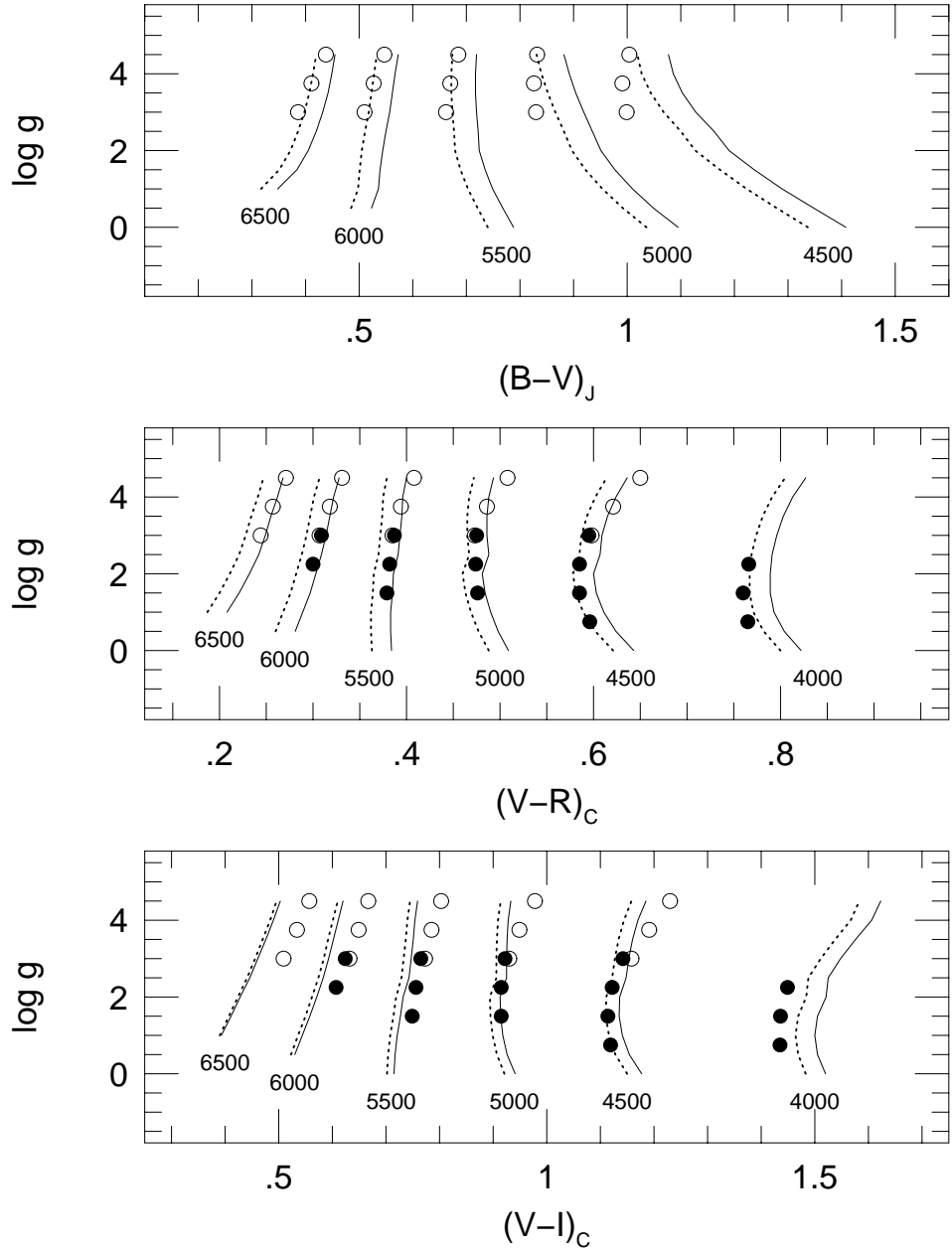}
\vspace*{-0.5in}
\caption{The Johnson B--V, Cousins V--R and Cousins V--I colors
of the current and previously-published MARCS/SSG models are compared
in the upper, middle and lower panels, respectively.  The uncalibrated colors
of our solar-metallicity models of a given effective temperature are shown
as dotted lines; solid lines give the corresponding calibrated colors.  Each
set of models is labeled with the \teff\ of the isotherm.  Colors taken
from Bell \& Gustafsson (1989; BG89) are shown as filled circles and those of
VandenBerg \& Bell (1985; VB85) as open circles.}
\label{opmodels}
\end{figure}

In the upper panel of Figure~\ref{opmodels}, we compare our CT relations and
the B--V colors calculated by \cite{vb85} using the filter-transmission-profiles
of \cite{ms63}.  The calibrated B--V colors are always redder than their previous
counterparts, but the uncalibrated colors of the new models do not differ appreciably
from those of \cite{vb85} for \teff~$\gtrsim$~5000~K; the main improvement
in the B--V colors of these hotter stars is due to the color calibrations.
At 4500~K, especially as log~g decreases, the improved low-temperature opacity
data used here becomes important, and this becomes the dominant
improvement for cooler stars, although the color corrections inferred from the
calibrations are still substantial.

The V--R colors of \cite{vb85} and \cite{bg89}, measured with the R-band filter
of \cite{cuz80},
are compared to our new, solar-metallicity color grid in the middle panel of
Figure~\ref{opmodels}.  Surprisingly, for effective temperatures hotter than
about 5000~K, the colors of the older models are in good agreement with those
of the newer, {\it calibrated} models; either the color differences
produced by the filter-transmission-profiles adopted by each group are being
almost exactly offset
by the effects of the color calibrations which we have derived, or the newer
opacities used here have moved the V--R colors of the models to the blue.
For the cool giants, however, the calibrated V--R colors are significantly
redder than those previously published.

For V--I, the new and old models are shown in the lower panel of
Figure~\ref{opmodels}.
This diagram is similar to the V--R plot in that it is difficult to disentangle
the color changes produced by the use of different filter profiles (\cite{vb85}
and \cite{bg89} use the I-band filter of Cousins 1980) and those
resulting from improvements in the stellar modelling.  It is apparent, however,
that the previous MARCS/SSG V--I colors are too red at hotter temperatures
(\teff~$\gtrsim$~5500~K) and too blue at cooler temperatures.

\subsubsection{The Near-Infrared Color-Temperature Relations}

\begin{figure}[p]
\epsfxsize=6.0in
\vspace*{-1.2in}
\hspace*{0.25in}
\epsfbox{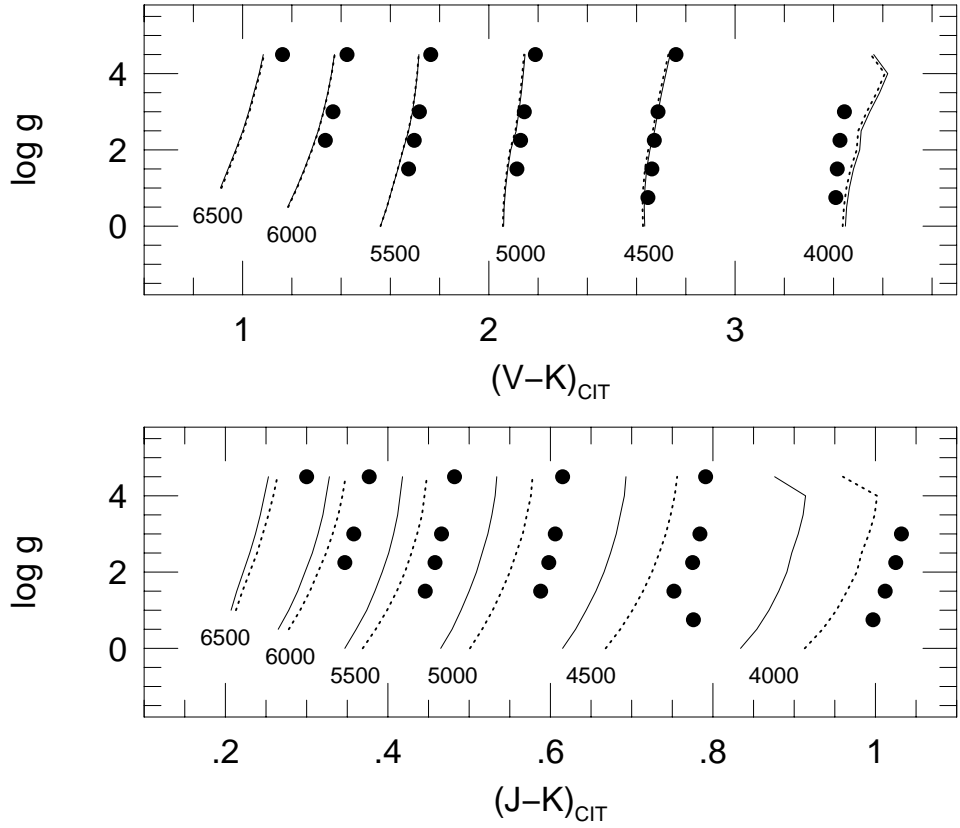}
\vspace*{-1.9in}
\caption{The CIT/CTIO V--K and J--K colors
of the current and previously-published MARCS/SSG models are compared
in the upper and lower panels, respectively.  The uncalibrated colors
of our solar-metallicity models of a given effective temperature are shown
as dotted lines; solid lines give the corresponding calibrated colors.  Each
set of models is labeled with the \teff\ of the isotherm.  Colors taken
from Bell \& Gustafsson (1989) are shown as filled circles.}
\label{irmodels}
\end{figure}

Figure~\ref{irmodels} (upper panel) shows that the synthetic V--K colors on the
CIT/CTIO system are not altered appreciably by calibration, at least with
respect to the color range of the models and the magnitude of the color
differences which exist between the newer and older models.
While the colors of our
models and those of \cite{bg89} are about the same at \teff~=~4500~K, the newer
models are redder at temperatures cooler than this and the older models are
redder at hotter temperatures.  Differences in the model atmospheres and
synthetic spectra are responsible for these effects, since \cite{bg89} used
the same filter-transmission-profiles that we have.

The CIT/CTIO J--K colors shown in the lower panel of Figure~\ref{irmodels} are
all significantly bluer than those previously published.  For the hotter models,
improvements in MARCS and the much better treatment of line absorption in SSG
are the major causes of these blueward shifts.  In fact, the magnitude of the
color differences between the older colors and the newer, uncalibrated
colors is about the same at all temperatures.  The color calibrations, however,
while also making the model J--K colors bluer, grow in importance as
\teff\ decreases; this effect begins to dominate the color shifts due to the
model improvements at \teff~$\sim$~4500~K.

\section{Conclusion}

We have calculated colors and bolometric corrections for grids of stellar
models having
4000~$\leq$~\teff~$\leq$~6500~K, 0.0~$\leq$~log~g~$\leq$~4.5 and
--3.0~$\leq$~[Fe/H]~$\leq$~0.0.  The synthetic colors which we measure
include Johnson U--V and B--V; Cousins V--R and V--I; Johnson-Glass V--K, J--K and
H--K; and CIT/CTIO V--K, J--K, H--K and CO.
These synthetic colors have been determined by convolving the most recent
estimates of the filter-transmission-profiles of the respective photometric
systems with newly-calculated synthetic spectra.  The synthetic spectra are
produced by the SSG spectral synthesis code, which has been updated
significantly, especially in terms of its low-temperature opacity data and
spectral line lists.  The MARCS model atmospheres which are used by SSG have
also been calculated using this new opacity data and improved opacity
distribution functions.

We have found that the initial synthetic colors require small corrections to
be put onto the observational systems.  Here, we have assembled a
set of field stars which have effective temperatures measured using the
infrared-flux method (IRFM), and we have used photometry of these stars to
derive the color calibrations needed to put the synthetic colors onto the
photometric systems.
The scale factors and zero points of these calibrations are gratifyingly
close to unity and small, respectively, but the UBVRI colors of the coolest
dwarfs (\teff\ $<$~5000~K) may require a more substantial correction.
Nevertheless, the specific
color calibrations which we adopt are only applicable to models calculated
with our versions of the MARCS and SSG computer codes.  We encourage
others to examine the need for such calibrations in their models as well.

Using the field-star data adopted in deriving the color calibrations, we
have also found that the recent angular diameter measurements of Mozurkewich
et al. (1991), Mozurkewich (1997) and Pauls et al. (1997) match the angular
diameters predicted by Bell \& Gustafsson (1989) very well, thus giving strong
confirmation of the latter's effective temperature scale, which we have adopted.
Since most of the synthetic colors are much more sensitive
to effective temperature than they are to the other model parameters, such as
metallicity and surface gravity, this infers that the differences
between the uncalibrated, synthetic colors and the field-star photometry
are not caused by errors in the temperatures of the field-star models.
The coolest field dwarfs may again prove an exception to this generalization,
as none of their angular diameters have been measured and their effective
temperatures are therefore more uncertain.
Knowing that the IRFM-derived temperatures are accurate has also allowed us to
determine empirical, color-temperature (CT) relations for these field F--K
stars.

After calibrating our synthetic colors, our newly-calculated, 4~Gyr, Z$_\odot$
isochrone is seen to be in good agreement with optical and near-infrared
color-magnitude diagrams of M67 from the upper main-sequence through the tip
of the red-giant branch.  For giants and for dwarfs hotter than about 5000~K,
our theoretical, solar-metallicity CT relations
and solar-metallicity isochrones are also found to match the empirical,
field-star color-temperature relations derived here and found elsewhere in the
literature.
For the coolest dwarfs (\teff\ $<$ 5000~K), there are some differences between
our empirical color-temperature relations and the
best-determined relations of others.  The calibrated colors of the cool-dwarf
models also remain
a bit bluer than the field-star relations (at a given \teff) and the M67
main-sequence stars (at a given absolute magnitude).  Because we have not been
able to determine whether the synthetic, optical colors of the cool dwarf models
are too blue because their adopted effective temperatures are too hot or the
synthetic spectra are in error, we present alternate UBVRI colors for our
models having \teff~$<$~5000~K and log~g~=~4.5.  These colors are based upon
separate color calibrations derived from only the cool field dwarfs and which
greatly improve the agreement between the calibrated, synthetic colors and
the empirical data.  We leave it to the reader to decide which model colors they
prefer.

This work would greatly benefit from further effective temperature
estimates of cool dwarfs and measurements of the corresponding angular
diameters needed to confirm them.  It would also be quite useful to acquire
additional photometry of the field stars used here to
derive the color calibrations.  Only a small subset of these stars, which
includes no cool dwarfs, has been observed on the Cousins system, making
the V--R and V--I color calibrations less well-determined than some of the others.
In addition, only Johnson near-infrared photometry was available for the
field stars.  Color transformations had to be applied to this
photometry before the Johnson-Glass and CIT/CTIO color calibrations could be
derived; this also prohibited the determination of a similar calibration for
the CIT/CTIO CO index.

Overall, the agreement between our theoretical CT relations and the empirical
data and between our 4 Gyr, solar-metallicity isochrone and the photometry of
M67 indicates that our synthetic spectra are generally providing a good
representation of
the spectral energy distributions of stars of the corresponding \teff, surface
gravity and metallicity.  This conclusion is further supported by our synthetic
spectra of M~giants, which are presented in a companion paper (Houdashelt
et al. 2000).  In a future paper (Houdashelt et al. 2001), we will use
similar synthetic spectra and newly-constructed isochrones as part of a program
to simulate the spectral energy distributions of early-type galaxies through
evolutionary synthesis.

\acknowledgments

We would like to thank the National Science Foundation (Grant AST93-14931)
and NASA (Grant NAG53028) for their support of this research.  We also
thank Ben Dorman for allowing us to use his isochrone-construction code
and Mike Bessell for providing many helpful suggestions on the manuscript.
MLH would like to express his gratitude to Rosie Wyse for providing support
while this work was completed.
The research has made use of the Simbad database, operated at CDS, Strasbourg,
France. The NSO/Kitt Peak FTS data used here were produced by NSF/NOAO.

%\clearpage

\begin{table}
\dummytable\label{angdiamtable}
\end{table}

\begin{table}
\dummytable\label{paramtable}
\end{table}

\begin{table}
\dummytable\label{colorcaltable}
\end{table}

\begin{table}
\dummytable\label{errortable}
\end{table}

\begin{table}
\dummytable\label{gridtable}
\end{table}

\begin{table}
\dummytable\label{fieldrelns}
\end{table}

\clearpage

\begin{figure}[p]
\vspace*{-1.5in}
\hspace*{-1.0in}
\epsfbox{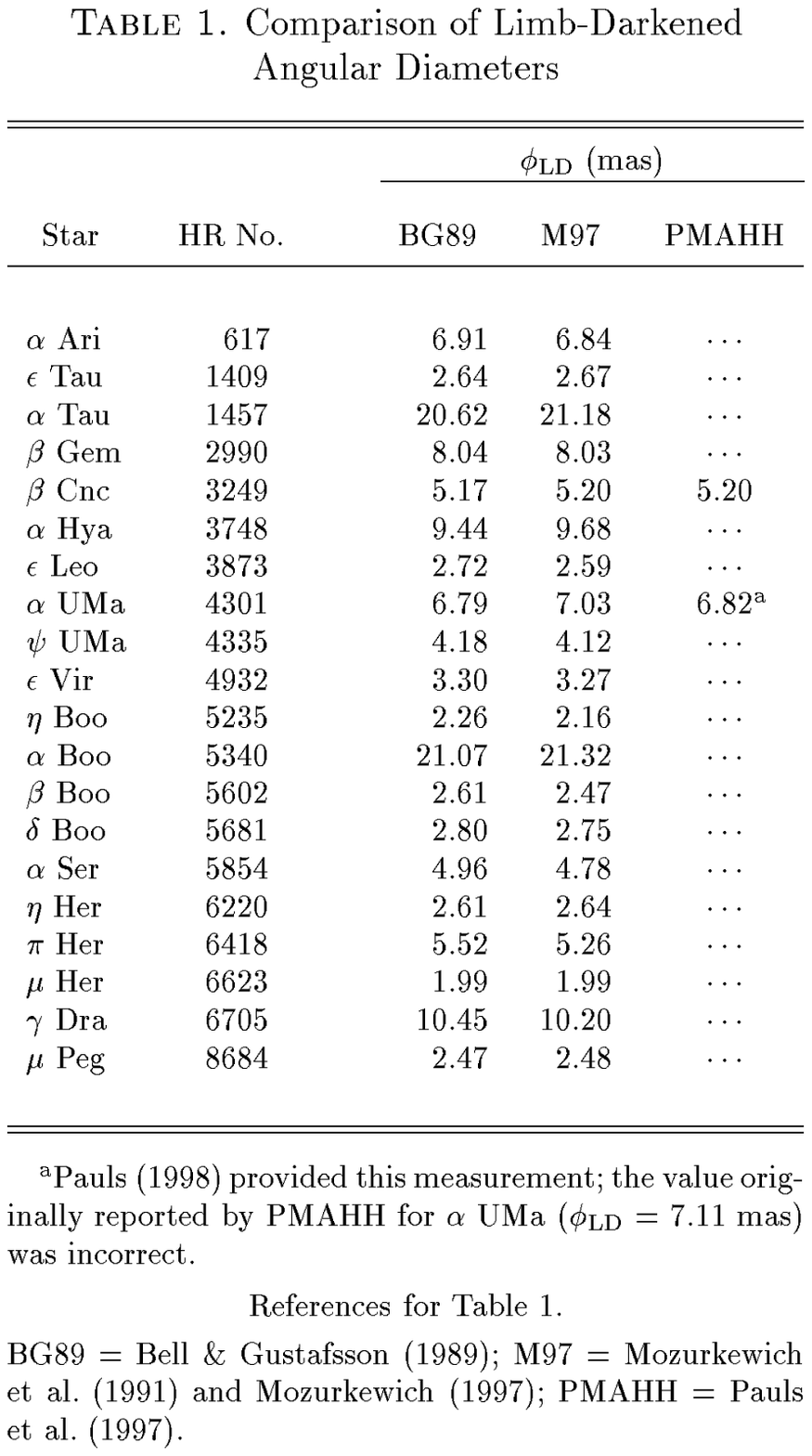}
\end{figure}

\begin{figure}[p]
\vspace*{-1.5in}
\hspace*{-1.0in}
\epsfbox{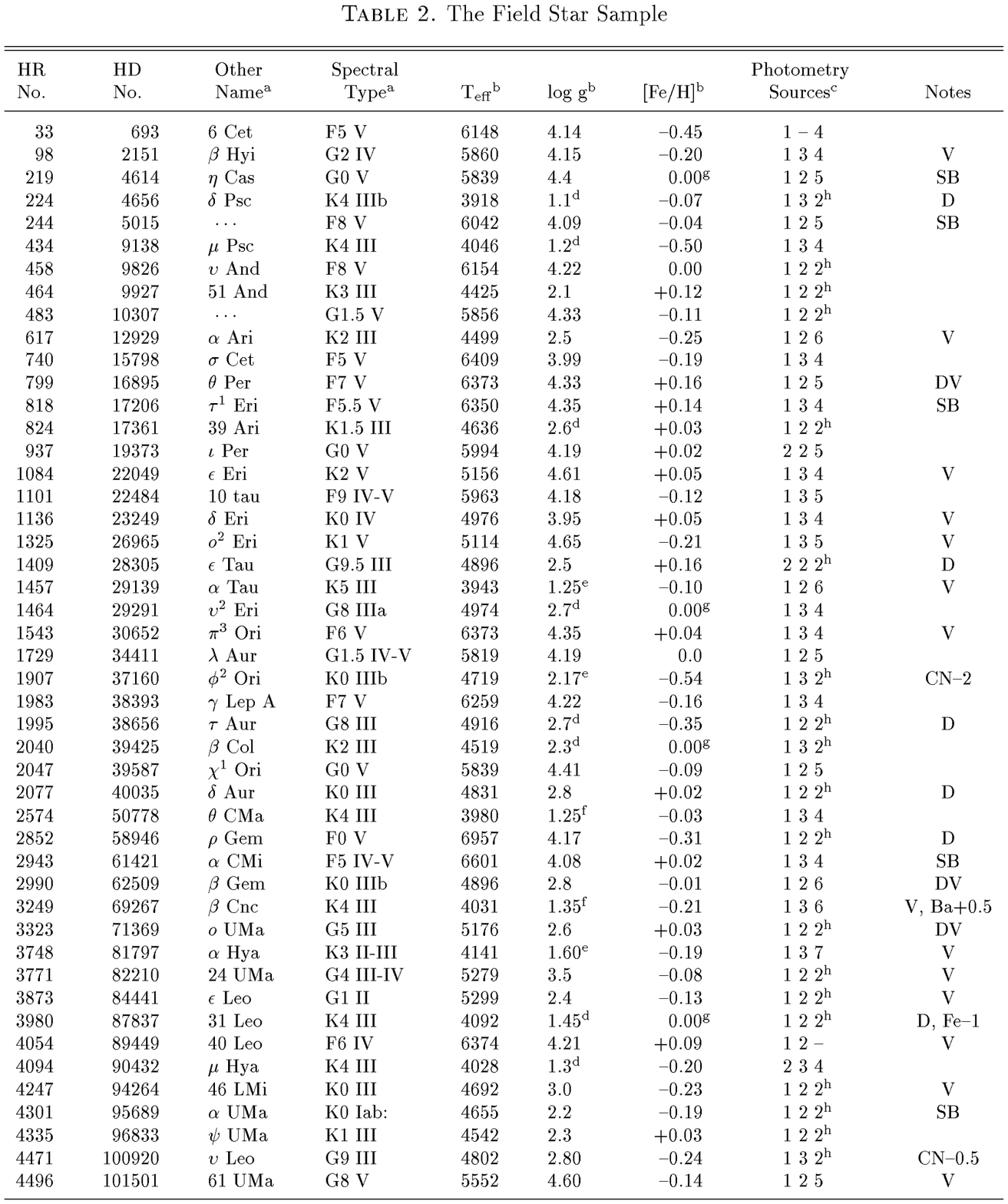}
\end{figure}

\begin{figure}[p]
\vspace*{-1.4in}
\hspace*{-1.0in}
\epsfbox{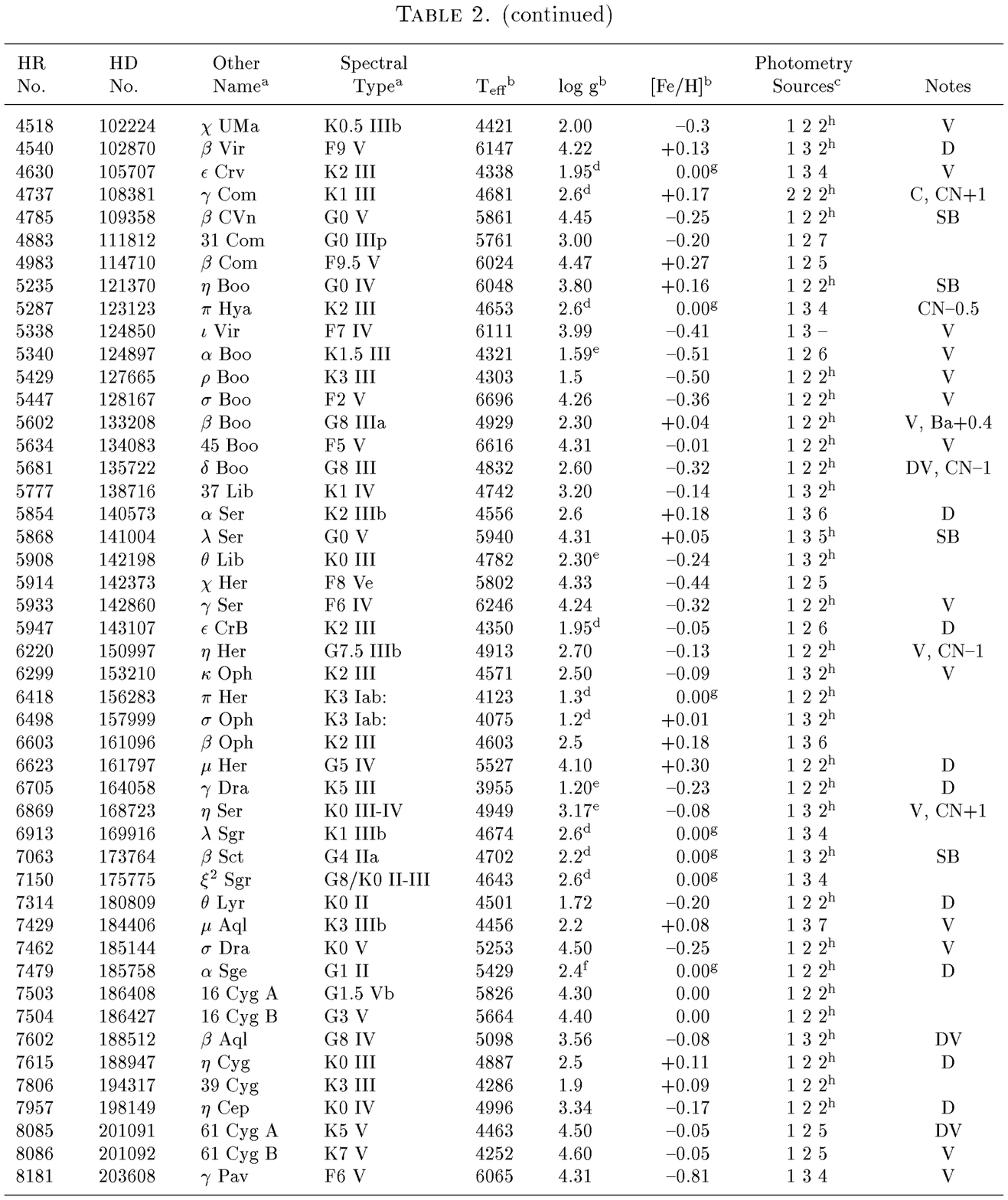}
\end{figure}

\begin{figure}[p]
\vspace*{-1.5in}
\hspace*{-1.0in}
\epsfbox{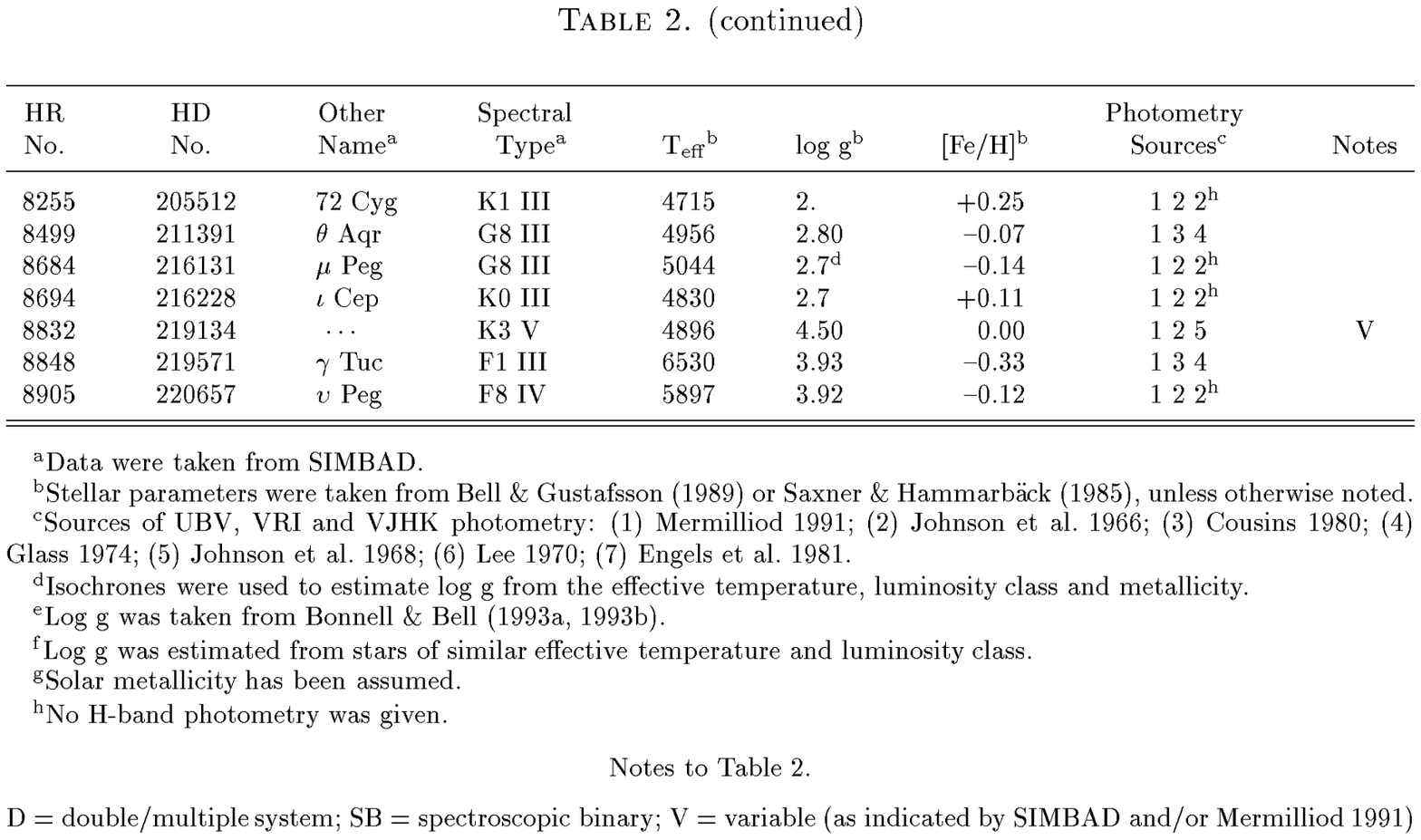}
\end{figure}

\begin{figure}[p]
\vspace*{-1.5in}
\hspace*{-1.0in}
\epsfbox{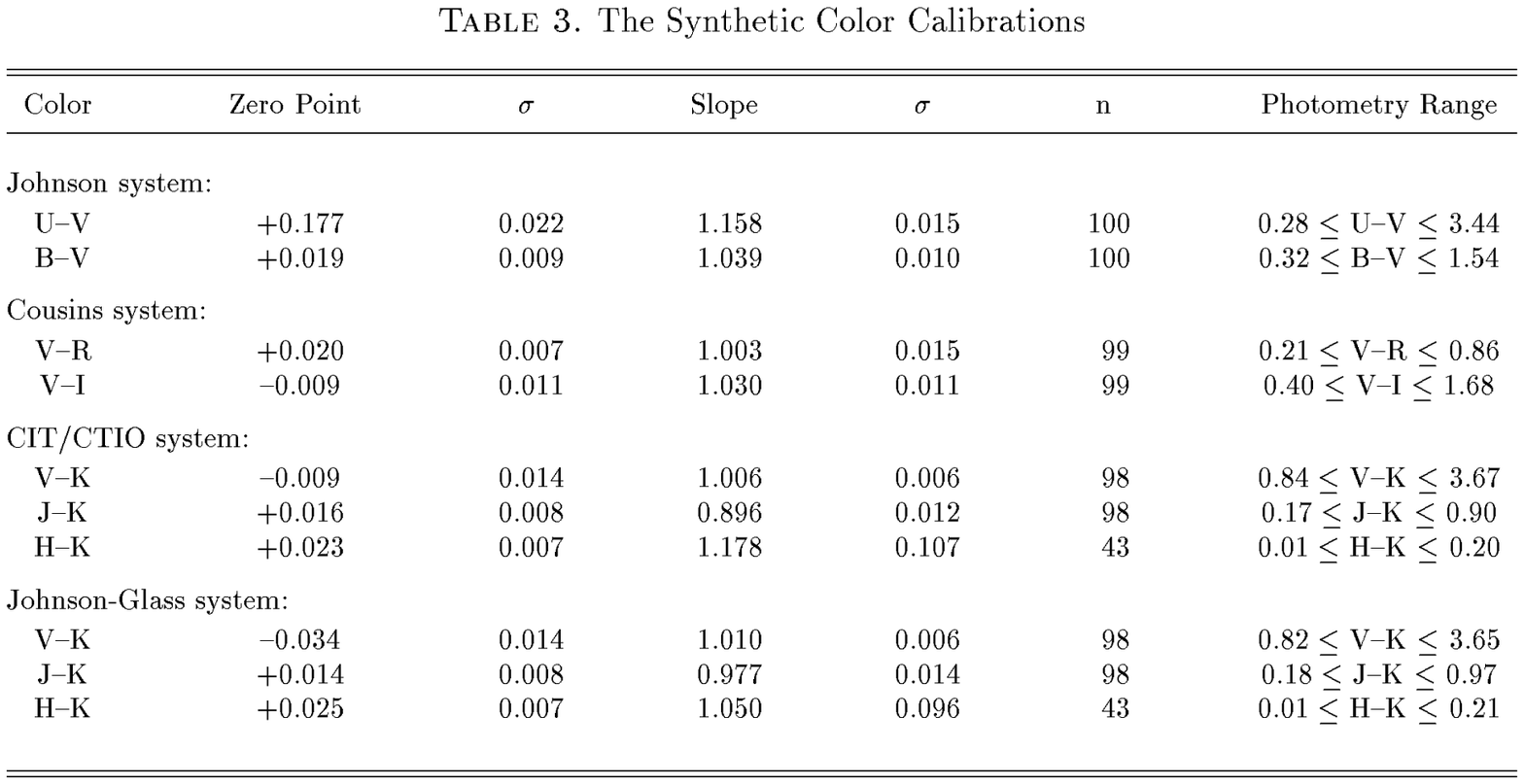}
\end{figure}

\begin{figure}[p]
\vspace*{-1.5in}
\hspace*{-1.0in}
\epsfbox{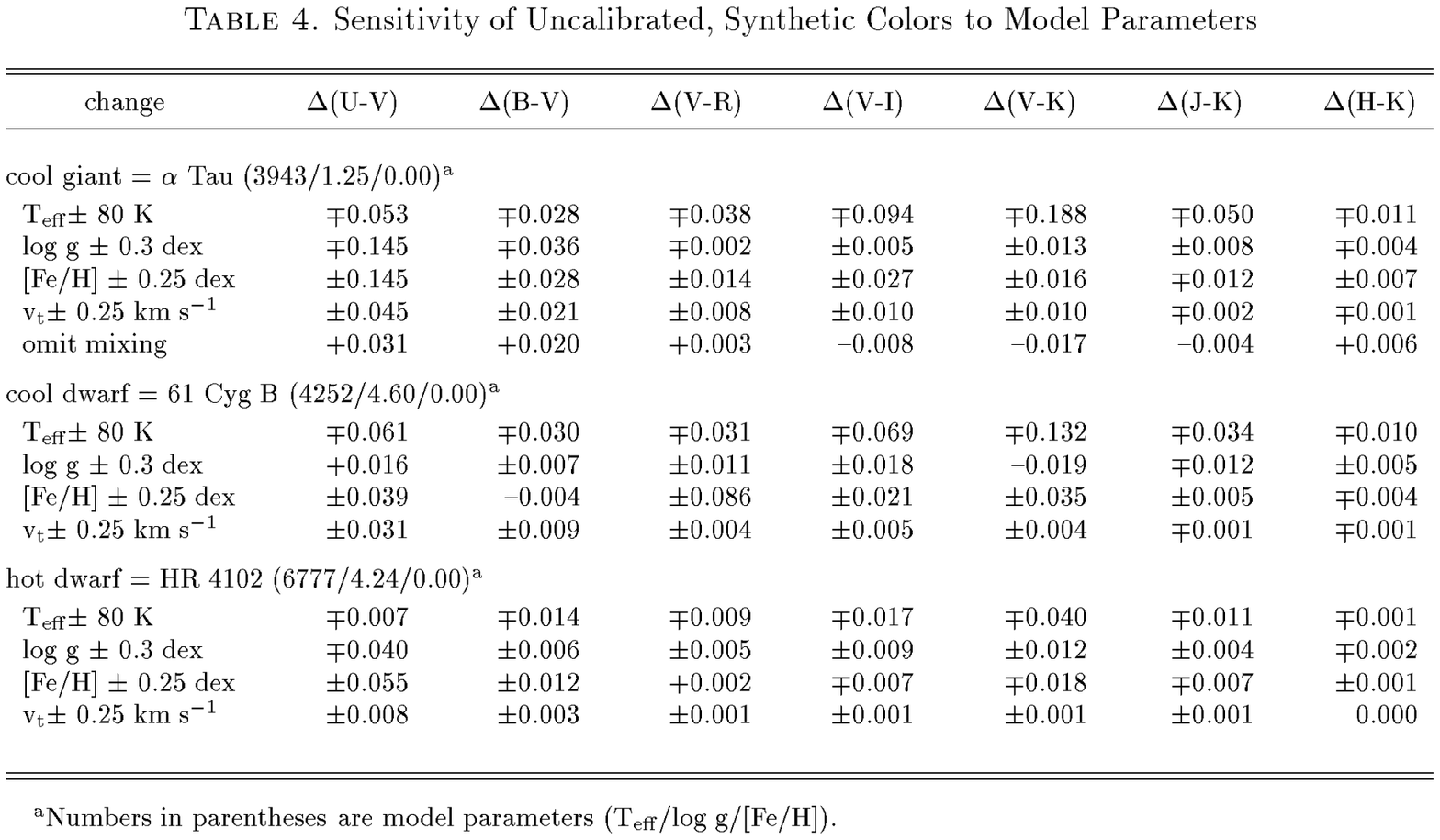}
\end{figure}

\begin{figure}[p]
\vspace*{-1.3in}
\hspace*{-1.0in}
\epsfbox{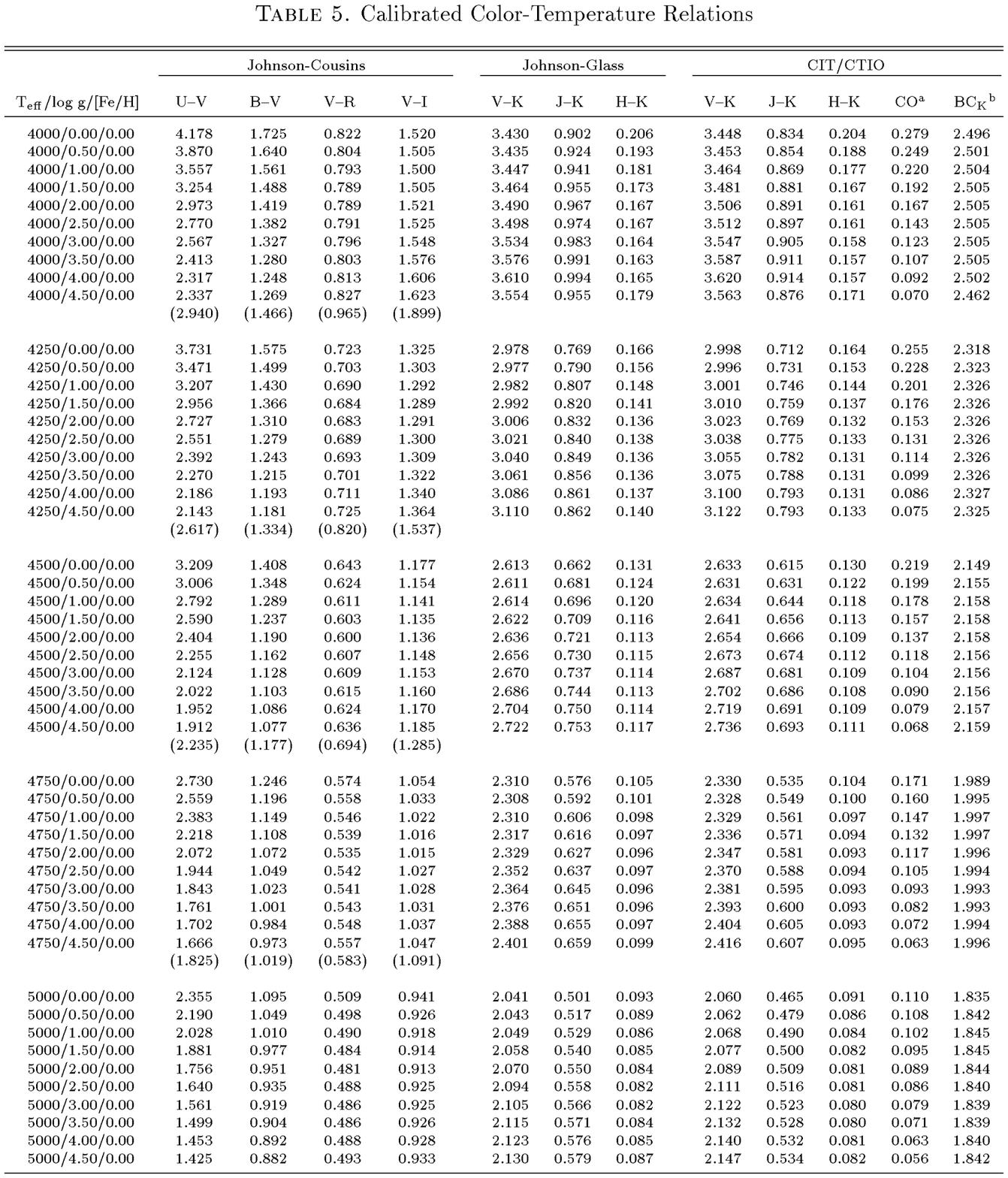}
\end{figure}

\begin{figure}[p]
\vspace*{-1.3in}
\hspace*{-1.0in}
\epsfbox{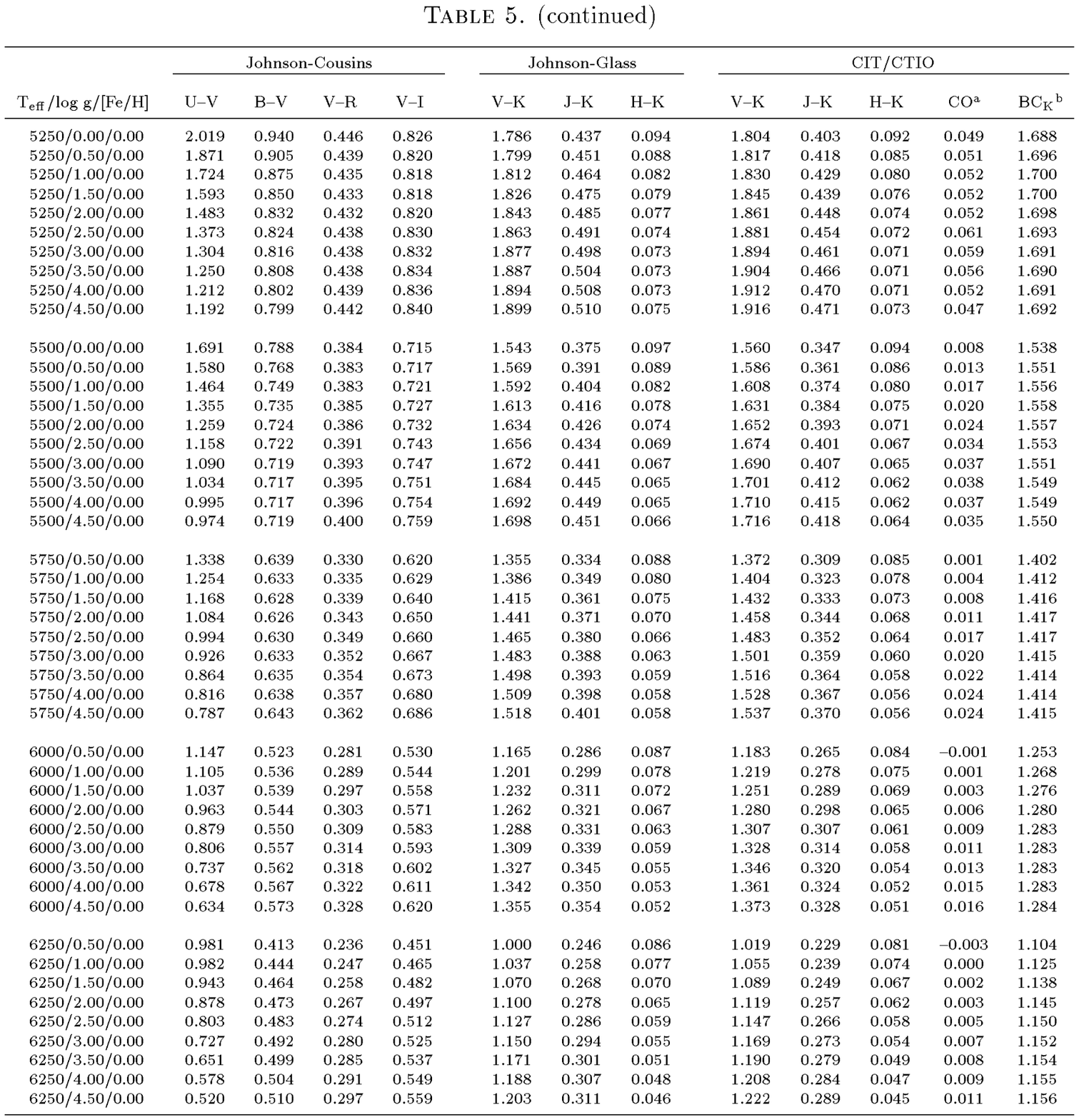}
\end{figure}

\begin{figure}[p]
\vspace*{-1.3in}
\hspace*{-1.0in}
\epsfbox{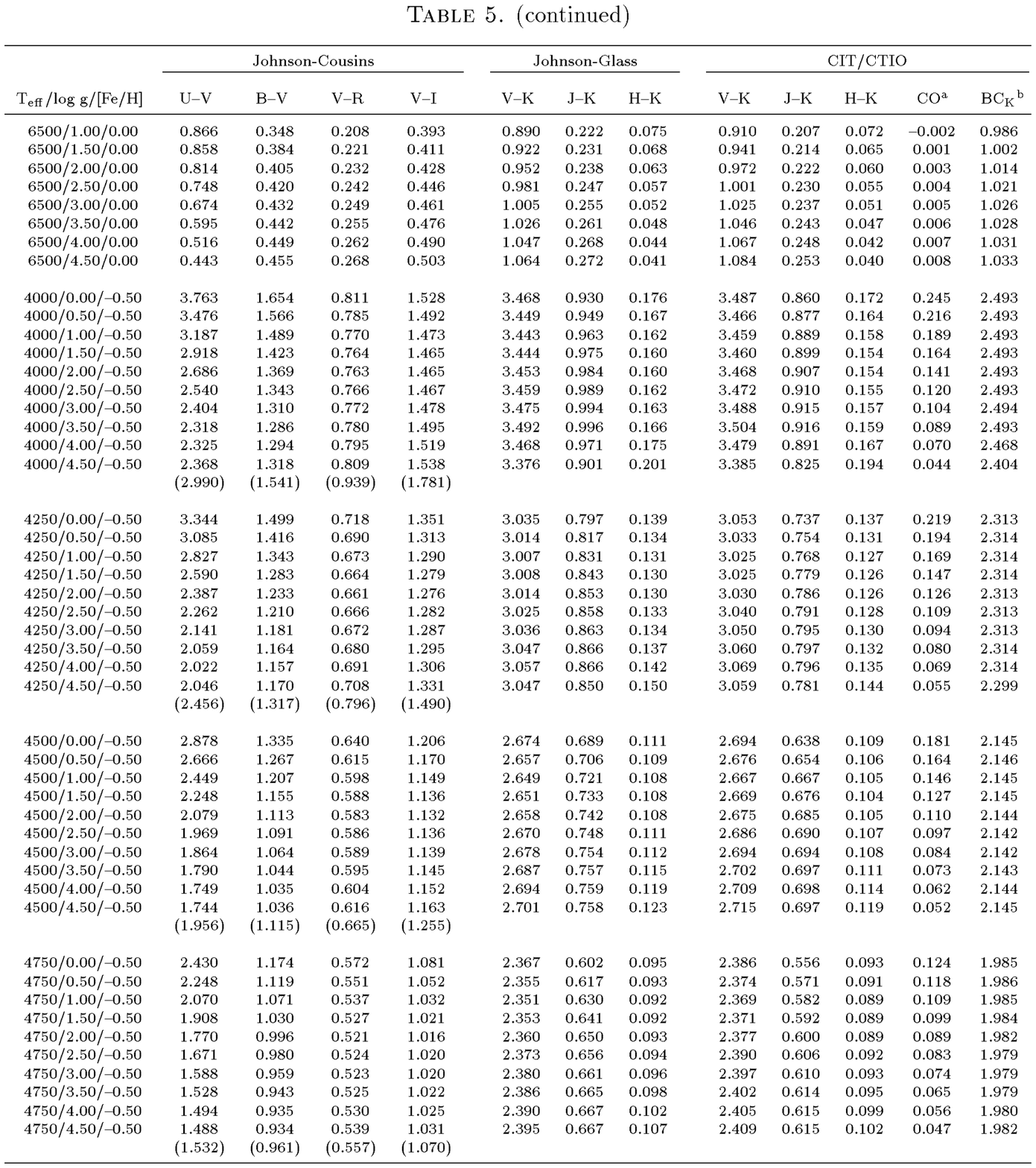}
\end{figure}

\begin{figure}[p]
\vspace*{-1.3in}
\hspace*{-1.0in}
\epsfbox{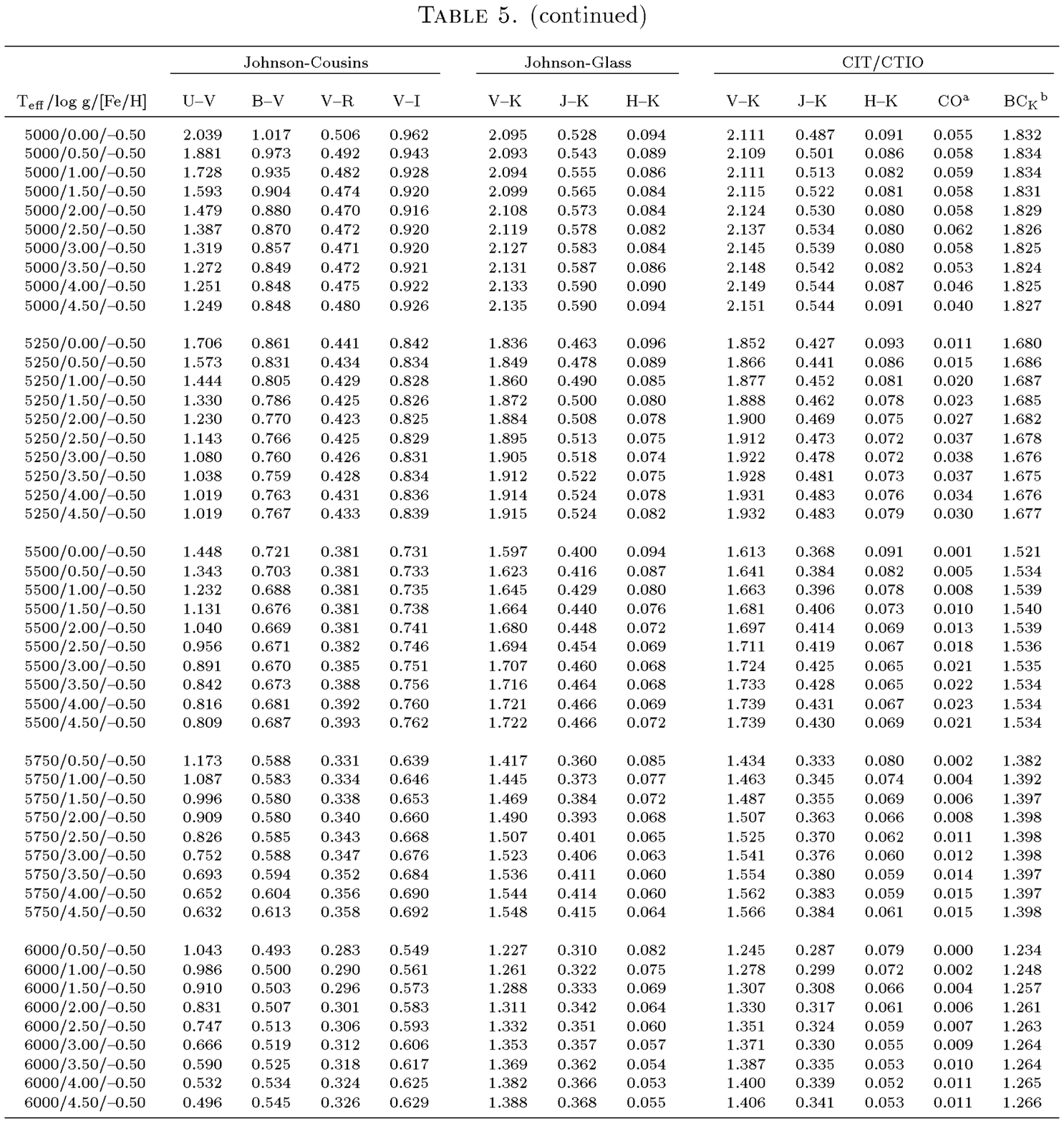}
\end{figure}

\begin{figure}[p]
\vspace*{-1.3in}
\hspace*{-1.0in}
\epsfbox{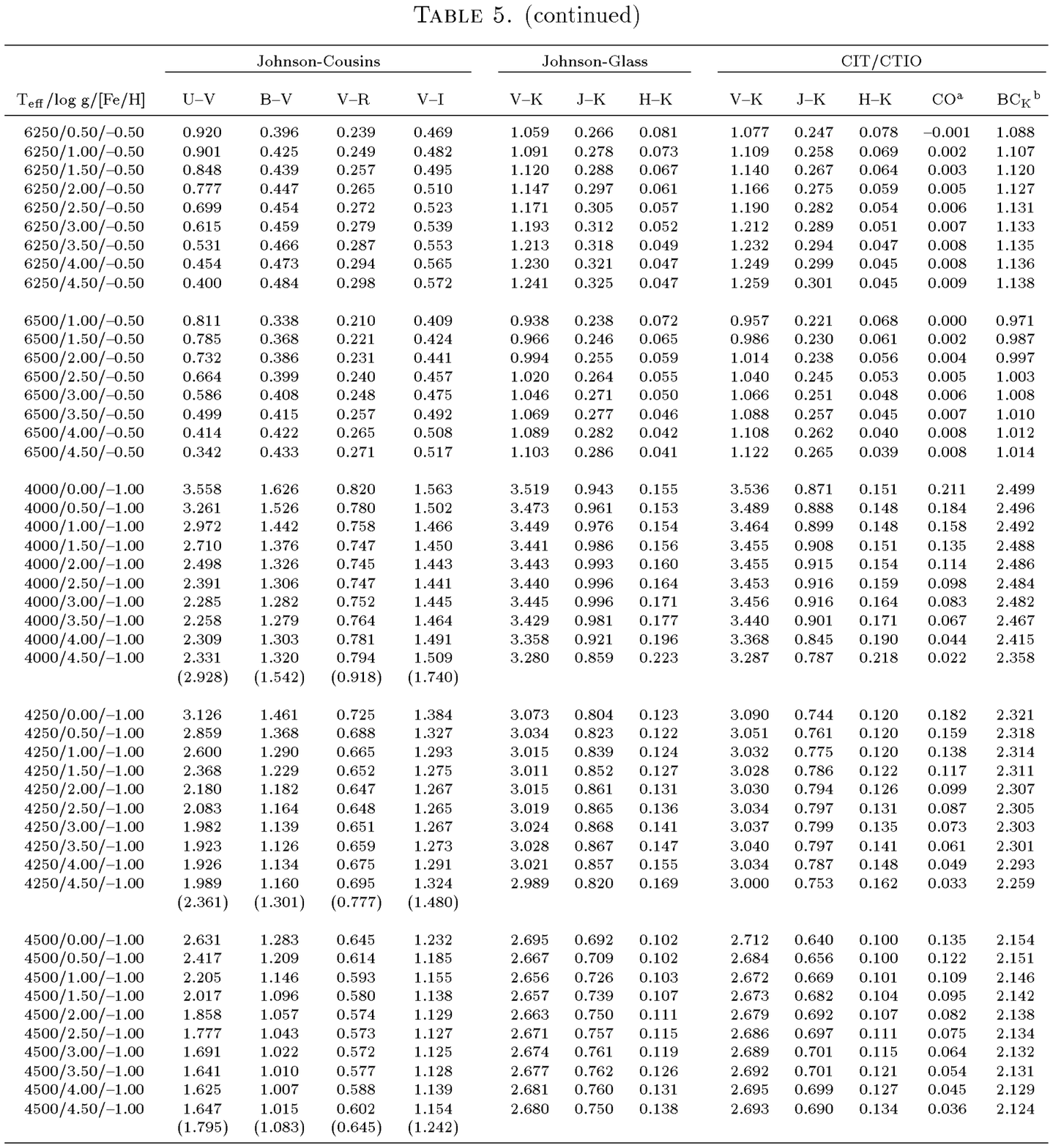}
\end{figure}

\begin{figure}[p]
\vspace*{-1.3in}
\hspace*{-1.0in}
\epsfbox{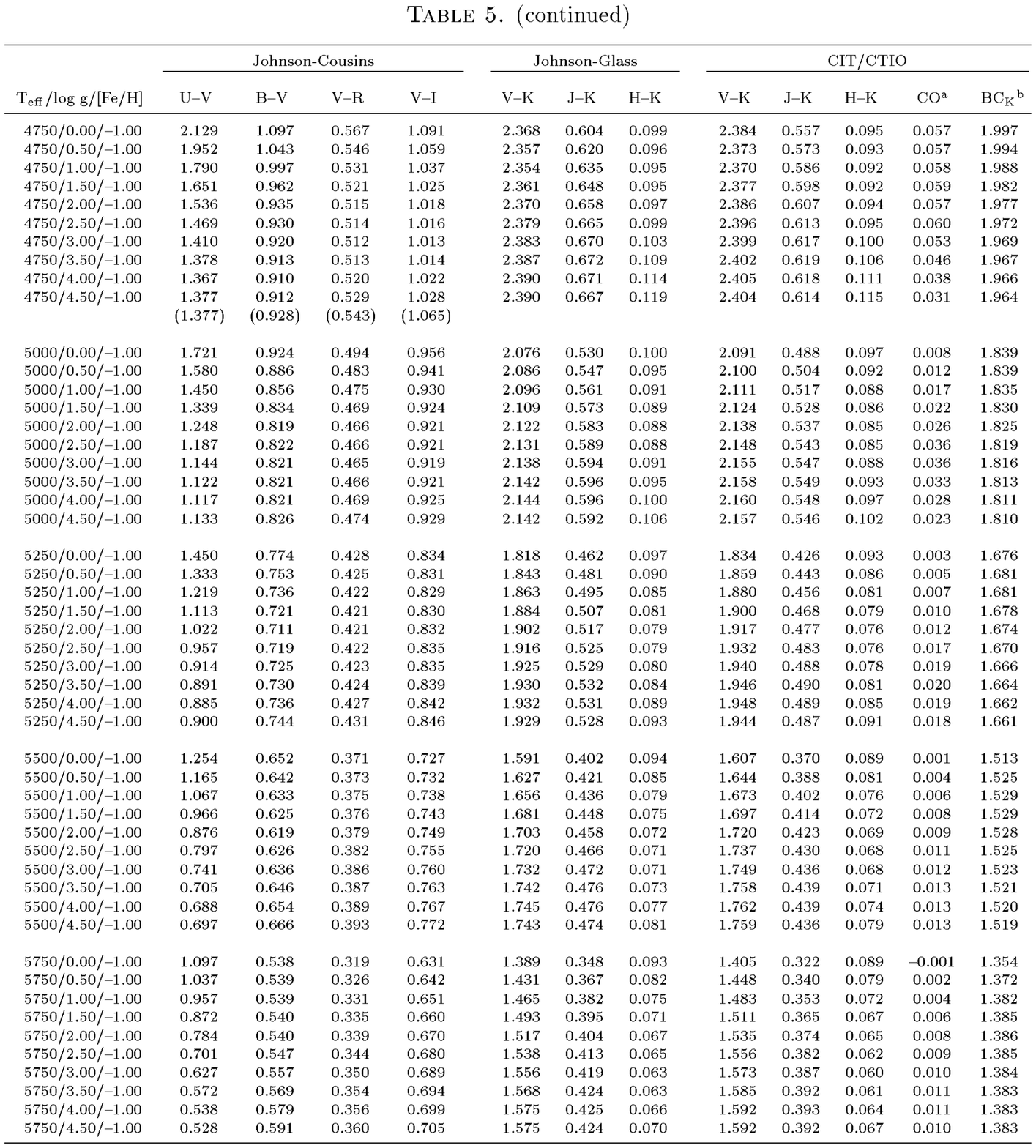}
\end{figure}

\begin{figure}[p]
\vspace*{-1.3in}
\hspace*{-1.0in}
\epsfbox{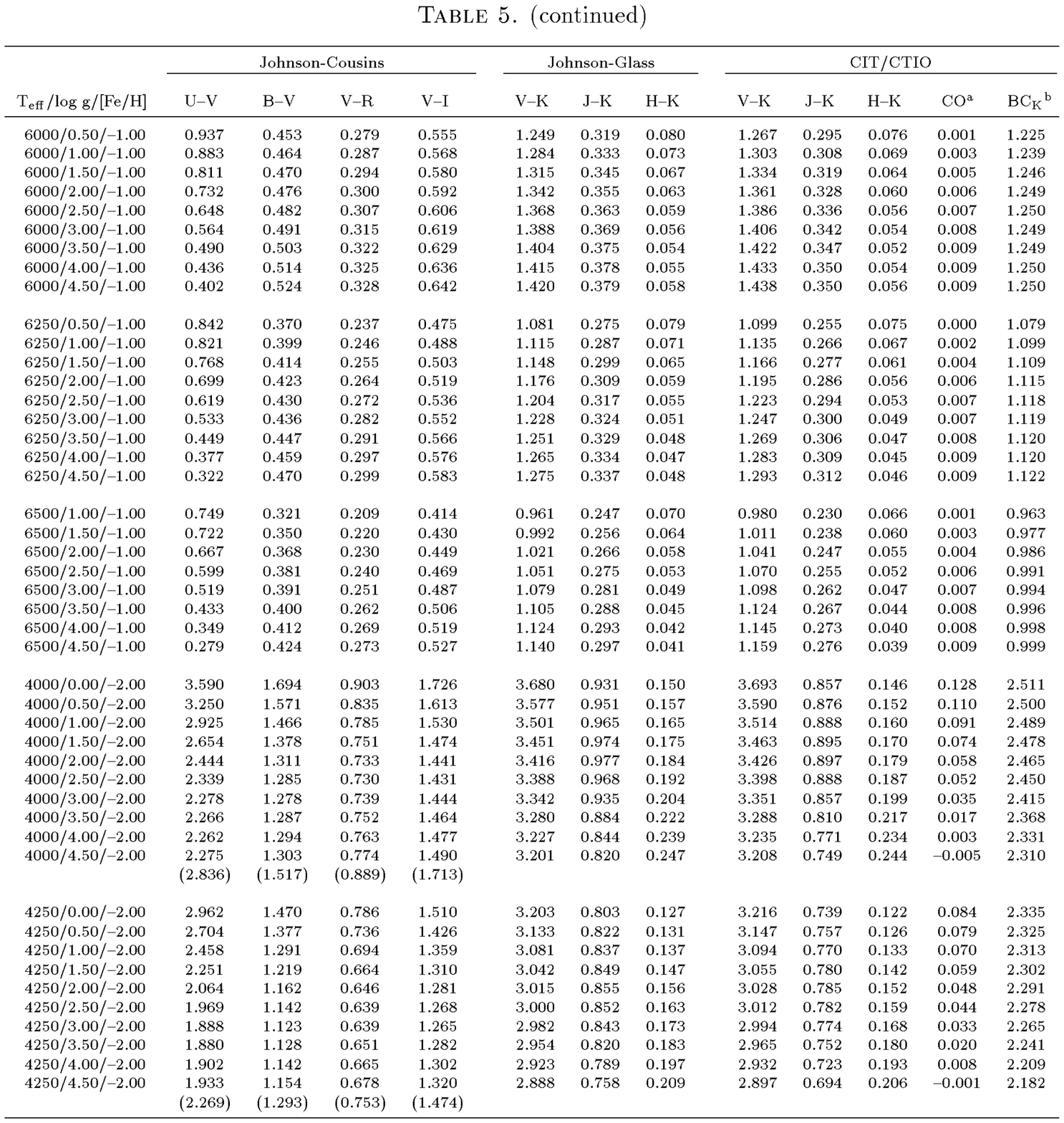}
\end{figure}

\begin{figure}[p]
\vspace*{-1.3in}
\hspace*{-1.0in}
\epsfbox{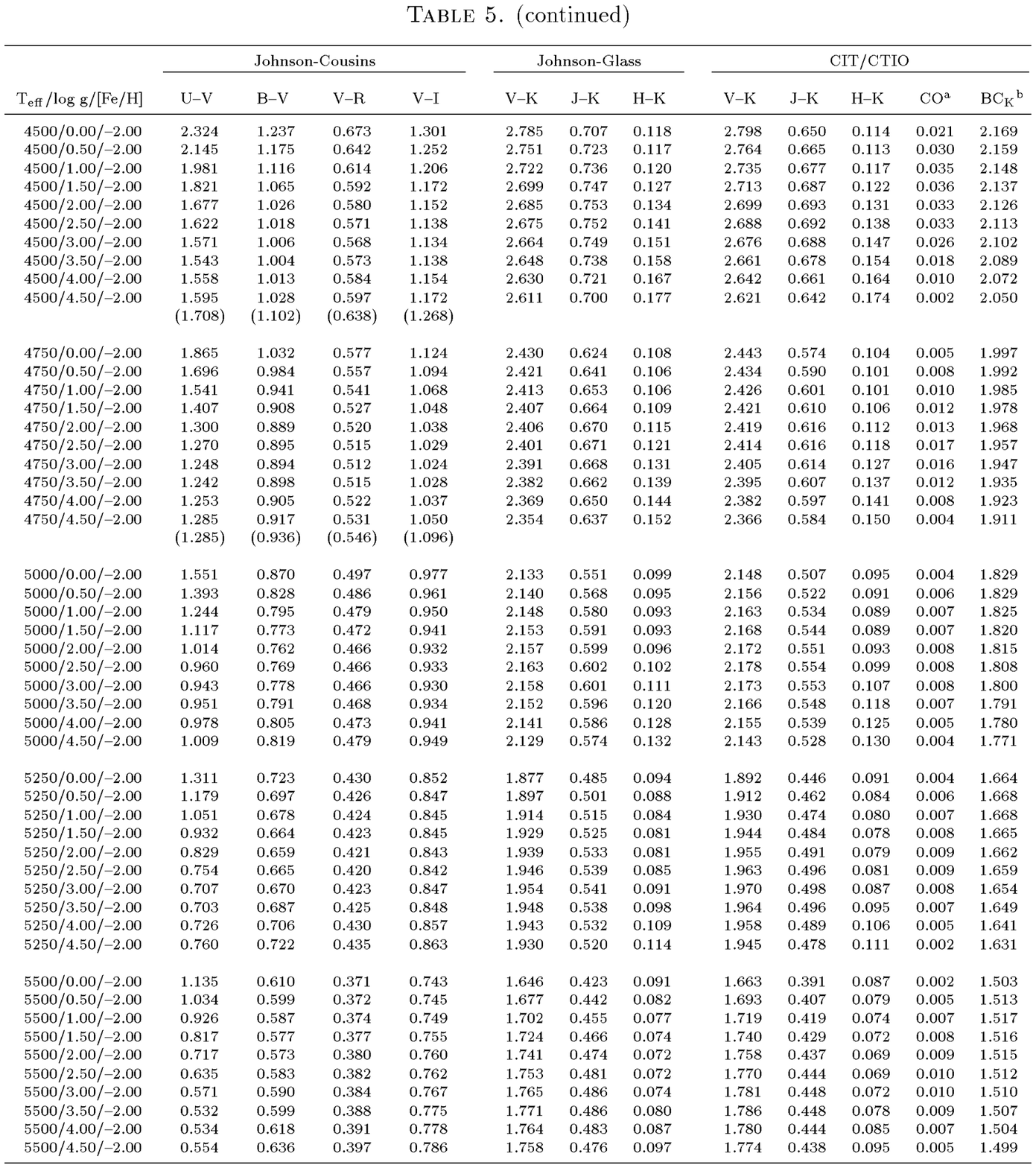}
\end{figure}

\clearpage

\begin{figure}[p]
\vspace*{-1.3in}
\hspace*{-1.0in}
\epsfbox{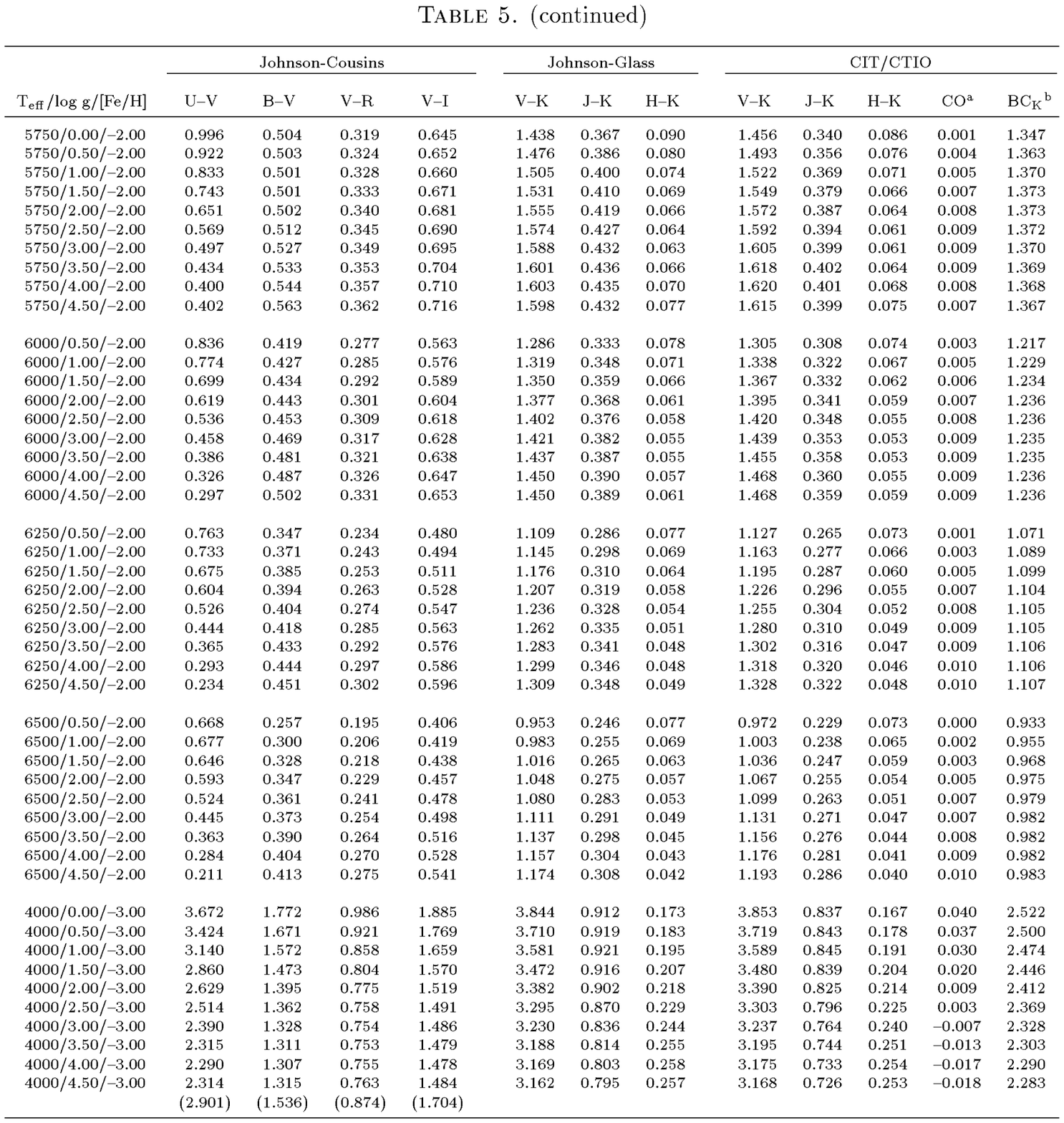}
\end{figure}

\begin{figure}[p]
\vspace*{-1.3in}
\hspace*{-1.0in}
\epsfbox{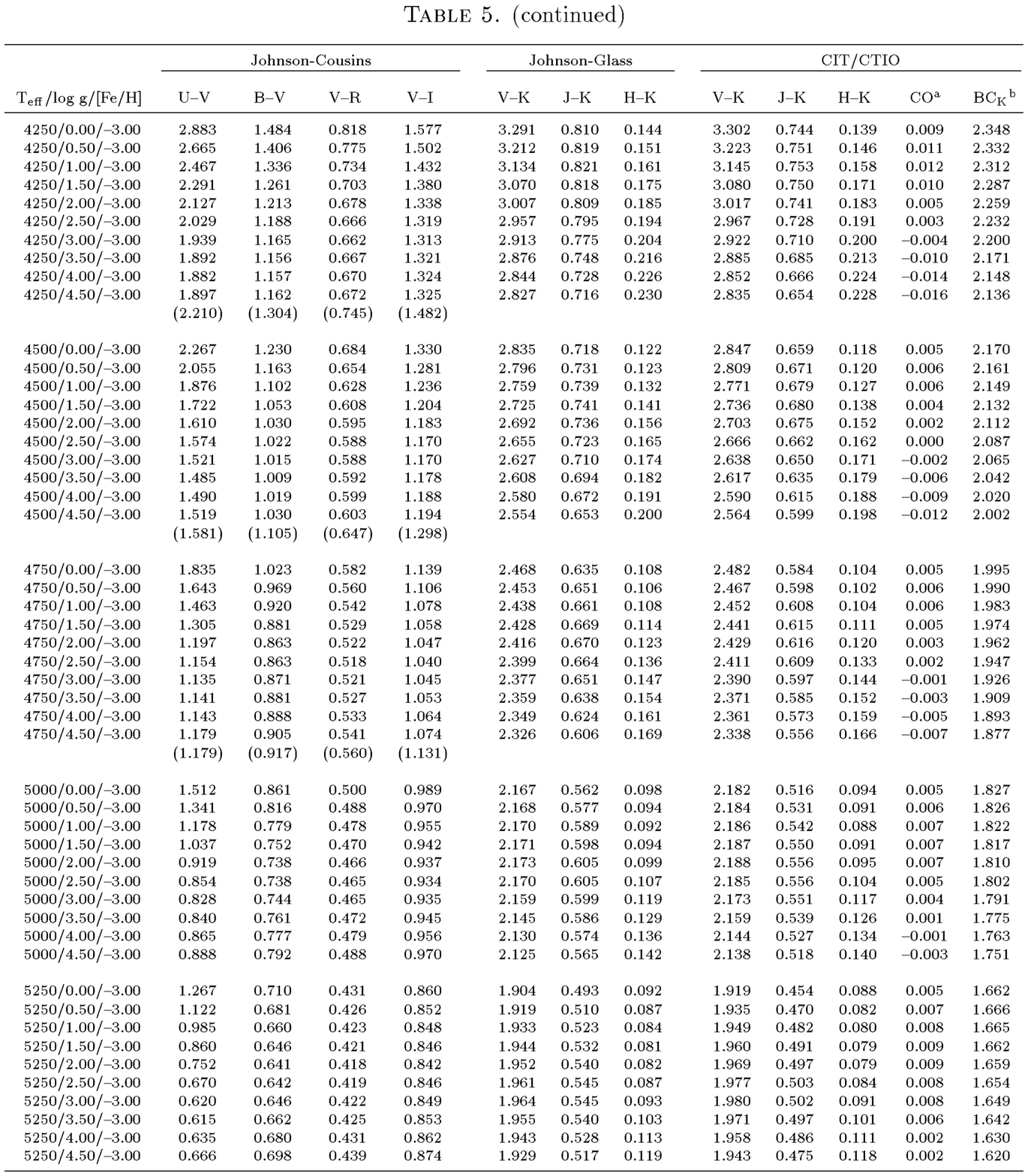}
\end{figure}

\begin{figure}[p]
\vspace*{-1.3in}
\hspace*{-1.0in}
\epsfbox{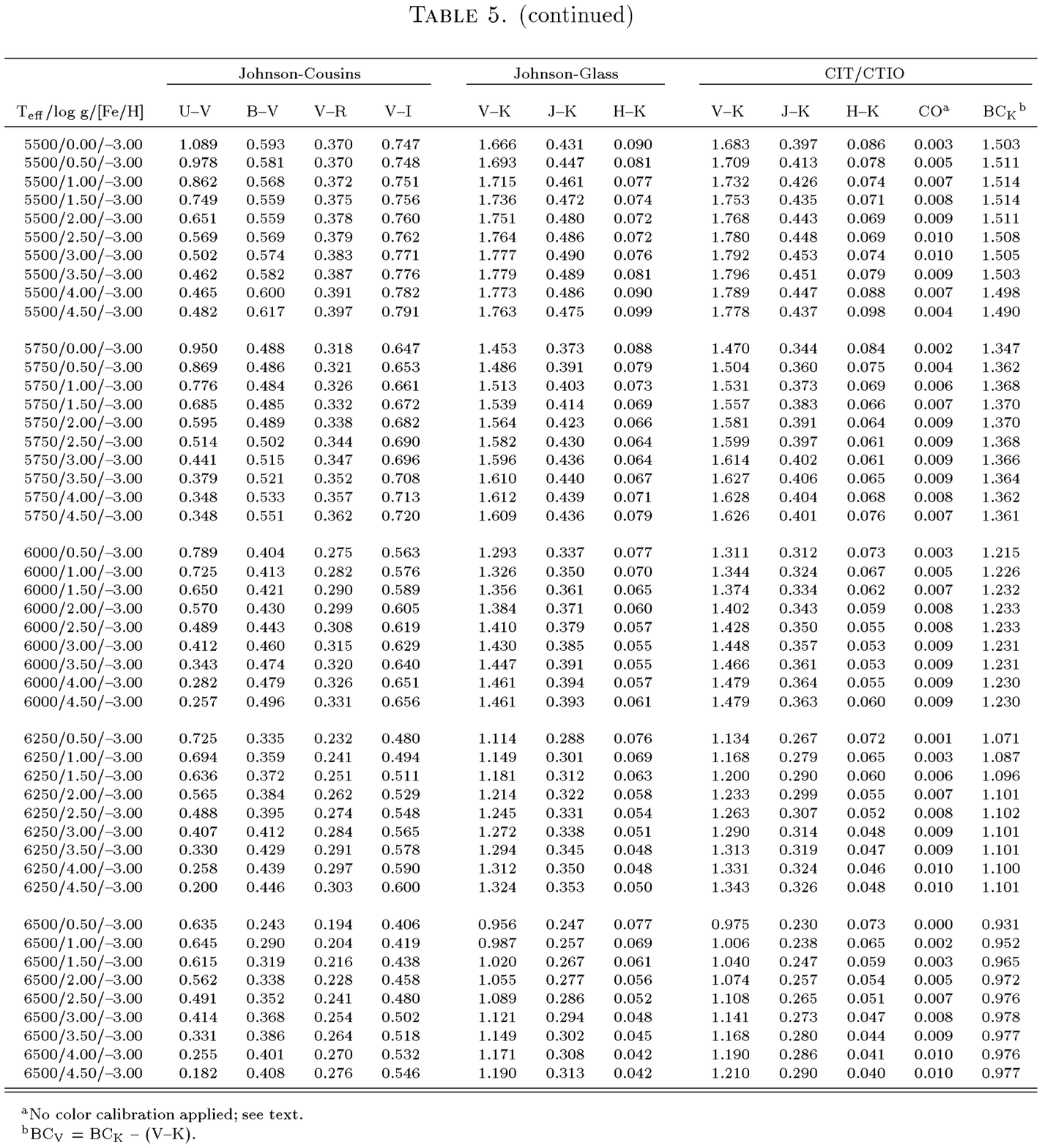}
\end{figure}

\begin{figure}[p]
\vspace*{-1.5in}
\hspace*{-1.0in}
\epsfbox{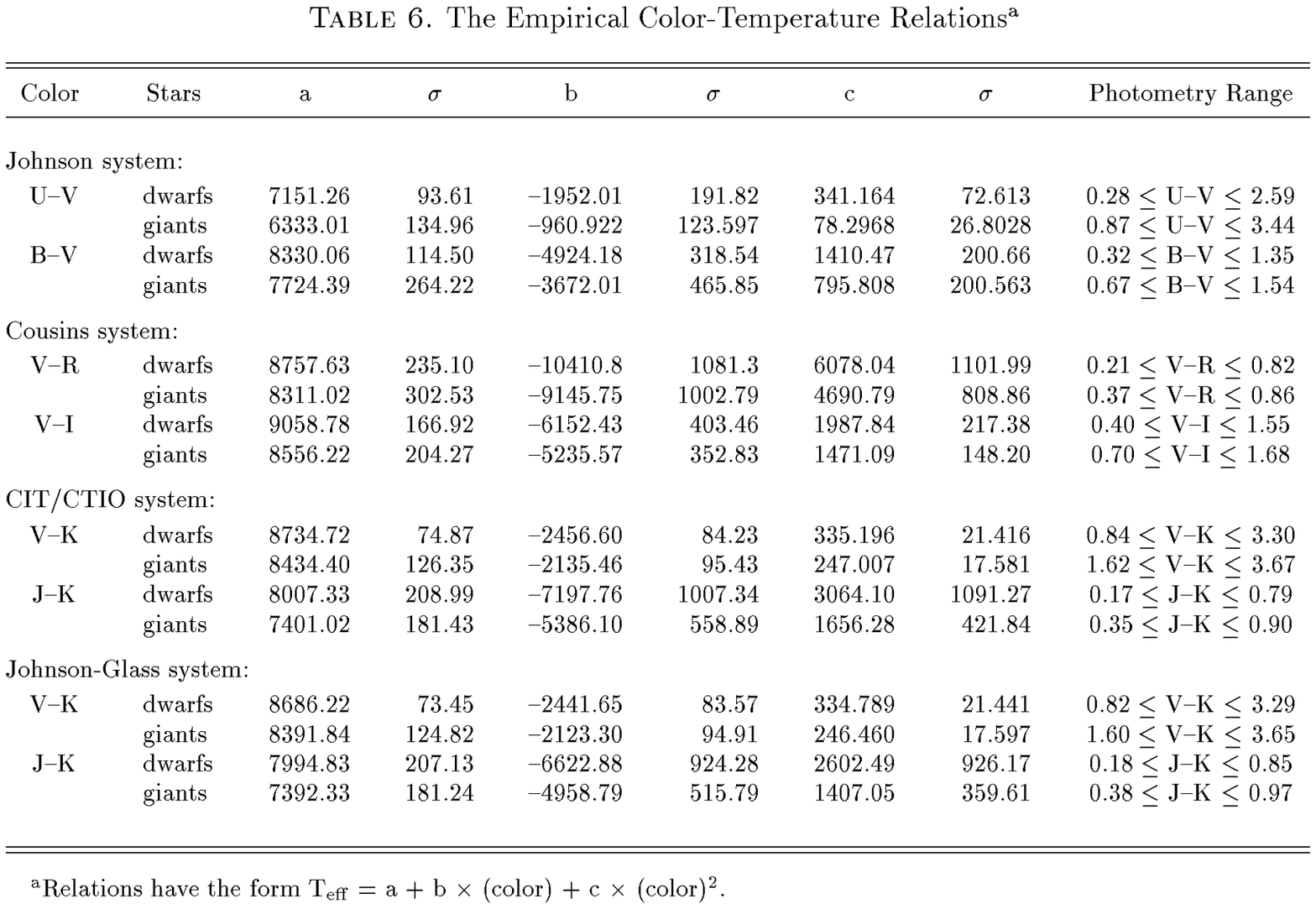}
\end{figure}


\begin{thebibliography}{}

\bibitem[Alonso et al.]{aam96} Alonso, A., Arribas, S., \& Mart\'{i}nez-Roger, C. 1996, \aaps, 117, 227
\bibitem[A\v{z}usienis \& Strai\v{z}ys 1969]{as69} A\v{z}usienis, A., \& Strai\v{z}ys, V. 1969, Soviet Astron. J., 13, 316
\bibitem[Balachandran \& Bell (1998)]{bb98} Balachandran, S. C., \& Bell, R. A. 1998, \nat, 392, 791
\bibitem[Bautista (1997)]{bautista} Bautista, M. A. 1997, \aaps, 122, 167
\bibitem[Bell \& Berrington (1987)]{bb87} Bell, K. L., \& Berrington, K. A. 1987, J. Phys. B, 20, 801
\bibitem[Bell et al. (1975)]{bkm75} Bell, K. L., Kingston, A. E., \& McIlveen, W. A. 1975, J. Phys. B, 8, 659
\bibitem[Bell (1997)]{bell97} Bell, R. A. 1997, in IAU Symp. 189, Fundamental Stellar Properties: The Interaction Between Observation and Theory, ed. T. R. Bedding, A. J. Booth, \& J. Davis (Dordrecht: Kluwer), 159
\bibitem[Bell et al. (1979)]{bell79} Bell, R. A., Dwivedi, P. H., Branch, D., \& Huffaker, J. N. 1979, \apjs, 41, 593
\bibitem[Bell et al. 1976]{begn} Bell, R. A., Eriksson, K., Gustafsson, B., \& Nordlund, \AA. 1976, \aaps, 23, 37
\bibitem[BG78]{bg78} Bell, R. A., \& Gustafsson, B. 1978, \aaps, 34, 229 (BG78)
\bibitem[BG89]{bg89} Bell, R. A., \& Gustafsson, B. 1989, \mnras, 236, 653 (BG89)
\bibitem[BPT94]{bpt94} Bell, R. A., Paltoglou, G., \& Tripicco, M. J. 1994, \mnras, 268, 771 (BPT94)
\bibitem[Bell \& Tripicco (1995)]{bt95} Bell, R. A., \& Tripicco, M. J. 1995, in ASP Conf. Ser. 78, Astrophysical Applications of Powerful New Databases, ed. S. J. Adelman, \& W. L. Wiese (San Francisco: ASP), 365
\bibitem[Bergbusch \& VandenBerg 1992]{bvb92} Bergbusch, P. A., \& VandenBerg, D. A. 1992, \apjs, 81, 163
\bibitem[B79]{bess79} Bessell, M. S. 1979, \pasp, 91, 589 (B79)
\bibitem[Bessell (1983)]{bess83} Bessell, M. S. 1983, \pasp, 95, 480
\bibitem[Bessell (1990)]{bsl90} Bessell, M. S. 1990, \pasp, 102, 1181
\bibitem[B95]{bess95} Bessell, M. S. 1995, in The Bottom of the Main Sequence -- And Beyond, ed. C. Tinney (Berlin: Springer-Verlag), 123 (B95)
\bibitem[B98]{bsl98} Bessell, M. S. 1998, private communication (B98)
\bibitem[Bessell \& Brett (1988)]{bb88} Bessell, M. S., \& Brett, J. M. 1988, \pasp, 100, 1134
\bibitem[BCP98]{bcp98} Bessell, M. S., Castelli, F., \& Plez, B. 1998, \aap, 333, 231 (BCP98)
\bibitem[Bi\'{e}mont et al. (1985a,]{bie85a} Bi\'{e}mont, E., Brault, J. W., Delbouille, L., \& Roland, G. 1985a, \aaps, 61, 107
\bibitem[1985b,]{bie85b} Bi\'{e}mont, E., Brault, J. W., Delbouille, L., \& Roland, G. 1985b, \aaps, 61, 185
\bibitem[1986)]{bie86} Bi\'{e}mont, E., Brault, J. W., Delbouille, L., \& Roland, G. 1986, \aaps, 65, 21
\bibitem[BLG94]{blg94} Blackwell, D. E., \& Lynas-Gray, A. E. 1994, \aap, 282, 899 (BLG94)
\bibitem[Blackwell et al. (1990)]{black90} Blackwell, D. E., Petford, A. D., Arribas, S., Haddock, D. J., \& Selby, M. J. 1990, \aap, 232, 396
\bibitem[Blackwell \& Shallis (1977)]{bs77} Blackwell, D. E., \& Shallis, M. J. 1977, \mnras, 180, 177
\bibitem[Bonnell \& Bell (1993a]{bb93a} Bonnell, J. T., \& Bell, R. A. 1993a, \mnras, 264, 319
\bibitem[1993b)]{bb93b} Bonnell, J. T., \& Bell, R. A. 1993b, \mnras, 264, 334
\bibitem[Briley et al. (1994)]{briley} Briley, M. M., Hesser, J. E., Bell, R. A., Bolte, M., \& Smith, G. H. 1994, \aj, 108, 2183
\bibitem[Buser \& Kurucz (1992)]{bk92} Buser, R., \& Kurucz, R. L. 1992, \aap, 264, 557
\bibitem[Charbonnel (1994,]{c94} Charbonnel, C. 1994, \aap, 282, 811
\bibitem[1995)]{c95} Charbonnel, C. 1995, \apjl, 453, L41
\bibitem[Charbonnel et al. (1998)]{c98} Charbonnel, C., Brown, J. A., \& Wallerstein, G. 1998, \aap, 332, 204
\bibitem[Chevalier \& Ilovaisky (1991)]{chev} Chevalier, C., \& Ilovaisky, S. A. 1991, \aaps, 90, 225
\bibitem[Cohen et al. (1981)]{cohen} Cohen, J. G., Frogel, J. A., Persson, S. E., \& Elias, J. H. 1981, \apj, 249, 481
\bibitem[Cousins (1980)]{cuz80} Cousins, A. W. J. 1980, South African Astron. Obs. Circ., 1, 234
\bibitem[Davis et al. (1978)]{dav78} Davis, D. S., Andrew, K. L., \& Verges, J. 1978, J. Opt. Soc. Am., 68, 235
\bibitem[Delbouille et al. (1973)]{del73} Delbouille, L., Roland, G., \& Neven, L. 1973, Photometric Atlas of the Solar Spectrum from $\lambda$~3000 to $\lambda~$10000, Institut d'Astrophysique de l'Universit\'{e} de Li\`{e}ge, Observatoire royal de Belgique
\bibitem[Demarque et al. 1992]{dem92} Demarque, P., Green, E. M., \& Guenther, D. B. 1992, \aj, 103, 151
\bibitem[di Benedetto \& Rabbia (1987)]{dib87} Di Benedetto, G. P., \& Rabbia, Y. 1987, \aap, 188, 114
\bibitem[Dinescu et al. 1995]{din95} Dinescu, D. I., Demarque, P., Guenther, D. B., \& Pinsonneault, M. H. 1995, \aj, 109, 2090
\bibitem[Doughty \& Fraser (1966)]{df66} Doughty, N. A., \& Fraser, P. A. 1966, \mnras, 132, 267
\bibitem[Dragon \& Mutschlecner (1980)]{dragon} Dragon, J. N., \& Mutschlecner, J. P. 1980, \apj, 239, 1045
\bibitem[Dreiling \& Bell 1980]{db80} Dreiling, L. A., \& Bell, R. A. 1980, \apj, 241, 736
\bibitem[Dyck et al. (1996)]{dbbr} Dyck, H. M., Benson, J. A., van Belle, G. T., \& Ridgway, S. T. 1996, \aj, 111, 1705
\bibitem[Edvardsson et al. (1993)]{edvard} Edvardsson, B., Andersen, J., Gustafsson, B., Lambert, D. L., Nissen, P. E., \& Tomkin, J. 1993, \aap, 275, 101
\bibitem[Elias et al. (1982)]{elias} Elias, J. H., Frogel, J. A., Matthews, K., \& Neugebauer, G.  1982, \aj, 87, 1029
\bibitem[Engels et al. (1981)]{eng81} Engels, D., Sherwood, W. A., Wamsteker, W., \& Schultz, G. V. 1981, \aaps, 45, 5
\bibitem[Fan et al. 1996]{fan} Fan, X., Burstein, D., Chen, J.-S., Zhu, J., Jiang, Z., Wu, H., Yan, H., Zheng, Z., Zhou, X., Fang, L.-Z., Chen, F., Deng, Z., Chu, Y., Hester, J. J., Windhorst, R. A., Li, Y., Lu, P., Sun, W.-H., Chen, W.-P., Tsay, W.-S., Chiueh, T.-H., Chou, C.-K., Ko, C.-M., Lin, T.-C., Guo, H.-J., \& Byun, Y.-I. 1996, \aj, 112, 628
\bibitem[Farmer \& Norton 1989]{fn89} Farmer, C. B., \& Norton, R. H. 1989, A High-Resolution Atlas of the Infrared Spectrum of the Sun and the Earth Atmosphere from Space, Vol. I: The Sun (NASA RP-1224) (Washington: NASA)
\bibitem[Forsberg (1991)]{fors91} Forsberg, P. 1991, \physscr, 44, 446
\bibitem[Frogel et al. (1978)]{fpam} Frogel, J. A., Persson, S. E., Aaronson, M., \& Matthews, K. 1978, \apj, 220, 75
\bibitem[Geller (1992)]{g92} Geller, M. 1992, A High-Resolution Atlas of the Infrared Spectrum of the Sun and the Earth Atmosphere from Space, Vol. III: Key to Identification of Solar Features (NASA RP-1224) (Washington: NASA)
\bibitem[Gilroy \& Brown 1991]{gilroy} Gilroy, K. K., \& Brown, J. A. 1991, \apj, 371, 578
\bibitem[Glass (1973)]{g73} Glass, I. S. 1973, \mnras, 164, 155
\bibitem[Glass (1974)]{g74} Glass, I. S. 1974, Mon. Notes Astron. Soc. South Africa, 33, 53
\bibitem[Goorvitch (1994)]{goor} Goorvitch, D. 1994, \apjs, 95, 535
\bibitem[Gratton (1998)]{gratton} Gratton, R. G. 1998, private communication
\bibitem[GCC96]{gcc96} Gratton, R. G., Carretta, E., \& Castelli, F. 1996, \aap, 314, 191 (GCC96)
\bibitem[Grevesse \& Noels (1993)]{grevesse} Grevesse, N., \& Noels, A. 1993, in Origin and Evolution of the Elements, ed. N. Pratzo, E. Vangioni-Flam, \& M. Casse (Cambridge: Cambridge), 15
\bibitem[GS83]{gs83} Gunn, J. E., \& Stryker, L. L. 1983, \apjs, 52, 121 (GS83)
\bibitem[Gustafsson \& Bell 1979]{gb79} Gustafsson, B., \& Bell, R. A. 1979, \aap, 74, 313
\bibitem[Gustafsson et al. 1975]{gben} Gustafsson, B., Bell, R. A., Eriksson, K., \& Nordlund, \AA. 1975, \aap, 42, 407
\bibitem[Hayes (1985)]{hayes} Hayes, D. S. 1985, in IAU Symp. 111, Calibration of Fundamental Stellar Quantities, ed. D. S. Hayes, L. E. Pasinetti, \& A. G. D. Philip (Dordrecht: Reidel), 225
\bibitem[Henyey et al. 1965]{henyey} Henyey, K., Vardya, M. S., \& Bodenheimer, P. 1965, \apj, 142, 841
\bibitem[Hinkle et al. 1995]{arcatlas} Hinkle, K., Wallace, L., \& Livingston, W. 1995, \pasp, 107, 1042
\bibitem[Hobbs \& Thorburn 1991]{hobbs} Hobbs, L. M., \& Thorburn, J. A. 1991, \aj, 102, 1070
\bibitem[Holweger (1970)]{holweger} Holweger, H. 1970, \aap, 4, 11
\bibitem[Holweger \& M\"{u}ller]{hm74} Holweger, H., \& M\"{u}ller, H. A. 1974, \solphys, 39, 19
\bibitem[Houdashelt et al. (1992)]{houdy1} Houdashelt, M. L., Frogel, J. A., \& Cohen, J. G. 1992, \aj, 103, 163
\bibitem[Paper~III]{houdy3} Houdashelt, M. L., Bell, R. A., \& Sweigart, A. V. 2001, in preparation (Paper~III)
\bibitem[Paper~II]{houdy2} Houdashelt, M. L., Bell, R. A., Sweigart, A. V., \& Wing, R. F. 2000, \aj, accepted (Paper~II)
\bibitem[Houk \& Cowley 1975]{mss} Houk, N., \& Cowley, A. P. 1975, Michigan Catalogue of Two-Dimensional Spectral Types for the HD Stars, Vol. 1 (Ann Arbor: Univ. Michigan Dept. Astron.)
\bibitem[Janes \& Smith 1984]{js84} Janes, K. A., \& Smith, G. H. 1984, \aj, 89, 487
\bibitem[Johansson \& Learner]{jl90} Johansson, S., \& Learner, R. C. M. 1990, \apj, 354, 755
\bibitem[Johnson (1965)]{j65} Johnson, H. L. 1965, \apj, 141, 923
\bibitem[Johnson et al. (1968)]{j68} Johnson, H. L., MacArthur, J. W., \& Mitchell, R. I. 1968, \apj, 152, 465
\bibitem[Johnson et al. (1966)]{j66} Johnson, H. L., Mitchell, R. I., Iriarte, B., \& Wi\'{s}niewski, W. Z. 1966, Comm. Lun. Plan. Lab, 4, 99
\bibitem[Joner \& Taylor (1990)]{jt90} Joner, M. D., \& Taylor, B. J. 1990, \pasp, 102, 1004
\bibitem[Kj{\ae}rgaard et al. 1982]{kgwh} Kj{\ae}rgaard, P., Gustafsson, B., Walker, G. A. H., \& Hultqvist, L. 1982, \aap, 115, 145
\bibitem[Kurucz (1991)]{k91} Kurucz, R. L. 1991, in Stellar Atmospheres: Beyond Classical Models, ed. L. Crivellari, I. Hubeny, \& D. G. Hummer (NATO ASI Ser. C, 341) (Dordrecht: Kluwer), 408
\bibitem[Kurucz et al. 1987]{kur87} Kurucz, R. L., van Dishoeck, E. F., \& Tarafdar, S. P. 1987, \apj, 322, 992
\bibitem[Lee (1970)]{lee70} Lee, T. A. 1970, \apj, 162, 217
\bibitem[Litzen et al. (1993)]{litz93} Litzen, U., Brault, J. W., \& Thorne, A. P. 1993, \physscr, 47, 628
\bibitem[Livingston \& Wallace 1991]{lw91} Livingston, W., \& Wallace, L. 1991, An Atlas of the Solar Spectrum in the Infrared from 1850 to 9000 cm$^{-1}$ (1.1 to 5.4 microns) (N.S.O. Technical Report \#91-001, July 1991)
\bibitem[Matthews \& Sandage (1963)]{ms63} Matthews, T. A., \& Sandage, A. R. 1963, \apj, 138, 30
\bibitem[Mermilliod (1991)]{merm91} Mermilliod, J.-C. 1991, ADC CD-ROM, Selected Astronomical Catalogs, Vol. 1
\bibitem[Meynet et al. 1993]{meynet} Meynet, G., Mermilliod, J.-C., \& Maeder, A. 1993, \aaps, 98, 477
\bibitem[Montgomery et al. 1993]{mont} Montgomery, K. A., Marschall, L. A., \& Janes, K. A. 1993, \aj, 106, 181
\bibitem[M97]{moz97} Mozurkewich, D. 1997, in Poster Proceedings of IAU Symposium 189 on Fundamental Stellar Properties: The Interaction Between Observation and Theory, ed. T. R. Bedding (Sydney: Univ. Sydney), 14 (M97)
\bibitem[Mozurkewich et al. (1991)]{moz91} Mozurkewich, D., Johnston, K. J., Simon, R. S., Bowers, P. F., \& Gaume, R. 1991, \aj, 101, 2207
\bibitem[Nave et al. (1994)]{nave94} Nave, G., Johansson, S., Learner, R. C. M., Thorne, A. P., \& Brault, J. W. 1994, \apjs, 94, 221
\bibitem[Nissen et al. 1987]{nissen} Nissen, P. E., Twarog, B. A., \& Crawford, D. L. 1987, \aj, 93, 634
\bibitem[Nordstr\"{o}m et al. 1997]{nord97} Nordstr\"{o}m, B., Andersen, J., \& Andersen, M. I. 1997, \aap, 322, 460
\bibitem[O'Brian et al. (1991)]{ob91} O'Brian, T. R., Wickliffe, M. E., Lawler, J. E., Whaling, W., \& Brault, J. W. 1991, J. Opt. Soc. Am. B, 8, 1185
\bibitem[Paltoglou \& Bell (1991)]{pb91} Paltoglou, G., \& Bell, R. A. 1991, \mnras, 253, 449
\bibitem[Pauls (1998)]{p98} Pauls, T. A. 1998, private communication
\bibitem[Pauls 1999]{pauls99} Pauls, T. A. 1999, private communication
\bibitem[PMAHH]{pauls} Pauls, T. A., Mozurkewich, D., Armstrong, J. T., Hajian, A. R., \& Hummel, C. A. 1997, \baas, 29, 1231 (PMAHH)
\bibitem[Perrin et al. (1998)]{perrin} Perrin, G., Coud\'{e} du Foresto, V., Ridgway, S. T., Mariotti, J.-M., Traub, W. A., Carleton, N. P., \& Lacasse, M. G. 1998, \aap, 331, 619
\bibitem[Persson (1980)]{pers} Persson, S. E. 1980, private communication
\bibitem[Ramsauer et al. (1995)]{rams95} Ramsauer, J., Solanki, S. K., \& Bi\'{e}mont, E. 1995, \aaps, 113, 71
\bibitem[Ridgway et al. (1980)]{rjww} Ridgway, S. T., Joyce, R. R., White, N. M., \& Wing, R. F. 1980, \apj, 235, 126
\bibitem[Rosvick \& VandenBerg 1998]{rv98} Rosvick, J. M., \& VandenBerg, D. A. 1998, \aj, 115, 1516
\bibitem[Rufener \& Nicolet 1988]{ruf88} Rufener, F., \&  Nicolet, B. 1988, \aap, 206, 357
\bibitem[SH85]{sh85} Saxner, M., \& Hammarb$\ddot{\rm a}$ck, G. 1985, \aap, 151, 372 (SH85)
\bibitem[Smith \& Lambert 1990]{sl90} Smith, V. V., \& Lambert, D. L. 1990, \apjs, 72, 387
\bibitem[Solanki et al. (1990)]{sol90} Solanki, S. K., Bi\'{e}mont, E., \& M\"{u}rset, U. 1990, \aaps, 83, 307
\bibitem[Sweigart (1997)]{sweigart} Sweigart, A. V. 1997, \apjl, 474, L23
\bibitem[Sweigart \& Mengel (1979)]{sweigart79} Sweigart, A. V., \& Mengel, J. G. 1979, \apj, 229, 624
\bibitem[Taklif (1990)]{tak90} Taklif, A. G. 1990, \physscr, 42, 69
\bibitem[Taylor \& Joner 1990]{tj90} Taylor, B. J., \& Joner, M. D. 1990, \aj, 100, 830
\bibitem[Tripicco \& Bell (1991)]{tb91} Tripicco, M. J., \& Bell, R. A. 1991, \aj, 102, 744
\bibitem[VandenBerg (1999)]{vb99} VandenBerg, D. A. 1999, private communication
\bibitem[VB85]{vb85} VandenBerg, D. A., \& Bell, R. A. 1985, \apjs, 58, 561 (VB85)
\bibitem[Woods et al. 1996]{woods} Woods, T. N., et al 1996, \jgr, D6, 9541
\bibitem[Worthey 1994]{worthey} Worthey, G. 1994, \apjs, 95, 107

\end{thebibliography}
\end{document}